\documentclass[12pt]{article}
\usepackage[utf8]{inputenc}
\usepackage[T1]{fontenc}
\usepackage{amsmath,amssymb,amsfonts,amsthm,mathtools,bm}
\usepackage{geometry}
\usepackage{booktabs,threeparttable,multirow,array}
\usepackage{caption,subcaption}
\usepackage{setspace}
\usepackage{natbib}
\usepackage{hyperref}
\usepackage{enumitem}
\usepackage{pgfplots}
\usepackage{tikz}
\usepackage{float}
\usepackage{graphicx}
\usepackage{adjustbox}
\usepackage{tabularx}
\usepackage{rotating}
\usepackage{pdflscape}
\usepackage{setspace}
\pgfplotsset{compat=1.17}
\newcommand{\safeincludegraphics}[2][]{  \IfFileExists{#2}{\includegraphics[#1]{#2}}{    \fbox{\parbox{0.88\linewidth}{\centering Missing external figure file: \texttt{\detokenize{#2}}.\\ The original \texttt{figs\_tabs.tex} source referenced this file.}}  }}
\graphicspath{{figures/}{beta_varies/}{likelihood_ratio_plots/}}
\hypersetup{hypertexnames=false,
colorlinks=true,
linkcolor=blue,
citecolor=blue,
urlcolor=blue
}
\newtheorem{theorem}{Theorem}
\newtheorem{lemma}{Lemma}
\newtheorem{proposition}{Proposition}

\newtheorem{assumption}{Assumption}

\newcommand{\vect}{\operatorname{vec}}
\DeclareMathOperator{\rank}{rank}
\newcommand{\adj}{\operatorname{adj}}

\providecommand{\func}[1]{\operatorname{#1}}

\begin{document}

\title{An Algebraic Approach to Continuously Updating GMM: Estimation and Inference}
\author{Marcelo J. Moreira\thanks{
FGV EPGE. Email: mjmoreira@fgv.br.} \and Whitney K. Newey\thanks{
MIT Department of Economics. Email: wnewey@mit.edu.} \and Mahrad
Sharifvaghefi\thanks{
University of Pittsburgh. Email: sharifvaghefi@pitt.edu.}}
\date{\today }
\maketitle

\begin{abstract}
\normalsize
The continuously updating GMM estimator (CUE) minimizes a nonconvex objective. We give an algebraic solution in linear instrumental variables models with a fixed joint variance estimate that allows for heteroskedasticity, autocorrelation, or clustering. With one endogenous regressor and a positive definite variance estimate, the
objective is a ratio of polynomials. Its global minimum is obtained by comparing the objective at no more than  $4k-2$ real stationary points and at infinity, with $k$ instruments. The roots can be computed as eigenvalues of a companion matrix. We extend the results to singular variance estimates and characterize the multivariate minimum using finitely many stationary objective values. These results provide methods for computing CUE, overidentification statistics, and likelihood ratio-based tests. Simulations show that local optimization can lead to substantially different estimates and test decisions. The algebraic results are exact, although their numerical implementation requires care in computing polynomial coefficients and roots.
\end{abstract}

\noindent\textbf{Keywords:} continuously updating GMM, instrumental
variables, weak instruments, conditional likelihood ratio, polynomial roots,
companion matrix, Gr\"{o}bner basis.

\noindent\textbf{JEL codes:} C12, C26, C36.

\clearpage

\section{Introduction}

The continuously updating GMM estimator (CUE) minimizes the GMM objective function simultaneously over the parameters in the moment functions and in the weighting matrix.
In instrumental variable (IV) settings, CUE is known to have lower bias than two step GMM, which minimizes the GMM objective over just the parameters in the moment functions. 
This CUE was given and its low bias property shown in important examples in \citet{HansenHeatonYaron96}.
The CUE was shown to have smaller asymptotic bias than two step GMM when there are many instrumental variables by \cite{NeweySmith04} and to be asymptotically normal under many weak instruments by \cite{NeweyWindmeijer09}. 
\cite{Kleibergen05} gave an approach to inference under weak identification based on the CUE. 
All of these results suggest how the CUE could be a useful alternative to two step GMM.

Unfortunately, the CUE can be difficult to compute, with an objective function that is very flat over a wide range of parameter values (see \citet{HansenHeatonYaron96}). 
In this paper we solve this problem using an algebraic approach for linear IV models. 
We show that for one endogenous variable the CUE objective function is a ratio of polynomials, so all flat points will be a root of a polynomial.
We use modern, fast algorithms for computing polynomial roots to find all flat points, and compute the CUE as the minimizer of the objective function over these and its value at extremes.  
In this way we solve the long standing problem of using gradient methods to try to compute the CUE that may fail to find the global minimum and so not have the relatively small bias of the CUE.  

In the classical linear instrumental variables model, conditional
homoskedasticity means that the reduced-form errors have a common variance
matrix that does not depend on the instruments. When these errors are also
conditionally uncorrelated across observations, the variance matrix of the
reduced-form model has a Kronecker product structure. Using the
usual reduced-form residual variance estimator, the continuously updating
GMM estimator coincides with the limited information maximum
likelihood estimator (LIML) of \citet{AndersonRubin49}; see also %
\citet{HansenHeatonYaron96}. The objective is a generalized Rayleigh
quotient. The minimum of the objective is given by the smallest generalized eigenvalue, and the estimated coefficients are given by the corresponding normalized eigenvector; see \citet{MagnusNeudecker88}. Section S-2 in the supplemental appendix provides this familiar characterization and discusses the case in which the minimizing direction is at infinity.

When the errors are heteroskedastic, autocorrelated, or clustered, the variance matrix of the reduced-form model need not follow a Kronecker product structure. The GMM framework of \citet{Hansen1982} allows more general variance matrices, and \citet{HansenHeatonYaron96} introduce CUE by updating the weighting matrix at each candidate parameter value. Without the Kronecker structure, the objective
need not be a generalized Rayleigh quotient, and local optimization may fail
to find its global minimum. This computational problem is particularly important with weak or many instruments, where conventional asymptotic approximations may be unreliable, as in \citet{StaigerStock97} or \citet{Bekker94}.

This paper provides an algebraic characterization of the global minimum. We first
consider a scalar structural coefficient and a fixed positive definite
estimate of the reduced-form variance matrix. The corresponding variance matrix of the sample moments is quadratic in the coefficient. The sample moments are also affine in the coefficient. So the CUE objective is a ratio of polynomials. From the first-order condition, we derive a polynomial equation with roots that correspond to the finite stationary points of the CUE objective. We find the maximum degree of this polynomial in terms of the number of instruments. Comparing the objective values at its real roots and at infinity characterizes the global minimum. This 
characterization does not require starting values
for local optimization or an arbitrarily bounded search region.

Computationally, finding these polynomial roots corresponds to finding the
eigenvalues of a companion matrix. This connects the general problem to the
classical LIML calculation, but the eigenvalues play different roles. In the homoskedastic problem, the smallest generalized eigenvalue is the minimized objective value, and the corresponding normalized eigenvector is the estimator of the coefficients. In the more general case where the errors are not necessarily homoskedastic, the real eigenvalues of the companion matrix are candidate coefficient values. A CUE is determined by comparing its objective at these eigenvalues and at infinity.

Classical algebraic characterizations of constrained extrema provided in \citet[Chapters~11 and~17]{MagnusNeudecker88} concern fixed quadratic forms subject to normalization or orthogonality restrictions. A general CUE objective does not follow this structure. Matrix inequalities, including Cauchy--Schwarz, H\"{o}lder, and Minkowski, provide bounds or alternative representations, but their equality conditions alone do not identify the structural coefficient that globally minimizes the CUE objective. In this paper, we provide an algebraic characterization of this global minimum and methods for computing it.

\citet{HansenLee21} study convergence of iterated GMM and inference under
misspecification. Each iteration minimizes an objective with the weight held
fixed at the preceding estimate. The limit is a fixed point of this map.
Under independence of observations and suitable moment and smoothness conditions, the sample map is a contraction with probability approaching to one. Their
asymptotic normality and inference results also require a nonsingular
variance matrix and do not cover weak identification. This fixed-point characterization concerns a different estimator. In the CUE objective, both the moments and the weight matrix vary as the coefficients change. Thus, there are terms that generally enter the first-order conditions for CUE but do not enter the first-order conditions for the iterated GMM estimator, for which the weight matrix is fixed at each iteration.
Consequently, convergence of iterated GMM need not deliver CUE, even when
the contraction conditions hold. In this paper, we  characterize the global minimum of the CUE objective
for the observed sample without requiring a contraction or strong
identification. The singular-variance extension discussed below also falls
outside their positive-definiteness conditions.

The correspondence between the stationary points of the CUE objective and the roots of the polynomial equation also allows us to determine whether a general formula using arithmetic operations and radicals can replace numerical root finding. While polynomial equations of degree at most four admit formulas using arithmetic operations and radicals, those of higher degree do not admit such formulas in general. However, degree alone does not rule out such a formula for a particular family of CUE objectives. Ruling out a formula using arithmetic operations and radicals requires an analysis of the Galois group.
Proposition S.2 considers an example of the CUE objective with a scalar coefficient and two instruments. The proposition shows that the polynomial derived from the first-order condition has degree six, is irreducible, and has the full symmetric Galois group. Every global
minimizer in this example is finite and cannot be expressed by radicals.
Thus there is no formula using radicals representing CUE for every admissible variance
matrix, even in this small model. Exact descriptions by polynomial equations
and exact algorithms for isolating real roots remain available.

Numerical eigenvalue computation is already part of the classical approach.
In the homoskedastic model, the degree of the characteristic polynomial for the generalized eigenvalue problem
is one greater than the number of endogenous regressors. Thus, formulas by radicals cover models with at most three endogenous regressors. For models with four or more endogenous regressors, the characteristic polynomial has degree at least five. No general formula by radicals is available for an unrestricted eigenvalue problem of that order. The relevant dimension is
that of the structural coefficient, not the number of instruments. The
eigenvector characterization itself remains valid in every dimension.
Likewise, in more general settings in which the errors are not necessarily homoskedastic, the roots of the polynomial derived from the first-order condition correspond to the eigenvalues of the companion matrix.
The value of this representation is that it reduces computation to a well-studied 
problem with established numerical methods; see \citet{AndersonEtAl1999}.

Finding the global minimum also allows inference. \citet{Moreira03} develops
the conditional likelihood ratio (CLR) test for linear IV models, and %
\citet{AndrewsMoreiraStock06} study its power relative to the envelope for
invariant similar tests in the homoskedastic Gaussian model. %
\citet{Kleibergen05} develops identification-robust GMM tests, while %
\citet{AndrewsMikusheva16} give a conditional inference framework for
general moment condition models. For the likelihood criterion considered
here, the LR statistic subtracts the global minimum of the CUE objective from its value under the null hypothesis. The same minimization must be performed for the
simulated processes used to obtain conditional critical values. %
\citet{MoreiraMoreira19} study conditional testing with heteroskedastic and
autocorrelated errors. In this setting, computing the full likelihood
criterion is relevant to the power comparisons in %
\citet{MoreiraRidderSharifvaghefi26}. The consequences of replacing a
continuous search by a finite grid are studied by %
\citet{MoreiraSharifvaghefi26}. Our results solve the minimization step.

The same minimum is the overidentification statistic for testing whether the
moment restrictions hold for some structural coefficient. When the conventional strong-identification conditions hold, the statistic under the null hypothesis has an asymptotic chi-square distribution with degrees of freedom equal to the number of overidentifying restrictions; see \citet{Hansen1982}. Weak identification can invalidate this asymptotic approximation, but
we can use a conservative chi-square critical value with degrees of freedom equal to the number of
moments. A local minimum can overstate the overidentification statistic and
cause the test to reject the null hypothesis too often. Therefore, it is important to compute the global minimum for both parameter inference and specification testing.

The scalar characterization extends to singular variance matrices when the
objective is defined with the Moore--Penrose inverse. \citet{NeweySmith04}
already allow generalized inverses in their definition of CUE. We show how to globally minimize the CUE objective based on the Moore--Penrose inverse when the rank can vary with the coefficient. On intervals of constant rank, the objective is rational.
Its values where rank falls must be evaluated separately, along with the
limit at infinity, since they need not agree with the limiting rational
expression; see Section \ref{sec:singular-univariate}.

We also establish an analogue of partialling out in homoskedastic LIML. For
a fixed positive definite full reduced-form variance estimate, minimizing
over the coefficients on included exogenous covariates gives exactly the
CUE criterion based on projected moments, provided the variance estimate
is projected correspondingly. The two formulations have the same finite
minimizing structural coefficients and the same infimum. Thus exogenous covariates do
not add dimensions to the remaining global search; see Section S-1 in the supplemental appendix.

The contribution is to connect an important, well-known econometric
optimization problem to theoretical and computational algebra. This
connection allows computational methods for CUE to be assessed against
established methods for solving polynomial equations. The polynomial
representation identifies all candidates for the scalar minimum, and companion
matrices provide a numerical implementation. Polynomial interpolation
recovers the coefficients without symbolic matrix inversion. With several
endogenous regressors, the first-order conditions become a system of
polynomial equations. We discuss methods for this extension, including the
treatment of stationary components and boundary directions. The multivariate
problem requires additional computation beyond the scalar root calculation.

Section \ref{sec:setup} introduces the model and its statistical
applications. Section \ref{sec:implementation} develops the polynomial
characterization and computation. Section \ref{sec:multivariate} treats
multiple endogenous regressors. Section \ref{sec:simulation} presents
numerical results. The supplemental appendix establishes the equivalence of CUE
based on the full and partialled-out objectives. It also
presents the homoskedastic benchmark, the interpolation
method, additional lemmas and propositions, and their proofs. 
The simulation supplement provides additional simulation results.

\section{Problem setup}

\label{sec:setup}

Let $y_1$ be the $n\times1$ outcome vector, $y_2$ be the $n\times1$ vector
of observations on an endogenous regressor, and $X$ be an $n\times p$ full-column-rank matrix
of observations on $p$ exogenous covariates. The
structural equation is
\begin{equation}
	y_1=y_2\beta^*+X\delta^*+u.  \label{eq:structural}
\end{equation}
The parameter of interest is $\beta^*$. Let $\widetilde{Z}$ be an $n\times k$
matrix of observations on $k$ instrumental variables (IVs). The reduced form
for $y_2$ is
\begin{equation}
	y_2=\widetilde{Z}\pi+X\varphi+v_2.  \label{eq:rf-y2}
\end{equation}
Substituting \eqref{eq:rf-y2} into \eqref{eq:structural} gives
\begin{equation}
	y_1=\widetilde{Z} \pi \beta^{\ast} + X \gamma^{\ast}+v_1, \qquad
	\gamma^{\ast}=\delta^{\ast}+\varphi \beta^{\ast}, \qquad
	v_1=u+v_2\beta^{\ast}.  \label{eq:rf-y1}
\end{equation}
Writing $Y=(y_1,y_2)$, $V=(v_1,v_2)$, $a^{\ast}=(\beta^{\ast},1)^{\prime}$,
and $\Gamma=(\gamma^{\ast}, \varphi)$, the reduced form is
\begin{equation}
	Y=\widetilde{Z}\pi a^{*\prime}+X\Gamma+V.  \label{eq:matrix-rf}
\end{equation}

Let
\begin{equation*}
	Z=M_X\widetilde{Z}, \qquad
	M_X=I_n-X\left(X^{\prime}X\right)^{-1}X^{\prime}.
\end{equation*}
Each column of $Z$ contains the variation in the
corresponding instrumental variable that is orthogonal
to the exogenous covariates in $X$. The instrumental-variable exclusion restrictions are $\mathbb{E}%
[Z^{\prime}u]=0$. Because $Z^{\prime}X=0$, the sample moments for $\beta$
can be written as
\begin{align}
	g_n(\beta) &=n^{-1/2}Z^{\prime}(y_1-\beta y_2) =n^{-1/2}Z^{\prime}Yb(\beta),
	\label{eq:moment} \\
	&=\left(b(\beta)^{\prime}\otimes I_k\right) \func{vec}\left(n^{-1/2}Z^{%
		\prime}Y\right),  \notag
\end{align}
where $b(\beta)=(1,-\beta)^{\prime}$.

Let $\Sigma_n$ denote the conditional variance matrix of
$\func{vec}\left(n^{-1/2}Z^{\prime}Y\right)$ given
$\widetilde{Z}$ and $X$. Then, the conditional variance matrix of
$g_n(\beta)$ is
\begin{equation}
	\Omega_n(\beta)
	=\left(b(\beta)^{\prime}\otimes I_k\right)\Sigma_n
	\left(b(\beta)\otimes I_k\right).
\end{equation}
Let $\widehat{\Sigma}_{n}$ be an estimate of $\Sigma_{n}$, computed once
and held fixed as $\beta$ varies. Variance estimation may allow for
heteroskedasticity, autocorrelation, or clustering; see
\citet{MacKinnonWhite85}, \citet{NeweyWest1987}, and
\citet{CameronGelbachMiller11}, respectively. Consistency requires the
sampling and regularity conditions appropriate to the chosen estimator.
The corresponding estimate of $\Omega_n(\beta)$ is
\begin{equation}
	\widehat{\Omega}_n(\beta) =\left(b(\beta)^{\prime}\otimes I_k\right)
	\widehat{\Sigma}_{n} \left(b(\beta)\otimes I_k\right).  \label{eq:omega-hat}
\end{equation}

Our minimization results are deterministic statements about the observed
moments and $\widehat{\Sigma}_n$. While our results rely on the quadratic representation
\eqref{eq:omega-hat}, we do not require a central limit theorem or strong
identification. The choice of variance estimator determines the objective
being minimized. Different variance estimators need not
produce the same CUE. A weighting rule that depends on $\beta$ is covered
only when it follows the representation \eqref{eq:omega-hat} with a fixed
$\widehat{\Sigma}_n$.

We first impose the following condition on this realized matrix.

\begin{assumption}
	\label{ass:sigma-pd}  The matrix $\widehat{\Sigma}_n$ is symmetric positive
	definite.
\end{assumption}
This assumption implies that $\widehat{\Omega}_n(\beta)$ is
positive definite for every $\beta\in\mathbb{R}$, as established by Lemma S.1 in the supplemental appendix. This ensures that the CUE objective,
\begin{equation}  \label{eq:cue-objective}
	Q_n(\beta) =g_n(\beta)^{\prime} \widehat{\Omega}_n(\beta)^{-1} g_n(\beta),
\end{equation}
is a continuous function of $\beta$ and has no finite poles. Assumption \ref{ass:sigma-pd} is a restriction on the variance
estimate, not on the strength of instrumental variables.
In Section \ref{sec:singular-univariate}, we extend our results to allow for the estimate of the variance matrix to be symmetric positive semidefinite and use the Moore--Penrose inverse to define the CUE objective.

Section S-1 in the supplemental appendix shows
that minimizing this partialled-out CUE objective gives the same finite
estimates of $\beta$ as jointly minimizing the corresponding full CUE
objective over the endogenous and exogenous coefficients.

In the linear IV model, the CUE objective function depends on $\beta$
only through the vector $b(\beta)$. So, we can write $Q_n(\beta)=Q_n\left(b(%
\beta)\right)$. For $\beta\in\mathbb{R}$, let $c(\beta)=1/\left\|
b(\beta)\right\|>0$ and define
\begin{equation}
	\overline{b}(\beta)=c(\beta) . b(\beta)=\left(\frac{1}{\left\| b(\beta)\right\|}%
	,-\frac{\beta}{\left\| b(\beta)\right\|}\right)^{\prime}.
\end{equation}
Note that $Q_n\left(b(\beta)\right)$ is homogeneous of degree zero in $%
b(\beta)$; that is, for any constant $c\neq0$, $Q_n\left(c\,b(\beta)%
\right)=Q_n\left(b(\beta)\right)$. Therefore, $Q_n\left(b(\beta)\right)=Q_n%
\left(\overline{b}(\beta)\right)$.

As $\beta\to\pm\infty$, $\overline{b}(\beta)\to\mp e_2$,
where $e_2=(0,1)^{\prime}$. Therefore, the normalized sample moments
converge to $\mp n^{-1/2}Z^{\prime}y_2$, and their
estimated variance matrix converges to
$\widehat{\Sigma}_{22,n}$, the lower-right $k\times k$ block
of $\widehat{\Sigma}_n$. Under Assumption \ref{ass:sigma-pd}, $\widehat{\Sigma}_{22,n}$ is positive definite. 
By continuity of matrix inversion,
\[
\lim_{\beta\to-\infty}Q_n(\beta)
=\lim_{\beta\to+\infty}Q_n(\beta)
=\frac{y_2^{\prime}Z\widehat{\Sigma}_{22,n}^{-1}
	Z^{\prime}y_2}{n}.
\]
We define
\begin{equation}\label{eq:cu-gmm-end_point-value}
	Q_n(-\infty)=Q_n(+\infty)
	=\frac{y_2^{\prime}Z\widehat{\Sigma}_{22,n}^{-1}
		Z^{\prime}y_2}{n}.
\end{equation}

By Lemma S.1,
$\widehat{\Omega}_n(\beta)$ is positive definite for every
finite $\beta$. Therefore, continuity of the sample moments and matrix
inversion implies that $Q_n(\beta)$ is continuous
on $\mathbb{R}$. Together with the endpoint limits established
above, this gives a continuous extension to
$\overline{\mathbb{R}}=\mathbb{R}\cup\{-\infty,+\infty\}$.
Compactness of this space and the extreme value theorem
yield the following result.

\begin{lemma}
	\label{lem:cue-existence} Under Assumption \ref{ass:sigma-pd}, the CUE
	objective $Q_n(\beta)$, with endpoint values defined by %
	\eqref{eq:cu-gmm-end_point-value}, is continuous on the extended real line $%
	\overline{\mathbb{R}}$ and attains its minimum there.
\end{lemma}

While the global minimum may occur at the infinite endpoints, this event has
probability zero if the conditional distribution of $\func{vec}%
\left(n^{-1/2}Z^{\prime}Y\right)$ given $\widehat{\Sigma}_n$ is absolutely
continuous with respect to Lebesgue measure almost surely; see Proposition %
S.1 in the supplemental appendix. In particular, under the weak-instrument asymptotics of
\citet{StaigerStock97}, the normalized reduced-form moments
have a Gaussian limit with a known reduced--form variance matrix $\Sigma$. When $\Sigma$
is positive definite, Proposition S.1
implies that every global minimizer of the limiting CUE
objective is finite with probability one.

In the just-identified case, where $k=1$, the minimum value of the CUE
objective $Q_n(\beta)$ over $\overline{\mathbb{R}}$ is zero. If $%
Z^{\prime}y_2\neq0$, the continuously updating estimator (CUE) that
minimizes $Q_n(\beta)$ over $\overline{\mathbb{R}}$ is
\begin{equation*}
	\widehat{\beta}_{\mathrm{CUE}}=\frac{Z^{\prime}y_1}{Z^{\prime}y_2},
\end{equation*}
so that $g_n\left(\widehat{\beta}_{\mathrm{CUE}}\right)=0$ and $Q_n\left(%
\widehat{\beta}_{\mathrm{CUE}}\right)=0$. Since the objective is
nonnegative, this is its global minimum. If $Z^{\prime}y_2=0$, the minimum
is also zero because $Q_n(\pm\infty)=0$.

In the rest of the paper, we focus on the overidentified case, where $k>1$.
Finding the global minimum of the CUE objective over $\overline{\mathbb{R}%
}$ generally requires solving a non-convex optimization problem. The main
text considers a general setting that allows for heteroskedasticity,
clustering, or autocorrelation in the errors. Section S-2
in the supplemental appendix discusses the simpler homoskedastic case.

The global minimum determines three statistical objects: the CUE,
the overidentification statistic, and the LR statistic underlying the CLR test.

The CUE minimizes $Q_n(\beta)$ over $\overline{\mathbb{R}}$:
\begin{equation}
	Q_n\left(\widehat{\beta}_{\mathrm{CUE}}\right) =\min_{\beta\in\overline{%
			\mathbb{R}}}Q_n(\beta).  \label{eq:cue-estimator}
\end{equation}
A boundary minimizer represents an infimum approached by finite
coefficients; it is not a finite coefficient estimate. Keeping that case
in the definition makes the minimization result apply to every sample.
When the CUE objective uses the homoskedastic variance specification, CUE
coincides with LIML; see Section S-2 in the supplemental appendix. 
With a more general robust variance estimator, the
two estimators need not coincide, even when the underlying errors are
homoskedastic.

Under standard regularity conditions, with a fixed number of valid
instruments and strong identification, 2SLS, LIML, and CUE are consistent
and asymptotically normal. Their properties
differ when the number of instruments grows with the sample size. Under the
homoskedastic many-instrument asymptotics of \citet{Bekker94}, with $%
k/n\to\alpha\in(0,1)$ and the maintained identification conditions, 2SLS can
be inconsistent, whereas LIML remains consistent and asymptotically normal.
Allowing for heteroskedasticity and many weak moment conditions, %
\citet{NeweyWindmeijer09} establish consistency of CUE when $k^2/n\to0$, and
asymptotic normality when $k^3/n\to0$. These results also require their
regularity and identification conditions. Applying them to a computed
estimator requires matching their variance specification and parameter
space, in addition to finding the global minimum.

The $J$-statistic for overidentification tests can be written as
\begin{equation}
	J_n=\min_{\beta\in\overline{\mathbb{R}}}Q_n(\beta).  \label{eq:overid-stat}
\end{equation}
The minimum is attained by Lemma \ref{lem:cue-existence}. The null
hypothesis is that the moment restrictions hold for some finite coefficient $%
\beta^*$. Under the usual strong-identification regularity conditions, with
a fixed number $k>1$ of instruments, $J_n\overset{d}{\longrightarrow}%
\chi^2_{k-1}$. Therefore, the usual overidentification test compares $J_n$
with a $\chi^2_{k-1}$ critical value. The loss of one degree of freedom
reflects estimation of the scalar coefficient $\beta$; see \citet[Section
3]{NeweyWindmeijer09}.

Under weak identification, the $J_n$ statistic does not converge to the usual $\chi^2_{k-1}$ null distribution. A conservative alternative is to compare $J_n$ with the $(1-\alpha)$ quantile of the $\chi^2_k$ distribution, where $\alpha$ is the desired upper bound on the asymptotic size of the test. To justify this comparison, suppose that, under the null hypothesis, the sample moments satisfy a central limit theorem with a positive definite asymptotic variance matrix. For fixed $k$, these conditions yield the following null asymptotic distribution for the Anderson--Rubin statistic:
\begin{equation*}
	AR_n(\beta^*)=Q_n(\beta^*) \overset{d}{\longrightarrow}\chi^2_k,
\end{equation*}
without requiring strong identification. Since $J_n\leq Q_n(\beta^*)$, as in \citet{MoreiraRidderSharifvaghefi26}, we have
\begin{equation}
	\Pr\left(J_n>\chi^2_{k,1-\alpha}\right)
	\leq\Pr\left(Q_n(\beta^*)>\chi^2_{k,1-\alpha}\right) \longrightarrow\alpha,
	\label{eq:overid-size-bound}
\end{equation}
where $\chi^2_{k,1-\alpha}$ denotes the $(1-\alpha)$ quantile of the $\chi^2_k$ distribution. Thus the test has asymptotic rejection probability at most $\alpha$ along
any null sequence for which this AR approximation holds. Uniform validity
over a class of null models requires a corresponding uniform AR
approximation. Under strong identification, the larger $\chi^2_k$ critical
value makes this test conservative relative to the usual $\chi^2_{k-1}$ test.

Suppose that, conditional on the IVs and exogenous
covariates, $\vect\left(n^{-1/2}Z^{\prime}Y\right)$ is normally
distributed with known positive definite variance matrix $\Sigma_n$.
Using $\Sigma_n$ in the CUE objective, the likelihood ratio
statistic underlying the CLR test of $H_0:\beta^*=\beta_0$ is
\begin{equation}\label{eq:lr-stat}
	LR_n(\beta_0)=Q_n(\beta_0)-J_n.
\end{equation}

Under weak-instrument asymptotics, consider the sequence
$\pi_n=\mu/\sqrt{n}$, where $\mu\in\mathbb{R}^k$ is the local first-stage
parameter. The null distribution of $LR_n(\beta_0)$ generally
depends on $\mu$, which cannot be consistently estimated.
Let $a_0=(\beta_0,1)^{\prime}$ and define
\begin{equation}
	T_n=
	\left[
	(a_0^{\prime}\otimes I_k)\Sigma_n^{-1}(a_0\otimes I_k)
	\right]^{-1/2}
	(a_0^{\prime}\otimes I_k)\Sigma_n^{-1}
	\vect\left(n^{-1/2}Z^{\prime}Y\right).
\end{equation}
Under $H_0$, $T_n$ is sufficient for $\mu$, so the conditional
null distribution of the LR statistic given $T_n$ does not
depend on $\mu$. Let $c_{\alpha}(t;\Sigma_n)$ denote the
$(1-\alpha)$ quantile of this distribution given $T_n=t$.
Following the conditioning argument of \citet{Moreira03},
this conditional critical value controls the size of the
CLR test at level $\alpha$. 

The conditional critical value
$c_{\alpha}(T_{n};\Sigma_n)$
can be estimated using the simulation method described
in Section \ref{sec: MC-CLR}. This requires repeated solution
of the global minimization problem for simulated data.

With nonnormal observations and unknown $\Sigma_n$, the same calculation
can be applied using $\widehat{\Sigma}_n$. Statistical validity then requires a Gaussian approximation,
such as that provided by the weak-instrument asymptotics of
\citet{StaigerStock97}. The results below solve the minimization problem required for the CLR test. They do not establish a new distributional approximation for a feasible CLR test.

The global minimizers of the CUE objective determine the set of CUE estimates, while its global minimum value determines the overidentification statistic. This minimum value also enters the LR statistic and the simulated statistics used to obtain its conditional critical value. Evaluating the objective at a local minimum or over a restricted grid can result in an overstated $J_n$ statistic that need not be bounded above by $Q_n(\beta^*)$. This can lead to overrejection of the overidentifying restrictions, and the size guarantee in \eqref{eq:overid-size-bound} does not automatically carry over.

In HAC IV models, \citet{MoreiraRidderSharifvaghefi26} distinguish the CLR test that builds on the full reduced-form likelihood from procedures based on AR and LM statistics. Their power comparisons motivate the CLR test. A separate issue is whether a numerical grid approximation of the CUE objective preserves the desirable theoretical properties of the CLR test. \citet{MoreiraSharifvaghefi26} give designs in which grid approximation fails to preserve the size and power properties of the CLR test, even when the number of grid points grows polynomially with sample size. This is not a failure of every grid. The accuracy of the grid approximation depends on whether the grid region contains the global minimizer and whether its points are sufficiently close together near that minimizer. Our contribution is the algebraic characterization of the required global minimum in linear IV.

Computing the CUE, the overidentification statistic, and the CLR test
requires a reliable algorithm for global minimization. With a general
variance matrix, $Q_n(\beta)$ need not be convex, so local optimization and
grid search do not in general guarantee the global minimum. The next section
shows that, with one endogenous regressor, $Q_n(\beta)$ is a ratio of
polynomials. Thus, its global minimum can be computed by finding all real
roots of the polynomial equation obtained from the first-order condition and
then evaluating the CUE objective at those roots and at either infinite
endpoint. This procedure makes the computed CUE, overidentification
statistic, and CLR statistic coincide with the objects defined above, up to
numerical error in coefficient recovery and root finding.

\section{Global minimization of the CUE objective}

\label{sec:implementation}

Section \ref{sec:poly-ratio-representation} expresses the CUE objective
as a ratio of polynomials. Section \ref{sec:poly-approach} characterizes its
global minimum using the real roots of the first-order condition and the
objective value at infinity. Section \ref{sec:companion} explains why
finding these roots generally requires numerical computation and shows how
to obtain them through an eigenvalue problem. Section \ref%
{sec:singular-univariate} extends these results to positive semidefinite
variance matrices using the Moore--Penrose inverse.

\subsection{Polynomial ratio representation}

\label{sec:poly-ratio-representation}

The vector of sample moments $g_n(\beta)=n^{-1/2}Z^{\prime}\left(y_1-\beta
y_2\right)$, defined in Section 2, is an affine function of $\beta$.
Moreover, by partitioning $\widehat{\Sigma}_n$ as
\begin{equation*}
	\widehat{\Sigma}_n=
	\begin{pmatrix}
		\widehat{\Sigma}_{11,n} & \widehat{\Sigma}_{12,n} \\
		\widehat{\Sigma}_{21,n} & \widehat{\Sigma}_{22,n}%
	\end{pmatrix}%
	,
\end{equation*}
where each block is $k\times k$, we can write the estimated variance matrix
of the sample moments as
\begin{equation}  \label{eq:sample_moment_variance_expanded}
	\widehat{\Omega}_n(\beta) =\widehat{\Sigma}_{11,n} -\beta\left(\widehat{%
		\Sigma}_{12,n}+\widehat{\Sigma}_{21,n}\right) +\beta^2\widehat{\Sigma}%
	_{22,n}.
\end{equation}
Each element of $\widehat{\Omega}_n(\beta)$ is a polynomial in $\beta$ of
degree at most two. Under Assumption \ref{ass:sigma-pd}, the adjugate
formula gives
\begin{equation}
	\widehat{\Omega}_n(\beta)^{-1} =\frac{\func{adj}\left\{\widehat{\Omega}%
		_n(\beta)\right\}} {\det\left\{\widehat{\Omega}_n(\beta)\right\}}.
\end{equation}
Therefore, the CUE objective can be written as
\begin{equation}  \label{eq:cue-objective-adjugate}
	Q_n(\beta) =\frac{g_n(\beta)^{\prime}\func{adj}\left\{\widehat{\Omega}%
		_n(\beta)\right\}g_n(\beta)} {\det\left\{\widehat{\Omega}_n(\beta)\right\}}.
\end{equation}
The determinant has degree at most $2k$. Its coefficient on $\beta^{2k}$
is $\det\left\{\widehat{\Sigma}_{22,n}\right\}>0$, so its degree is exactly
$2k$. Each entry of the adjugate has degree at most $2k-2$; multiplying
on both sides by the affine moment vector gives a numerator of degree
at most $2k$. Its coefficient on $\beta^{2k}$ is
\[
\frac{1}{n}y_2^{\prime}Z
\adj\left(\widehat{\Sigma}_{22,n}\right)Z^{\prime}y_2
=\det\left\{\widehat{\Sigma}_{22,n}\right\}Q_n(+\infty).
\]
Thus the numerator also has degree $2k$ when $Z^{\prime}y_2\neq0$.
If $Z^{\prime}y_2=0$, it has lower degree or is identically zero.
The following proposition summarizes the representation.

\begin{proposition}
	\label{prop:ratio-formal}  Under Assumption \ref{ass:sigma-pd}, the CUE
	objective function in \eqref{eq:cue-objective-adjugate} can be written as
	\begin{equation}  \label{eq:ratio-formal}
		Q_n(\beta)=\frac{p_n(\beta)}{q_n(\beta)},
	\end{equation}
	where $p_n(\beta)$ and $q_n(\beta)$ are the numerator and denominator in %
	\eqref{eq:cue-objective-adjugate}, respectively, and (i) $p_n(\beta)$ is a
	polynomial of degree at most $2k$ and $q_n(\beta)$ is a polynomial of degree
	exactly $2k$, (ii) $q_n(\beta)>0$ for every finite $\beta$, and (iii) both
	polynomials have degree $2k$ if $Z^{\prime}y_2\neq0$.
\end{proposition}

Our original proof of the rational representation in Proposition
\ref{prop:ratio-formal} used the Sherman--Morrison--Woodbury formula
to obtain the inverse of $\widehat{\Omega}_n(\beta)$.
This approach requires verifying that each intermediate matrix is
invertible. Moreover, repeated application of the formula can introduce common polynomial factors in the numerator and denominator that must be cancelled to achieve bounds on the degree of polynomials stated in the proposition. The adjugate
representation establishes these bounds directly.

The degrees in Proposition \ref{prop:ratio-formal} refer to the determinant
and adjugate representation. Common factors can reduce these degrees. Under
the homoskedastic variance structure, a factor of degree $2k-2$ cancels,
leaving a ratio of quadratic forms; see Section S-2 in
the supplemental appendix.

\subsection{Global minimization through polynomial roots}

\label{sec:poly-approach} Under Assumption \ref{ass:sigma-pd}, the CUE
objective $Q_n(\beta)$ is a continuously differentiable function of $\beta$
on $\mathbb{R}$. Using the representation $Q_n(\beta)=p_n(\beta)/q_n(\beta)$
from Section \ref{sec:poly-ratio-representation}, its derivative with
respect to $\beta$ is
\begin{equation}
	\frac{\partial Q_n(\beta)}{\partial\beta} =\frac{\frac{\partial p_n(\beta)}{%
			\partial\beta}q_n(\beta) -p_n(\beta)\frac{\partial q_n(\beta)}{\partial\beta}%
	}{q_n(\beta)^2}.
\end{equation}
Since $q_n(\beta)>0$ for every finite $\beta$, the finite stationary points
of $Q_n(\beta)$ are exactly the solutions to
\begin{equation}  \label{eq:foc}
	h_n(\beta) =\frac{\partial p_n(\beta)}{\partial\beta}q_n(\beta) -p_n(\beta)%
	\frac{\partial q_n(\beta)}{\partial\beta} =0.
\end{equation}

Each product in \eqref{eq:foc} has degree at most $4k-1$.
If the numerator and denominator both have degree $2k$, their leading
derivative terms cancel. If the numerator has lower degree, each product
already has degree at most $4k-2$. Thus $h_n(\beta)$ is identically zero
or has degree at most $4k-2$.

If $h_n(\beta)$ is not identically zero, the stationary points of $Q_n(\beta)
$ are the real roots of a polynomial of degree at most $4k-2$. We know that
such a polynomial has at most $4k-2$ distinct real roots. Therefore, to find
the global minimum of $Q_n(\beta)$ over the extended real line $\overline{%
	\mathbb{R}}=\mathbb{R}\cup\{-\infty,+\infty\}$, we need to evaluate the
objective at no more than $4k-1$ points, that are at most $4k-2$ stationary
points and the endpoint $\beta=+\infty$. It is enough to evaluate one
endpoint because $Q_n(-\infty)=Q_n(+\infty)$.

If $h_n(\beta)$ is identically zero, the derivative of $Q_n(\beta)$ is zero
for every finite $\beta$. So, $Q_n(.)$ is constant on $\mathbb{R}$ and, by
continuity at the endpoints, on $\overline{\mathbb{R}}$. In this case,
evaluating the objective at any one point gives its global minimum. Theorem %
\ref{thm:global-min-scalar} formalizes these findings.

\begin{theorem}
	\label{thm:global-min-scalar}  Suppose Assumption \ref{ass:sigma-pd} holds,
	and let $h_n(\beta)$ be defined by \eqref{eq:foc}. If $h_n(\beta)$ is not
	identically zero, let
	\begin{equation*}
		\mathcal{C}_n =\left\{\beta\in\mathbb{R}:h_n(\beta)=0\right\}\cup\left\{+%
		\infty\right\}.
	\end{equation*}
	Then $\mathcal{C}_n$ contains at most $4k-1$ points, and
	\begin{equation}  \label{eq:min-candidate-formal}
		\min_{\beta\in\overline{\mathbb{R}}}Q_n(\beta) =\min_{\beta\in\mathcal{C}%
			_n}Q_n(\beta).
	\end{equation}
	If $h_n(\beta)$ is identically zero, then $Q_n(.)$ is constant on $\overline{%
		\mathbb{R}}$, and every point in $\overline{\mathbb{R}}$ is a global
	minimizer.
\end{theorem}

\subsection{Computing polynomial roots using companion matrices}

\label{sec:companion}

Suppose $h_n(\beta)$ is not identically zero. It cannot be a nonzero
constant: otherwise $Q_n(\beta)$ would be strictly monotone on
$\mathbb{R}$, contradicting its equal limits at the two endpoints. Write
\begin{equation}  \label{eq:foc-polynomial}
	h_{n}(\beta)= \delta_{m,n} \beta^{m} + \delta_{m-1,n} \beta^{m-1}+\cdots+
	\delta_{1,n} \beta + \delta_{0,n},
\end{equation}
where the coefficients $\delta_{i,n}$, $i = 0, 1, \ldots, m$, depend on the
observed data, $\delta_{m,n}\neq0$, and the degree $m$ is at most $4k-2$.

For polynomial equations of degree greater than four, Galois theory implies
that there is no general formula expressing their roots in terms of the
coefficients using arithmetic operations and radicals; see \citet[Section
14.7, Corollary 40]{DummitFoote2004}. Although particular polynomials may
admit such formulas, the proof of Proposition S.2 in the
supplemental appendix gives a CUE example with two instruments in which every global
minimizer is finite and cannot be expressed by radicals. This motivates a
numerical approach to computing the roots of $h_n(\beta)$. These roots can
be obtained as the eigenvalues of a companion matrix and computed using the
QR algorithm. To construct the companion matrix, consider the following
monic version of the polynomial $h_n(\beta)$:
\begin{equation*}
	\overline{h}_{n}(\beta)= \beta^{m} + \overline{\delta}_{m-1,n}
	\beta^{m-1}+\cdots+ \overline{\delta}_{1,n} \beta + \overline{\delta}_{0,n},
\end{equation*}
where, for $i = 0, 1, \ldots, m-1$,
\begin{equation}
	\overline{\delta}_{i,n} = \frac{\delta_{i,n}}{\delta_{m,n}}.
\end{equation}
Define the companion matrix of $\overline{h}_{n}(\beta)$ as
\begin{equation}  \label{eq:companion}
	C_{\overline{h}}=
	\begin{pmatrix}
		0 & 1 & 0 & \cdots & 0 \\
		0 & 0 & 1 & \cdots & 0 \\
		\vdots & \vdots & \vdots & \ddots & \vdots \\
		0 & 0 & 0 & \cdots & 1 \\
		-\overline{\delta}_{0,n} & -\overline{\delta}_{1,n} & -\overline{\delta}%
		_{2,n} & \cdots & -\overline{\delta}_{m-1,n}%
	\end{pmatrix}%
	.
\end{equation}
Normalization does not change the roots, so the eigenvalues of $C_{\overline{%
		h}}$ are the roots of $h_n(\beta)$. Retaining the real eigenvalues gives the
following characterization of the finite stationary points.

\begin{proposition}
	\label{prop:companion}  Under Assumption \ref{ass:sigma-pd}, suppose $%
	h_n(\beta)$ is not identically zero. The finite stationary points of the
	CUE objective function $Q_{n}(\beta)$ given by \eqref{eq:cue-objective}
	are exactly the real eigenvalues of the companion matrix $C_{\overline{h}}$
	given by \eqref{eq:companion}.
\end{proposition}

Section S-2 in the supplemental appendix gives the
generalized eigenvalue characterization under homoskedastic errors.
Proposition \ref{prop:companion} provides an eigenvalue approach for general
variance structures. Together with Theorem \ref{thm:global-min-scalar}, it
reduces global minimization to finding all real eigenvalues of the companion
matrix and comparing the objective values there and at infinity.

The roots of $h_n(\beta)$ can also be computed using other methods, such as
the Jenkins--Traub algorithm or Ehrlich--Aberth iterations. Each implementation must find all real roots needed for the comparison
in Theorem \ref{thm:global-min-scalar}. A stationary point obtained from a
single starting value does not suffice.

For implementation, the coefficients of $p_n(\beta)$ and $q_n(\beta)$ can be
computed symbolically from \eqref{eq:cue-objective-adjugate} or recovered by
polynomial interpolation as described in Section S-3 in the supplemental appendix. Once the real roots
of $h_n(\beta)$ have been computed, evaluate the original CUE objective
at these roots and at either infinite endpoint, as prescribed by Theorem \ref%
{thm:global-min-scalar}\footnote{The polynomial characterization provides a systematic
	way to compute the global minimum. Numerical accuracy can be improved by using suitable scaling and higher-precision arithmetic.}.

\subsection{CUE with singular variance matrices}

\label{sec:singular-univariate}

Throughout this subsection, we allow $\widehat{\Sigma}_n$ to be symmetric
positive semidefinite. As a result, $\widehat{\Omega}_n(\beta)$ is symmetric
positive semidefinite for every $\beta\in\mathbb{R}$, with rank between zero
and $k$. Its rank may vary with $\beta$. For every $\beta \in \mathbb{R} $,
define
\begin{equation}  \label{eq:singular-objective}
	Q_n(\beta)=g_n(\beta)^{\prime} \widehat{\Omega}_n(\beta)^{\dagger}g_n(\beta),
\end{equation}
where $\dagger$ denotes the Moore--Penrose inverse. This inverse assigns zero weight to the projection of the sample moment vector onto the null space of the estimated variance matrix. It specifies the CUE objective to be minimized but does not by itself justify inference when the estimated rank changes.
See
\citet{Andrews87b} for conditions under which
test statistics using generalized inverse weighting matrices
keep their usual asymptotic chi-square distributions.\footnote{Alternatively, we can use the ridge-regularized variance estimator $\widehat{\Sigma}_n+\lambda_n I_{2k}$, where $\lambda_n>0$ does not depend on $\beta$. This matrix is positive definite, so the corresponding CUE objective is a ratio of polynomials and the results of the preceding subsections apply. This regularization generally defines a different objective from \eqref{eq:singular-objective}.} For a symmetric positive semidefinite matrix, this inverse reciprocates its positive eigenvalues and leaves its zero eigenvalues equal to zero. It coincides with the ordinary inverse whenever the matrix is nonsingular.

Set $q_{0,n}(\beta)=1$. For $j=1,\ldots,k$, define
\begin{equation}  \label{eq:singular-coefficients}
	q_{j,n}(\beta)= \sum_{\substack{J\subseteq\{1,\ldots,k\}\\ |J|=j}}
    \sum_{\substack{I\subseteq\{1,\ldots,k\}\\ |I|=j}}
	\det\left\{\left[\widehat{\Omega}_n(\beta)\right]_{I,J}\right\}^{2},
\end{equation}
where the subscripts $I,J$ select rows and columns of $\widehat{\Omega}%
_n(\beta)$, respectively. Thus $q_{j,n}(\beta)$ is the sum of the squared $%
j\times j$ minors. It is a polynomial of degree at most $4j$ and is positive
if and only if the rank is at least $j$. When the rank equals $r\geq1$,
Lemma \ref{lem:singular-inverse} gives
\begin{equation}  \label{eq:singular-inverse}
	\widehat{\Omega}_n(\beta)^{\dagger} =\frac{\displaystyle%
		\sum_{j=0}^{r-1}(-1)^j q_{r-1-j,n}(\beta) \widehat{\Omega}_n(\beta)^{2j+1}}{%
		q_{r,n}(\beta)}.
\end{equation}

\begin{lemma}
	\label{lem:singular-inverse} For every $\beta\in\mathbb{R}$ at which $\rank%
	\left\{\widehat{\Omega}_n(\beta)\right\}=r\geq1$, $\widehat{\Omega}%
	_n(\beta)^{\dagger}$ is given by \eqref{eq:singular-inverse}, with $%
	q_{r,n}(\beta)>0$. If the rank is zero, $\widehat{\Omega}_n(\beta)^{%
		\dagger}=0$ and $Q_n(\beta)=0$.
\end{lemma}

Equation \eqref{eq:singular-inverse} supplies a rational formula at every
positive rank, including ranks below $k$. Let
\begin{equation*}
	r_{\max}=\max_{\beta\in\mathbb{R}} \rank\left\{\widehat{\Omega}%
	_n(\beta)\right\}.
\end{equation*}
If $r_{\max}=0$, the objective is identically zero. Otherwise, define the
generic numerator and denominator by
\begin{align}
	p_n(\beta)&=\sum_{j=0}^{r_{\max}-1}(-1)^j q_{r_{\max}-1-j,n}(\beta)
	g_n(\beta)^{\prime} \widehat{\Omega}_n(\beta)^{2j+1}g_n(\beta),
	\label{eq:singular-generic-p} \\
	q_n(\beta)&=q_{r_{\max},n}(\beta) .  \label{eq:singular-generic-q}
\end{align}
Both polynomials have degree at most $4r_{\max}$. These bounds need not be
sharp. When $r_{\max}=k$, the formulas for the determinant and adjugate provide an alternative representation in which the numerator and denominator are polynomials of degree at most $2k$.

The real roots of $q_n(\beta)$ identify the rank-drop points. Away from
those points, the objective has the representation
\begin{equation}  \label{eq:singular-generic-ratio}
	Q_n(\beta)=\frac{p_n(\beta)}{q_n(\beta)} = \frac{\sum_{j=0}^{r_{%
				\max}-1}(-1)^j q_{r_{\max}-1-j,n}(\beta) g_n(\beta)^{\prime}\widehat{\Omega}%
		_n(\beta)^{2j+1}g_n(\beta) }{q_{r_{\max},n}(\beta)}
\end{equation}
The following proposition gives the corresponding partition.

\begin{proposition}
	\label{prop:singular-intervals}  Suppose $r_{\max}>0$. The polynomial $%
	q_n(\beta)$ given by \eqref{eq:singular-generic-q} is not identically  zero.
	Its distinct real roots $\beta_1<\cdots<\beta_m$, possibly none,  are
	exactly the points where the rank is below $r_{\max}$,  with $%
	m\leq4r_{\max}\leq4k$. The open intervals
	\begin{equation*}
		\left(-\infty,\beta_1\right),\quad
		\left(\beta_1,\beta_2\right),\quad\ldots,\quad
		\left(\beta_m,+\infty\right),
	\end{equation*}
	together with the singleton sets $\{\beta_1\},\ldots,\{\beta_m\}$,
	partition $\mathbb{R}$. On every interval, the rank is $r_{\max}$, and $%
	Q_n(\beta)$  has the representation \eqref{eq:singular-generic-ratio}  and
	is infinitely differentiable. If $m=0$, the rational  representation holds
	on all of $\mathbb{R}$.
\end{proposition}

At the rank-drop points, the objective values are computed directly from %
\eqref{eq:singular-objective}. To ensure that all these points are identified, we must compute the real roots of $q_n(\beta)$ in \eqref{eq:singular-generic-ratio} before cancelling its common factors with the numerator. For example, if $\widehat{\Sigma}_n=\func{diag}(0,1,1,0)$
and $g_n(\beta)=\left(\beta,0\right)^{\prime}$, then $\widehat{\Omega}%
_n(\beta)=\func{diag}(\beta^2,1)$. The generic objective is $%
\beta^2/\beta^2=1$ for $\beta\neq0$, whereas $Q_n(0)=0$.

On all open intervals in Proposition \ref{prop:singular-intervals},
\begin{equation}  \label{eq:singular-foc}
	\frac{\partial}{\partial \beta}Q_n(\beta)=\frac{h_n(\beta)}{q_n(\beta)^2}%
	,\qquad h_n(\beta)= \frac{\partial p_n(\beta)}{\partial \beta} q_n(\beta) -
	p_n(\beta) \frac{\partial q_n(\beta)}{\partial \beta} .
\end{equation}
Rank-drop points and infinite endpoints are evaluated separately.

We extend the objective to the infinite endpoints by defining
\begin{equation}  \label{eq:singular-endpoints}
	Q_n(-\infty)=\lim_{\beta\to-\infty}Q_n(\beta),\qquad
	Q_n(+\infty)=\lim_{\beta\to+\infty}Q_n(\beta).
\end{equation}

Outside finitely many points where the rank of $\widehat{\Omega}_n(\beta)$
falls below its maximum, the objective equals the same nonnegative rational
function. Comparing its leading terms shows that the limits at positive and negative infinity coincide. This common value is either finite and nonnegative or $+\infty$. When $\widehat{\Sigma}_{22,n}$ is positive definite, the common
limit is
\begin{equation}  \label{eq:singular-endpoints-pd}
	Q_n(-\infty)=Q_n(+\infty) =\frac{y_2^{\prime}Z\widehat{\Sigma}%
		_{22,n}^{-1}Z^{\prime}y_2}{n}.
\end{equation}
When $\widehat{\Sigma}_{22,n}$ is singular, replacing its inverse in this
expression by its Moore--Penrose inverse need not give the limit. If $%
r_{\max}>0$, the endpoint values are obtained from the limits of the
rational expression in \eqref{eq:singular-generic-ratio}. If $r_{\max}=0$,
both endpoint values are zero.

The following theorem shows that, while $Q_n(\beta)$ may be discontinuous at rank-drop points, it is lower semicontinuous. Therefore, its value at each
finite rank-drop point is no larger than either one-sided limit.
Lower semicontinuity and compactness of the extended real line ensure that
the extended objective attains its minimum. Defining the endpoint values by %
\eqref{eq:singular-endpoints} makes this minimum equal to $\inf_{\beta\in%
	\mathbb{R}}Q_n(\beta)$.

\begin{theorem}
	\label{thm:singular-global}  Suppose that $\widehat{\Sigma}_n$ is symmetric
	positive semidefinite,  and define $Q_n(\beta)$ on $\overline{\mathbb{R}}$
	by  \eqref{eq:singular-objective} and \eqref{eq:singular-endpoints}. Then
	
	\begin{itemize}
		\item[(i)] The extended objective is lower semicontinuous and attains  a
		finite minimum equal to  $\inf_{\beta\in\mathbb{R}}Q_n(\beta)$.
		
		\item[(ii)] If $r_{\max}>0$ and $h_n(\beta)$ is not identically zero, the
		set
		\begin{equation}  \label{eq:singular-candidates}
			\mathcal{C}_n= \left\{\beta\in\mathbb{R}:h_n(\beta)=0,\ q_n(\beta)>0\right\}
			\cup\left\{\beta_1,\ldots,\beta_m\right\} \cup\left\{-\infty,+\infty\right\}
		\end{equation}
		is finite and contains every global minimizer. In particular,
		\begin{equation}  \label{eq:singular-global-min}
			\inf_{\beta\in\mathbb{R}}Q_n(\beta) =\min_{\beta\in\overline{\mathbb{R}}%
			}Q_n(\beta) =\min_{\beta\in\mathcal{C}_n}Q_n(\beta).
		\end{equation}
		
		\item[(iii)] If $r_{\max}>0$ and $h_n(\beta)$ is identically zero, the
		generic  objective has the same constant value on all the open intervals.
		Replacing the first set in \eqref{eq:singular-candidates} by any one  finite
		point with $q_n(\beta)>0$ still gives  \eqref{eq:singular-global-min} and at
		least one global minimizer.
		
		\item[(iv)] If $r_{\max}=0$, every point of $\overline{\mathbb{R}}$ is a
		minimizer  and the minimum is zero.
	\end{itemize}
\end{theorem}

\section{Linear IV models with multiple endogenous variables}

\label{sec:multivariate}

In the linear IV model with $q$ endogenous regressors, let $%
Y_2=\left(y_2,\ldots,y_{q+1}\right)$. The exclusion restrictions associated
with $k>q$ instruments are
\begin{equation}
	\mathbb{E}\left[Z^{\prime}\left(y_{1} - Y_{2} \beta^{\ast}\right)\right]=0,
	\label{eq:general-population-moments}
\end{equation}
and the corresponding sample moments are
\begin{align}
	g_n(\beta) &=n^{-1/2}Z^{\prime}\left(y_{1} -Y_{2} \beta \right)
	=n^{-1/2}Z^{\prime} Y b(\beta),  \label{eq:general-sample-moments} \\
	&=\left(b(\beta)^{\prime}\otimes I_k\right) \func{vec}\left(n^{-1/2}Z^{%
		\prime}Y\right),  \notag
\end{align}
where $\beta\in\mathbb{R}^q$, $Y=\left(y_1,Y_{2}\right)$ and $%
b(\beta)=\left(1,-\beta^{\prime}\right)^{\prime}$. As in Section \ref%
{sec:setup}, let $\widehat{\Sigma}_n$ be a consistent estimator of the
asymptotic variance matrix of $\func{vec}\left(n^{-1/2}Z^{\prime}Y\right)$.
The corresponding estimator of the variance matrix of $g_n(\beta)$ is
\begin{equation}
	\widehat{\Omega}_n(\beta) =\left(b(\beta)^{\prime}\otimes I_k\right)
	\widehat{\Sigma}_n \left(b(\beta)\otimes I_k\right).
	\label{eq:general-induced-variance}
\end{equation}
We impose Assumption \ref{ass:sigma-pd} on the enlarged $k(q+1)\times k(q+1)$
matrix $\widehat{\Sigma}_n$. This assumption implies that $\widehat{\Omega}%
_n(\beta)$ is positive definite for every $\beta \in \mathbb{R}^{q}$.
Therefore, the CUE objective is well defined by
\begin{equation}
	Q_n(\beta)=g_n(\beta)^{\prime} \widehat{\Omega}_n(\beta)^{-1}g_n(\beta).
	\label{eq:general-cue-objective}
\end{equation}

The Moore--Penrose criterion can also be defined for a positive
semidefinite matrix, but the results in this section maintain positive
definiteness. The singular extension in Section
\ref{sec:singular-univariate} concerns a scalar coefficient.

As in Section \ref{sec:setup}, the objective depends on $\beta$ only through
$b(\beta)$ and is homogeneous of degree zero in $b(\beta)$, $%
Q_n(cb(\beta))=Q_n(b(\beta))$ for every constant $c\neq0$. Thus, defining
\begin{equation}
	\overline{b}(\beta) =\frac{b(\beta)}{\left\| b(\beta)\right\|} =\frac{%
		\left(1,-\beta^{\prime}\right)^{\prime}} {\sqrt{1+\left\|\beta\right\|^2}},
	\label{eq:general-normalized-direction}
\end{equation}
we have $Q_n(\beta)=Q_n\left(\overline{b}(\beta)\right)$.

Let $\widehat{\Sigma}_{22,n}$ denote the positive definite
lower-right $kq\times kq$ block of $\widehat{\Sigma}_n$, corresponding to
the estimated variance matrix of $\func{vec}\left(n^{-1/2}Z^{\prime}Y_2%
\right)$.

For $q>1$, the limit at infinity can depend on the direction of divergence.
If $\left\|\beta\right\|\to\infty$ and
$\beta/\left\|\beta\right\|\to d$, where $\left\|d\right\|=1$, then
$\overline{b}(\beta)\to\left(0,-d^{\prime}\right)^{\prime}$.
The induced variance matrix at this limit is positive definite. Therefore, continuity
of matrix inversion gives the boundary value
\begin{equation}
\begin{split}
Q_n(\infty_d)
&:=\lim_{\substack{\left\|\beta\right\|\to\infty\\
\beta/\left\|\beta\right\|\to d}}Q_n(\beta)\\
&=\frac{1}{n}d^{\prime}Y_2^{\prime}Z
\left[\left(d^{\prime}\otimes I_k\right)\widehat{\Sigma}_{22,n}
\left(d\otimes I_k\right)\right]^{-1}Z^{\prime}Y_2d.
\end{split}
\label{eq:general-boundary-objective}
\end{equation}
Here $\infty_d$ denotes the boundary point in direction $d$.
Opposite directions have the same value:
$Q_n(\infty_d)=Q_n\left(\infty_{-d}\right)$.

To keep these directions, define the extended parameter space as
\begin{equation}
	\overline{\mathbb{R}}^{q} =\mathbb{R}^q\cup \left\{\infty_{d}:d\in\mathbb{R}%
	^q,\ \left\| d\right\|=1\right\}.  \label{eq:general-extended-space}
\end{equation}
To define the topology on this extended space, map each finite $\beta$ to $%
\beta/\sqrt{1+\left\|\beta\right\|^2}$ in the open unit ball and each
direction at infinity $\infty_d$ to the boundary point $d$. With the
topology associated to this mapping, the extended space is homeomorphic to
the closed unit ball in $\mathbb{R}^q$ and therefore it is compact.

Equivalently, normalized vectors with nonnegative first coordinate form
the closed upper unit hemisphere. The associated moments and variance
matrix are continuous there, and the variance matrix is everywhere
positive definite. Their quadratic form consequently gives a continuous
extension of $Q_n(\beta)$ to this compact set. Since finite coefficients
are dense in the extension, its attained minimum equals the infimum over
finite coefficients.

\begin{lemma}
	\label{lem:general-cue-existence}  Under Assumption \ref{ass:sigma-pd} for
	the enlarged matrix  $\widehat{\Sigma}_n$, the objective in %
	\eqref{eq:general-cue-objective},  with the boundary values in %
	\eqref{eq:general-boundary-objective}, is  continuous on the compact space $%
	\overline{\mathbb{R}}^{q}$ defined in  \eqref{eq:general-extended-space}. It
	attains its minimum, and
	\begin{equation*}
		\min_{\beta\in\overline{\mathbb{R}}^{q}}Q_n(\beta)  =\inf_{\beta\in\mathbb{R}%
			^q}Q_n(\beta).
	\end{equation*}
\end{lemma}

We define the CUE over this extended space by
\begin{equation}
	Q_n\left(\widehat{\beta}_{\mathrm{CUE}}\right) =\min_{\beta\in\overline{%
			\mathbb{R}}^{q}}Q_n(\beta).  \label{eq:general-cue-estimator}
\end{equation}
The minimum may occur at a finite parameter vector or at a direction at
infinity. Opposite boundary directions are distinct points of the extended
space but have the same objective value.

As in the case with one endogenous regressor, the probability that a global
minimizer occurs at a direction at infinity is zero if the
distribution of $\func{vec}\left(n^{-1/2}Z^{\prime}Y\right)$ conditional on $\widehat{%
	\Sigma}_n$ is absolutely continuous with respect to Lebesgue measure on $%
\mathbb{R}^{k(q+1)}$ almost surely; see Proposition S.3 in the supplemental appendix. The same conclusion holds for
the limiting objective under a Gaussian weak-instrument
approximation with a fixed positive definite variance matrix.

\subsection{Polynomial ratio representation}

\label{subsec:general-polynomial-ratio}

We now follow Section \ref{sec:poly-ratio-representation} to write the
CUE objective as a ratio of polynomials in $\beta$.
Throughout this subsection, degree means total degree, that is the degree of
a monomial is the sum of its exponents, and the degree of a nonzero
polynomial is the largest degree among its monomials with nonzero
coefficients.

Partition the enlarged variance matrix according to $Y=(y_1,Y_2)$ as
\[
\widehat{\Sigma}_n=
\begin{pmatrix}
	\widehat{\Sigma}_{11,n}&\widehat{\Sigma}_{12,n}\\
	\widehat{\Sigma}_{21,n}&\widehat{\Sigma}_{22,n}
\end{pmatrix},
\]
where the blocks have dimensions $k\times k$, $k\times kq$,
$kq\times k$, and $kq\times kq$, respectively, and
$\widehat{\Sigma}_{21,n}=\widehat{\Sigma}_{12,n}^{\prime}$. Then
\begin{equation}
		\widehat{\Omega}_n(\beta)
		=\widehat{\Sigma}_{11,n}
		-\widehat{\Sigma}_{12,n}\left(\beta\otimes I_k\right)
		-\left(\beta^{\prime}\otimes I_k\right)\widehat{\Sigma}_{21,n}
		+\left(\beta^{\prime}\otimes I_k\right)
		\widehat{\Sigma}_{22,n}\left(\beta\otimes I_k\right).
	\label{eq:general-variance-expanded}
\end{equation}
So, each entry of $\widehat{\Omega}_n(\beta)$ is a polynomial of degree at
most two. Under Assumption \ref{ass:sigma-pd}, this matrix is positive
definite for every finite $\beta$. Therefore, the adjugate formula gives
\begin{equation}
	Q_n(\beta) =\frac{g_n(\beta)^{\prime} \func{adj}\left\{\widehat{\Omega}%
		_n(\beta)\right\}g_n(\beta)} {\det\left\{\widehat{\Omega}_n(\beta)\right\}}
	\label{eq:general-polynomial-ratio}
\end{equation}

For every nonzero direction $d\in\mathbb{R}^q$, the coefficient on $\tau^2$
in $\widehat{\Omega}_n( \tau d)$ is the positive definite matrix $%
\left(d^{\prime}\otimes I_k\right)\widehat{\Sigma}_{22,n}
\left(d\otimes I_k\right)$. Therefore, its determinant is the strictly
positive coefficient on $\tau^{2k}$ in $\det\left\{\widehat{\Omega}_n(\tau
d)\right\}$. Thus, $\det\left\{\widehat{\Omega}_n(\beta)\right\}$ has degree
exactly $2k$.

Each entry of the adjugate has total degree of at most $2(k-1)$. Since the
sample moments are linear in $\beta$, the numerator of $Q_n(\beta)$ has
total degree at most $2k$.

As in the scalar case, the values at infinity determine whether the
numerator reaches this degree. If $Z^{\prime}Y_2\neq0$, there is a unit
vector $d$ such that $Z^{\prime}Y_2d\neq0$, and %
\eqref{eq:general-boundary-objective} gives $Q_n(\infty_{d})>0$. Along the
ray $\beta= \tau d$, the denominator has degree $2k$ in $\tau$, so the
numerator must also have degree $2k$ to give this positive limit. This
establishes the following extension of Proposition \ref{prop:ratio-formal}.

\begin{proposition}
	\label{prop:general-ratio} Under Assumption \ref{ass:sigma-pd} for the
	enlarged matrix $\widehat{\Sigma}_n$, the CUE objective can be written as
	\begin{equation}
		Q_n(\beta)=\frac{p_n(\beta)}{q_n(\beta)},
	\end{equation}
	where $p_n(\beta)$ and $q_n(\beta)$ are the numerator and denominator in %
	\eqref{eq:general-polynomial-ratio}, respectively, and (i) $p_n(\beta)$ is a
	polynomial of degree at most $2k$, including the possibility that it is
	identically zero, and $q_n(\beta)$ is a polynomial of degree exactly $2k$,
	(ii) $q_n(\beta)>0$ for every $\beta\in\mathbb{R}^q$, and (iii) both
	polynomials have degree $2k$ if $Z^{\prime}Y_2\neq0$.
\end{proposition}

The degree bounds depend on the number of instruments $k$, but not on the
number of endogenous regressors $q$. Nevertheless, the number of polynomial
coefficients and the difficulty of solving the resulting systems can
increase with the number of endogenous regressors.

As in the scalar case, common factors can reduce the degrees in Proposition %
\ref{prop:general-ratio}. The cancellation under homoskedastic errors
described in Section S-2 in the supplemental appendix also
applies here, with $Y=(y_1,Y_2)$ and $b(\beta)=(1,-\beta^{\prime})^{\prime}$%
, and reduces the objective to a ratio of quadratic forms.

\subsection{Global minimization through polynomial systems}

\label{subsec:general-polynomial-systems}

Under Assumption \ref{ass:sigma-pd} for the enlarged matrix $\widehat{\Sigma}%
_n$, the CUE objective is continuously differentiable on $\mathbb{R}^q$.
Using the representation in \eqref{eq:general-polynomial-ratio}, we have
\begin{equation}
	\nabla Q_n(\beta) =\frac{h_n(\beta)}{q_n(\beta)^2}, \qquad
	h_n(\beta)=q_n(\beta)\nabla p_n(\beta) -p_n(\beta)\nabla q_n(\beta).
	\label{eq:general-foc}
\end{equation}
Since $q_n(\beta)>0$ for every finite $\beta$, the finite stationary points
are exactly the real solutions of the polynomial system
\begin{equation}
	h_n(\beta)=0.  \label{eq:general-stationary-system}
\end{equation}
Each element of the $q \times 1$ vector $h_{n}(\beta)$ has total degree of
at most $4k-1$. For $q>1$, the cancellation that gives the bound $4k-2$ in
Section \ref{sec:poly-approach} need not occur. When the scalar numerator
and denominator both have degree $2k$, their highest-degree terms are
multiples of the same monomial. In the multivariate case, the homogeneous parts of highest degree in the numerator and denominator are not necessarily proportional. Thus, the same cancellation may not occur when taking partial derivatives.

Unlike a nonzero polynomial in one variable, a multivariate polynomial
system can have infinitely many real solutions. Thus, even when the
objective is not constant, \eqref{eq:general-stationary-system} need not
give a finite set of stationary points. If all $h_{j,n}(\beta)$ are zero for
every $\beta \in \mathbb{R}^{q}$, the gradient is zero everywhere on $%
\mathbb{R}^q$. The objective is then constant on $\mathbb{R}^q$ and, by
continuity, on the extended parameter space.

We must also consider minima at infinity. For $q>1$, the boundary contains a
sphere of directions, and its objective value can vary with the direction.
Let $p_{n,\infty}(d)$ and $q_{n, \infty}(d)$ be the homogeneous parts of
degree $2k$ of $p_n(\beta)$ and $q_n(\beta)$, evaluated at $d$.
Equivalently,
\begin{equation}
	p_{n,\infty}(d)=\lim_{\tau \to\infty}\frac{p_n(\tau d)}{\tau^{2k}}, \qquad
	q_{n,\infty}(d)=\lim_{\tau \to\infty}\frac{q_n(\tau d)}{\tau^{2k}}.
	\label{eq:general-leading-polynomials}
\end{equation}
Here, $p_{n,\infty}(d)$ is identically zero if $p_n(\beta)$ has degree less
than $2k$. By Proposition \ref{prop:general-ratio} and its proof,
\begin{equation*}
	q_{n,\infty}(d) =\det\left\{\left(d^{\prime}\otimes I_k\right) \widehat{%
		\Sigma}_{22,n}\left(d\otimes I_k\right)\right\}>0 \quad\text{for }%
	d\neq0.
\end{equation*}
Therefore, the boundary objective in \eqref{eq:general-boundary-objective}
can be written as
\begin{equation}
	Q_{n}(\infty_{d})=\frac{p_{n,\infty}(d)}{q_{n,\infty}(d)}, \qquad \left\|
	d\right\|=1.  \label{eq:general-boundary-ratio}
\end{equation}

At a stationary point on the unit sphere, the gradient of the boundary
ratio equals $\lambda d$ for a Lagrange multiplier $\lambda$. The ratio
is homogeneous of degree zero, so Euler's identity gives
\[
d^{\prime}\nabla\left\{\frac{p_{n,\infty}(d)}{q_{n,\infty}(d)}\right\}=0.
\]
Taking the inner product of the multiplier equation with $d$ yields
$\lambda=0$. Therefore, positivity of the denominator makes the stationary
boundary directions exactly the real solutions of
\begin{equation}
\begin{split}
h_{n,\infty}(d)
&=q_{n,\infty}(d)\nabla p_{n,\infty}(d)
-p_{n,\infty}(d)\nabla q_{n,\infty}(d)=0,\\
d^{\prime}d&=1.
\end{split}
\label{eq:general-boundary-system}
\end{equation}
Each element of the $q\times 1$ vector $h_{n,\infty}(d)$ is either
identically zero or a homogeneous polynomial of degree $4k-1$. The following
theorem extends Theorem \ref{thm:global-min-scalar}.

\begin{theorem}
	\label{thm:general-global-min} Suppose Assumption \ref{ass:sigma-pd} holds
	for the enlarged matrix $\widehat{\Sigma}_n$. Define
	\begin{equation}
		\mathcal{C}_n =\left\{\beta\in\mathbb{R}^q:h_n(\beta)=0\right\} \cup
		\left\{\infty_{d}: \left\| d\right\|=1, \ h_{n,\infty}(d)=0\right\}.
		\label{eq:general-candidate-set}
	\end{equation}
	Every global minimizer belongs to $\mathcal{C}_n$, and
	\begin{equation}
		\min_{\beta\in\overline{\mathbb{R}}^{q}}Q_n(\beta) =\min_{\beta\in\mathcal{C}%
			_n} Q_n(\beta)  \label{eq:general-min-candidates}
	\end{equation}
	If $h_{n}(\beta)$ is zero for all $\beta$, then $Q_n(\beta)$ is constant on
	the extended parameter space and every point is a global minimizer.
\end{theorem}

Thus, global minimization can be characterized by two polynomial systems.
One for finite stationary points and one for stationary directions at
infinity. Opposite boundary directions give the same objective value. Thus,
only one representative of each pair is needed. For $q=1$, the boundary
ratio is constant on the two directions $d=\pm1$, and the finite system
reduces to the scalar polynomial equation in Section \ref{sec:poly-approach}.

Even when there are infinitely many stationary points, there are only
finitely many distinct stationary objective values. The real solution sets
of \eqref{eq:general-stationary-system} and %
\eqref{eq:general-boundary-system} are defined by polynomial equations and
hence are semialgebraic. Therefore, they have finitely many connected
components, each semialgebraic; see \citet[Theorem
2.4.5]{BochnakCosteRoy1998}. Any two points in a component can be joined by
a continuous semialgebraic path in that component; see \citet[Proposition
2.5.13]{BochnakCosteRoy1998}. Moreover, the path is smooth on each open
subinterval of a finite partition of its parameter interval; see %
\citet[Proposition 2.9.10 and its proof]{BochnakCosteRoy1998}. Along each
smooth piece of a path $\beta(\tau)$ in the finite stationary set,
\begin{equation}
	\frac{\partial}{\partial \tau}Q_n\left(\beta(\tau)\right) =\nabla
	Q_n\left(\beta(\tau)\right)^{\prime} \frac{\partial}{\partial \tau}%
	\beta(\tau)=0.  \label{eq:general-constant-along-path}
\end{equation}
Thus, the objective is constant on each smooth piece. Continuity implies
that these constants agree across the partition points, so the objective is
constant on each connected component of the finite stationary set. The same
argument applies to the boundary ratio in \eqref{eq:general-boundary-ratio},
whose tangent derivatives are zero at every stationary direction. Thus, a
connected component can contain infinitely many stationary points but
contributes only one objective value.

\begin{proposition}\label{prop:general-finite-critical-values} 
	Under Assumption \ref{ass:sigma-pd} for the
	enlarged matrix $\widehat{\Sigma}_n$, the finite stationary set in %
	\eqref{eq:general-stationary-system} and the stationary direction set in %
	\eqref{eq:general-boundary-system} each have finitely many connected
	components. The corresponding objective is constant on each component.
	Consequently,
	\begin{equation*}
		\left\{Q_n(\beta):\beta\in\mathcal{C}_n\right\}
	\end{equation*}
	is a finite, nonempty set, where $\mathcal{C}_n$ is defined
	in \eqref{eq:general-candidate-set}. Its smallest element
	is the global minimum of the extended CUE objective.
\end{proposition}

The proposition reduces the global minimization problem to a finite
comparison of objective values, although finding a representative from each
real connected component can still be challenging.

\subsection{Computing stationary objective values}

\label{subsec:general-solving-systems}

Proposition \ref{prop:general-finite-critical-values} shows that the global
minimization requires comparing finitely many stationary objective values,
even when there are infinitely many stationary points. Therefore, we can
target these values directly rather than recover all stationary points. Both
the finite stationary system \eqref{eq:general-stationary-system} and the
boundary system \eqref{eq:general-boundary-system} must be considered.

Introduce a scalar $t$ representing the objective value and add
\begin{equation}
	p_n(\beta)-t q_n(\beta)=0  \label{eq:general-critical-value-equation}
\end{equation}
to \eqref{eq:general-stationary-system}. These are $q+1$ polynomial
equations in $\beta$ and $t$. While $q_n(\beta)>0$ for every real $\beta$,
it can be equal to zero at complex points. Algebraic elimination considers
complex solutions, so we also introduce an auxiliary variable $s$ and impose
\begin{equation}
	s q_n(\beta)-1=0.  \label{eq:general-denominator-exclusion}
\end{equation}
This equation excludes denominator zeros and preserves every real stationary
point, with $s=1/q_n(\beta)$.

Substituting $p_n(\beta)=tq_n(\beta)$ into \eqref{eq:general-foc} gives, on
this level set,
\begin{equation*}
	h_{n}(\beta) =q_n(\beta)\left( \nabla p_n(\beta) -t \nabla q_n(\beta)
	\right).
\end{equation*}
Since \eqref{eq:general-denominator-exclusion} ensures $q_n(\beta)\neq0$, we
can replace the stationary equations by
\begin{equation}
	\nabla p_n(\beta) -t \nabla q_n(\beta) = 0.
	\label{eq:general-value-derivatives}
\end{equation}
We use \eqref{eq:general-critical-value-equation}, %
\eqref{eq:general-denominator-exclusion}, and %
\eqref{eq:general-value-derivatives} for elimination. The derivative equations have degree at most $2k-1$ in $\beta$, compared with at most $4k-1$ for the original stationary equations, and are at most linear in $t$. Their degree in $(\beta,t)$ is at most $2k$.

A Gr\"{o}bner basis is a finite generating set for the ideal of polynomial
combinations of the equations that permits systematic polynomial division.
With a lexicographic order placing $\beta$ and $s$ before $t$, it permits
elimination of $\beta$ and $s$; see \citet[Theorem 1.4.1 and Corollary
1.4.2]{Hibi2013}. For this augmented stationary system, the resulting
elimination ideal contains a nonzero univariate polynomial
\begin{equation}
	\psi_n(t)=0.  \label{eq:general-value-eliminant}
\end{equation}
Every real stationary objective value is a root of $\psi_n(t)$.
Proposition S.4 in the
supplemental appendix establishes this claim by showing that there are
finitely many complex stationary values as well. The complex
stationary set where the denominator is nonzero has finitely
many connected components, and the rational objective is
constant on each component. This additional argument is needed
because Proposition \ref{prop:general-finite-critical-values}
concerns only real solutions, whereas elimination also considers
complex solutions. If the augmented system has no complex
solution, its elimination ideal contains $1$, indicating that
there are no candidates.

For boundary values, apply the same construction to %
\eqref{eq:general-boundary-system}, including $d^{\prime}d-1=0$, and add
\begin{equation}
	p_{n,\infty}(d)-t q_{n,\infty}(d)=0, \qquad s q_{n,\infty}(d)-1=0.
	\label{eq:general-boundary-value-equations}
\end{equation}
As in the finite case, replace the boundary stationary equations by
\begin{equation}
	\nabla p_{n,\infty}(d) -t\nabla q_{n,\infty}(d)=0.
	\label{eq:general-boundary-value-derivatives}
\end{equation}
These equations have degree at most $2k-1$ in $d$ and are at most linear in $%
t$. Together with $d^{\prime}d=1$ and %
\eqref{eq:general-boundary-value-equations}, they define the same augmented
solution set as the original boundary stationary equations. Eliminating $d$ and $s$ similarly gives a nonzero univariate polynomial with roots that include every real stationary boundary value.

Consider all distinct real roots of the two elimination polynomials and order them from smallest to largest. A real candidate value of the objective function can correspond to complex stationary points without being attainable at any real point. Thus, we need to start with the smallest candidate value of the objective function and check whether there is at least one real solution corresponding to this value. For a candidate $t_{i}$ from the finite stationary system,
check whether $p_n(\beta)-t_i q_n(\beta)=0$ admits a solution $\beta\in%
\mathbb{R}^q$. For a candidate $t_{i}$ from the boundary stationary system, check
whether $p_{n,\infty}(d)-t_i q_{n,\infty}(d)=0$ admits a real solution
satisfying $d^{\prime}d=1$. Since the global minimum is among the candidate values, the smallest candidate value for which a real solution exists is the global minimum. The stationary equations need not be included in these checks.

Cylindrical algebraic decomposition provides an exact method for checking
whether these equations have real solutions; see \citet{Collins1975} or %
\citet{BasuPollackRoy2006}. For a candidate value $t_i$ from the finite
stationary system, it partitions $\mathbb{R}^q$ into finitely many cells on
which $p_n(\beta)-t_i q_n(\beta)$ has constant sign and provides a sample
point in each cell. The equation has a real solution if and only if this
polynomial is equal to zero at one of these sample points. For a candidate
from the boundary stationary system, apply the same procedure to $%
p_{n,\infty}(d)-t_i q_{n,\infty}(d)$ and $d^{\prime}d-1$, requiring both
polynomials to be equal to zero at the same sample point. At the smallest
candidate value for which either check yields a real solution, the
corresponding sample point provides a finite CUE or a minimizing boundary
direction. When the coefficients are specified exactly as rational or real algebraic numbers, the characterization is a finite algebraic procedure. With floating-point coefficients, this procedure applies exactly to the rounded coefficients. The effects of rounding the coefficients must be assessed separately. Moreover, the cost of elimination and real-feasibility checks can grow rapidly with the number of endogenous variables and the degrees of the polynomial equations. Thus, the multivariate result neither provides the same simple eigenvalue calculation as the scalar result nor establishes computational efficiency in general.

\section{Numerical studies}\label{sec:simulation}

This section examines how computing the global minimum affects
estimation and testing. We first compare the polynomial and
quasi-Newton methods for computing CUE in simulations with one
and two endogenous regressors. We then compare CUE with 2SLS,
LIML, two-step GMM, and iterated GMM in simulations and five
empirical applications. These comparisons distinguish differences
arising from numerical implementation from those arising from
the choice of estimator. Finally, we examine how the computation
of the global minimum affects CLR statistics, conditional critical
values, and rejection decisions.

The simulations follow the designs in
\citet{HasmanNeweyWoutersenChaoSwanson12} and extend them to
two endogenous regressors. They show that failure to find
the global minimum can distort the reported finite-sample
behavior of CUE, even when quasi-Newton methods use starting
points close to the true parameter. In the reported designs,
CUE computed using the polynomial method generally has
smaller absolute median bias than the other estimators
and less dispersion than LIML.

Motivated by the theoretical results of \citet{NeweyWindmeijer09}
under many weak moment conditions and heteroskedasticity, we next
compare these estimators across five applications selected for
their prominence in the IV literature. Three are discussed in
the review article by \citet{AndrewsStockSun19}: the public
spending multiplier in \citet{AcconciaCorsettiSimonelli14},
returns to schooling in \citet{StephensYang14}, and the elasticity
of worker efficacy with respect to employment shares in
\citet{Young14}. We also consider two commonly referenced
applications: returns to schooling in \citet{AngristKrueger91}
and the intertemporal elasticity of substitution in \citet{Yogo04}.

The CLR simulations use the designs of
\citet{MoreiraRidderSharifvaghefi21a}, which allow for
heteroskedasticity and autocorrelation. The polynomial
and quasi-Newton implementations produce different
rejection decisions under both the null and alternative
hypotheses. These disagreements show the sensitivity
of the implemented test to the computation of the global
minimum. They do not necessarily translate into differences of the same magnitude in the size and power of the CLR test across implementations.

\subsection{Simulation studies: estimation}

We closely follow the simulation designs of
\citet{HasmanNeweyWoutersenChaoSwanson12}, and consider
\begin{equation*}
	y_{1i}=y_{2i}\beta^{\ast}+u_i
	\qquad\text{and}\qquad
	y_{2i}=z_{1i}\pi+v_{2i},
\end{equation*}
where $z_{1i}$ and $v_{2i}$ are independent standard normal
variables. The structural parameter is set to $\beta^{\ast} = 5$. Furthermore,
\begin{equation*}
	u_i=\rho v_{2i}
	+\sqrt{1-\rho^2}
	\left(w\xi_{1i}+\sqrt{1-w^2}\xi_{2i}\right),
	\qquad
	\xi_{1i}\mid z_{1i}\sim\mathcal{N}(0,z_{1i}^2),
	\qquad
	\xi_{2i}\sim\mathcal{N}(0,1).
\end{equation*}
Conditional on $z_{1i}$, the variables $v_{2i}$, $\xi_{1i}$,
and $\xi_{2i}$ are independent, and observations are independent
across $i$.

We set the sample size to $n=800$. The parameter $\rho$ controls
endogeneity: when $\rho=0$, the regressor is exogenous, whereas
larger values of $|\rho|$ increase the absolute correlation between
$v_{2i}$ and $u_i$. We consider $\rho=0.3,0.6,0.9$.
The parameter $w$ controls heteroskedasticity. When $w=0$,
the structural error is homoskedastic, whereas larger values
of $w$ increase the variation in its conditional variance.
We consider $w=0,1/2,\sqrt{2}/2,\sqrt{3}/2,1$.
We control the first-stage signal through the concentration
parameter $\mu^2=n\pi^2$, taking values $1$, $4$, and $16$.
\footnote{The normalization uses $\operatorname{Var}(z_{1i})=\operatorname{Var}(v_{2i})=1$.
	Under heteroskedasticity, $\mu^2$ alone does not fully characterize
	the information available for estimating $\beta^{\ast}$.}

We consider $k=1,4,10,30,60$ instruments. When $k=1$,
$z_{1i}$ is the only instrument. As discussed in
Section \ref{sec:setup}, in this just-identified case, CUE,
LIML, 2SLS, two-step GMM, and iterated GMM coincide whenever
the sample first-stage coefficient is nonzero. We include
this case as a diagnostic benchmark for the numerical procedures.
For $k=4$, the instruments are the first four powers of $z_{1i}$.
For $k=10,30,60$, we also include interactions with independent
Bernoulli variables:
\[
Z_i=\left(
z_{1i},z_{1i}^2,z_{1i}^3,z_{1i}^4,
D_{1i}z_{1i},\ldots,D_{(k-4)i}z_{1i}
\right)^{\prime},
\qquad
D_{ji}\sim\operatorname{Bernoulli}\left(\frac{1}{2}\right).
\]
The dummy variables are independent of the other underlying
variables. 

The reported results are based on 10,000 Monte Carlo replications.
In every replication, we treat the intercept as a deterministic
exogenous variable and partial it out by demeaning $y_{1i}$,
$y_{2i}$, and each component of $Z_i$. The tables report median
bias and, in parentheses, the 90\% interquantile range, defined
as the difference between the 95th and 5th percentiles of the
estimator's empirical distribution. Table \ref{tablelocalopt} shows the results only for $\mu^2=4$, $k=1,10,30,60$, $\rho=0.3$, and
$w=\sqrt{2}/2$. It illustrates that local optimization can fail to find the global minimum even when we initialize from the infeasible true parameter value or the 2SLS estimator. The full set of simulation results is reported in the simulation supplement.

\begin{table}[htbp]
	\centering
	\caption{Numerical computation of CUE:
		median bias and 90\% interquantile range}
	\label{tablelocalopt}
	\small
	\setlength{\tabcolsep}{6pt}
	\renewcommand{\arraystretch}{1}
	
	\begin{tabular}{ccccccc}
		\toprule\toprule
		& & \multicolumn{4}{c}{Quasi-Newton method with starting value:} & Polynomial \\
		\cmidrule(lr){3-6}
		$\mu^2$ & $k$ & $\widehat{\beta}_{\mathrm{2SLS}}$ & $5$ & $4.5$ & $0$ & method \\
		\midrule
		\multirow{8}{*}{$4$}
		& \multirow{2}{*}{$1$} & 0.0194 & 0.0029 & -0.0290 & -0.3430 & 0.0194 \\
		& & (3.5929) & (3.7701) & (4.8258) & (2852.2165) & (3.5929) \\
		\addlinespace[3pt]
		& \multirow{2}{*}{$10$} & 0.0701 & -0.0520 & -0.2716 & -1.8390 & 0.0686 \\
		& & (12.4874) & (132.1221) & (1205.7786) & (4395.3701) & (14.0699) \\
		\addlinespace[3pt]
		& \multirow{2}{*}{$30$} & 0.1210 & -0.1750 & -0.7229 & -4.1714 & 0.0677 \\
		& & (19.2901) & (674.1622) & (1671.0159) & (5393.9388) & (18.9725) \\
		\addlinespace[3pt]
		& \multirow{2}{*}{$60$} & 0.1706 & -0.3526 & -1.2976 & -7.3877 & 0.0779 \\
		& & (51.9800) & (1008.3832) & (1900.2972) & (6133.2933) & (27.0050) \\
		\bottomrule
	\end{tabular}
	
	\medskip
	\begin{minipage}{\linewidth}
		\footnotesize
		\textit{Notes:}
		Entries report median bias, with the difference between the
		95th and 5th percentiles in parentheses.
		Results are based on 10,000 replications with $n=800$,
		$\beta^{\ast}=5$, $\rho=0.3$, and $w=\sqrt{2}/2$.
	\end{minipage}
\end{table}

Table \ref{tablelocalopt} compares CUE computed using the polynomial
method with estimates obtained by the quasi-Newton method\footnote{We implement the quasi-Newton method in MATLAB using
	\texttt{fminunc}.}, starting from the 2SLS estimator and the fixed values $5$, $4.5$,
and $0$. Considering the infeasible benchmark of $\beta^{\ast}=5$ allows us to assess whether starting from the true
parameter value resolves the numerical difficulty. The other fixed
values assess sensitivity to starting points near and farther
from the true parameter.

Even in the just-identified case $k=1$, where an explicit solution
is available as discussed in Section \ref{sec:setup}, the
quasi-Newton procedure can fail to recover CUE. Starting at
the true parameter does not eliminate this problem. For example,
when $\mu^2=4$ and $k=10$, the 90\% interquantile range is
$132.1221$ when starting at $5$, compared with $14.0699$ for CUE
computed using the polynomial method. Starting at 2SLS recovers
CUE when $k=1$, but discrepancies arise when $k=10$, $30$, or $60$.
For $\mu^2=4$ and $k=60$, the ranges for the quasi-Newton method
started at 2SLS and the polynomial method are $51.9800$ and $27.0050$,
respectively.

These results show that the failure of a numerical method to recover the global minimum of the CUE objective can substantially change the reported finite-sample behavior of CUE \footnote{In the simulations reported in the simulation supplement,
	the objective value obtained by the polynomial method is never
	larger than that obtained by the quasi-Newton methods and can
	be substantially smaller.}. 
The theoretical properties established for CUE, including those under the many weak moment conditions of Newey and Windmeijer (2009), concern the estimator defined by global minimization of the CUE objective. They do not necessarily extend to numerical approximations that fail to recover that estimator.

\begin{table}[htbp]
	\centering
	\caption{Estimator comparison:
		median bias and 90\% interquantile range}
	\label{tableestimators}
	\small
	\setlength{\tabcolsep}{7pt}
	\renewcommand{\arraystretch}{1.05}
	
	\begin{tabular}{@{}ccccccc@{}}
		\toprule\toprule
		$\mu^2$ & $k$ & 2SLS & LIML & Two-step & Iterated & CUE \\
		\midrule
		\multirow{8}{*}{$4$}
		& \multirow{2}{*}{$1$} & 0.0194 & 0.0194 & 0.0194 & 0.0194 & 0.0194 \\
		& & (3.5929) & (3.5929) & (3.5929) & (3.5929) & (3.5929) \\
		\addlinespace[3pt]
		& \multirow{2}{*}{$10$} & 0.2084 & -0.3479 & 0.2052 & 0.2064 & 0.0686 \\
		& & (1.3698) & (27.6176) & (1.4856) & (1.4824) & (14.0699) \\
		\addlinespace[3pt]
		& \multirow{2}{*}{$30$} & 0.2624 & -1.0925 & 0.2621 & 0.2619 & 0.0677 \\
		& & (0.7935) & (48.8021) & (0.8874) & (0.8882) & (18.9725) \\
		\addlinespace[3pt]
		& \multirow{2}{*}{$60$} & 0.2803 & -1.7310 & 0.2800 & 0.2802 & 0.0779 \\
		& & (0.5473) & (70.7657) & (0.6570) & (0.6588) & (27.0050) \\
		\bottomrule
	\end{tabular}
	
	\medskip
	\begin{minipage}{\linewidth}
		\footnotesize
		\textit{Notes:}
		Entries report median bias, with the difference between the
		95th and 5th percentiles in parentheses.
		Results are based on 10,000 replications with $n=800$,
		$\beta^{\ast}=5$, $\rho=0.3$, and $w=\sqrt{2}/2$.
		CUE is computed using the polynomial method.
	\end{minipage}
\end{table}

Table \ref{tableestimators} compares CUE with 2SLS, LIML,
two-step GMM, and iterated GMM. For two-step GMM, we first
minimize the GMM objective using the identity weight matrix.
We then evaluate the estimated variance matrix at this preliminary
estimate and minimize the GMM objective using its inverse as
the fixed weight matrix. For iterated GMM, we repeat the
weight-update and minimization procedure 100 times.

In the reported overidentified designs, CUE has a larger
90\% interquantile range than 2SLS, two-step GMM, and iterated
GMM, but a smaller range than LIML. CUE also has smaller absolute
median bias than all these estimators. The combination of larger absolute median bias and
smaller dispersion shows that 2SLS, two-step GMM, and iterated
GMM can be tightly concentrated away from the true parameter.
This is a finite-sample comparison under heteroskedasticity, not a
verification of a many-instrument limit. Two-step and iterated GMM produce
results close to 2SLS in these designs. The iterated-GMM column uses the weight matrix obtained after 100 updates. Interpreting the result as a converged fixed point would require additional convergence checks. In any case, the weight matrix at each step of iterated GMM is fixed. Hence, its first-order conditions do not contain all the terms in the first-order conditions of CUE, where both the moments and the weight matrix vary with the parameters.

Numerical implementation may help explain the extreme dispersion reported for CUE in earlier studies. For example,
\citet[Table 5]{HasmanNeweyWoutersenChaoSwanson12} report a 90\% interquantile range of $1.5\times10^{12}$ for CUE, compared with $60.512$ for LIML, in a heteroskedastic design. An exact replication would require access to the original simulation code and additional details on the design and CUE implementation. In our closely related designs, while CUE computed using the polynomial method has considerably less dispersion than LIML, CUE computed using quasi-Newton implementations can have greater dispersion than LIML. These findings suggest that numerical optimization may contribute to the extreme dispersion reported previously.

Table \ref{tab:multivariate_case} considers an extension of the
designs in \citet{HasmanNeweyWoutersenChaoSwanson12} to two
endogenous regressors:
\begin{align*}
	y_{1i}
	&=y_{2i}\beta_1^{\ast}+y_{3i}\beta_2^{\ast}+u_i,\\
	y_{2i}&=z_{1i}\pi_1+v_{2i},
	\qquad
	y_{3i}=z_{2i}\pi_2+v_{3i},
\end{align*}
where $z_{1i}$, $z_{2i}$, $v_{2i}$, and $v_{3i}$ are independent
standard normal variables. The structural error is
\[
u_i=\rho_1v_{2i}+\rho_2v_{3i}
+\sqrt{1-\rho_1^2-\rho_2^2}
\left(w\xi_{1i}+\sqrt{1-w^2}\xi_{2i}\right),
\]
with
\[
\xi_{1i}\mid z_{1i},z_{2i}
\sim\mathcal{N}\left(0,\left(\frac{z_{1i}^2+z_{2i}^2}{2}\right)^2\right),
\qquad
\xi_{2i}\sim\mathcal{N}(0,1).
\]
Conditional on the instruments, $v_{2i}$, $v_{3i}$,
$\xi_{1i}$, and $\xi_{2i}$ are independent, and observations
are independent across $i$.

We set $n=800$, $\pi_1=\pi_2=\pi$ with $n\pi^2=4$,
$\beta_1^{\ast}=\beta_2^{\ast}=2$,
$\rho_1=\rho_2=0.4$, and $w=\sqrt{2}/2$.
The five instruments are $Z_i=\left(z_{1i},z_{1i}^2,z_{2i},z_{2i}^2,
z_{1i}z_{2i}\right)^{\prime}.$

The multivariate polynomial method requires solving systems
of polynomial equations, as discussed in
Section \ref{subsec:general-solving-systems}.
The results broadly agree with the univariate findings.
In the reported simulations, the objective value obtained
using the polynomial method is no larger than that obtained
using the quasi-Newton methods. Panel A of
Table \ref{tab:multivariate_case} shows that the numerical
method and starting point can substantially change the reported
median bias and dispersion of CUE. Starting from 2SLS produces
results close to those of the polynomial method in this design.
However, as the univariate results in Table \ref{tablelocalopt} show, 
starting from 2SLS does not always recover CUE. 

\begin{table}[htbp]
	\centering
	\caption{Multivariate estimation:
		median bias and 90\% interquantile range}
	\label{tab:multivariate_case}
	\small
	\setlength{\tabcolsep}{6pt}
	\renewcommand{\arraystretch}{1.05}
	
	\begin{tabular}{@{}cccccc@{}}
		\toprule \toprule
		\multicolumn{6}{@{}l}{\textit{Panel A: Numerical methods for computing CUE}} \\
		\addlinespace[4pt]
		& \multicolumn{4}{c}{Quasi-Newton method with starting values:} & Polynomial \\
		\cmidrule(lr){2-5}
		Parameter & $\widehat{\beta}_{\mathrm{2SLS}}$
		& $(2,2)^{\prime}$ & $(1.5,1.5)^{\prime}$ & $(0,0)^{\prime}$ & method \\
		\midrule
		$\beta_1$ & 0.0533 & -0.0351 & -0.1270 & -0.3624 & 0.0608 \\
		& (12.4382) & (165.0510) & (74777.2079)
		& (246259.2421) & (13.3977) \\
		\addlinespace[4pt]
		$\beta_2$ & 0.0552 & -0.0326 & -0.1134 & -0.3965 & 0.0527 \\
		& (11.8000) & (61.8557) & (67203.5305)
		& (269375.2046) & (12.5874) \\
		\midrule
		
		\multicolumn{6}{@{}l}{\textit{Panel B: Comparison across estimators}} \\
		\addlinespace[4pt]
		Parameter & 2SLS & Two-step & Iterated & LIML & CUE \\
		\midrule
		$\beta_1$ & 0.1951 & 0.1679 & 0.1737 & -0.1387 & 0.0608 \\
		& (2.7174) & (2.6391) & (2.5934) & (23.0126) & (13.3977) \\
		\addlinespace[4pt]
		$\beta_2$ & 0.1930 & 0.1698 & 0.1694 & -0.1564 & 0.0527 \\
		& (2.6814) & (2.6573) & (2.6254) & (22.1940) & (12.5874) \\
		\bottomrule
	\end{tabular}
	
	\medskip
	\begin{minipage}{\linewidth}
		\footnotesize
		\textit{Notes:}
		Entries report median bias, with the difference between the
		95th and 5th percentiles in parentheses.
		The true coefficients are $\beta_1^{\ast}=\beta_2^{\ast}=2$.
		In Panel A, $\widehat{\beta}_{\mathrm{2SLS}}$ denotes the
		vector of 2SLS estimates used to initialize the quasi-Newton
		method. In both panels, the final column reports CUE computed
		using the polynomial method.
	\end{minipage}
\end{table}

Panel B compares CUE with the other estimators. For both
coefficients, CUE has the smallest absolute median bias.
Its 90\% interquantile range is smaller than that of LIML,
but larger than those of 2SLS, two-step GMM, and iterated GMM.

\subsection{Empirical applications}
\label{subsec:empirical-applications}

We compare CUE with 2SLS, LIML, and iterated GMM across the five
applications introduced above, retaining the specifications considered
in each application. Our comparison includes 20 specifications from
\citet{AcconciaCorsettiSimonelli14}, each with two instruments;
30 from \citet{StephensYang14}, each with three instruments; and
44 from \citet{Yogo04}, each with four instruments. We also consider
40 specifications from \citet{Young14}, with 21--59 instruments,
and 24 from \citet{AngristKrueger91}, with 3--180 instruments across
different census samples. The purpose is to assess how the choice of estimator affects the reported coefficients, rather than to reassess
the validity of the instruments or the specifications.

\begin{figure}[tbp]
	\centering
	\begin{subfigure}[t]{0.49\textwidth}
		\centering
		\includegraphics[trim = 0mm 0mm 0mm 0mm, clip, width=\linewidth]{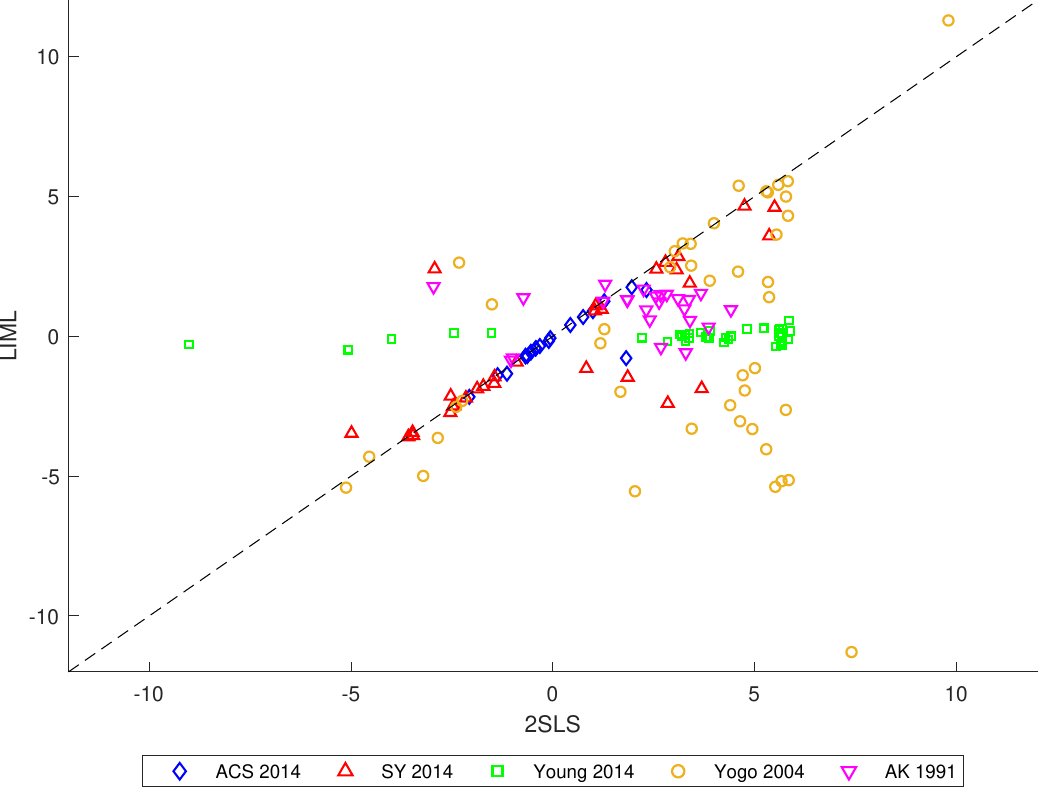}
	\end{subfigure}
	\begin{subfigure}[t]{0.49\textwidth}
		\centering
		\includegraphics[trim = 0mm 0mm 0mm 0mm, clip, width=\linewidth]{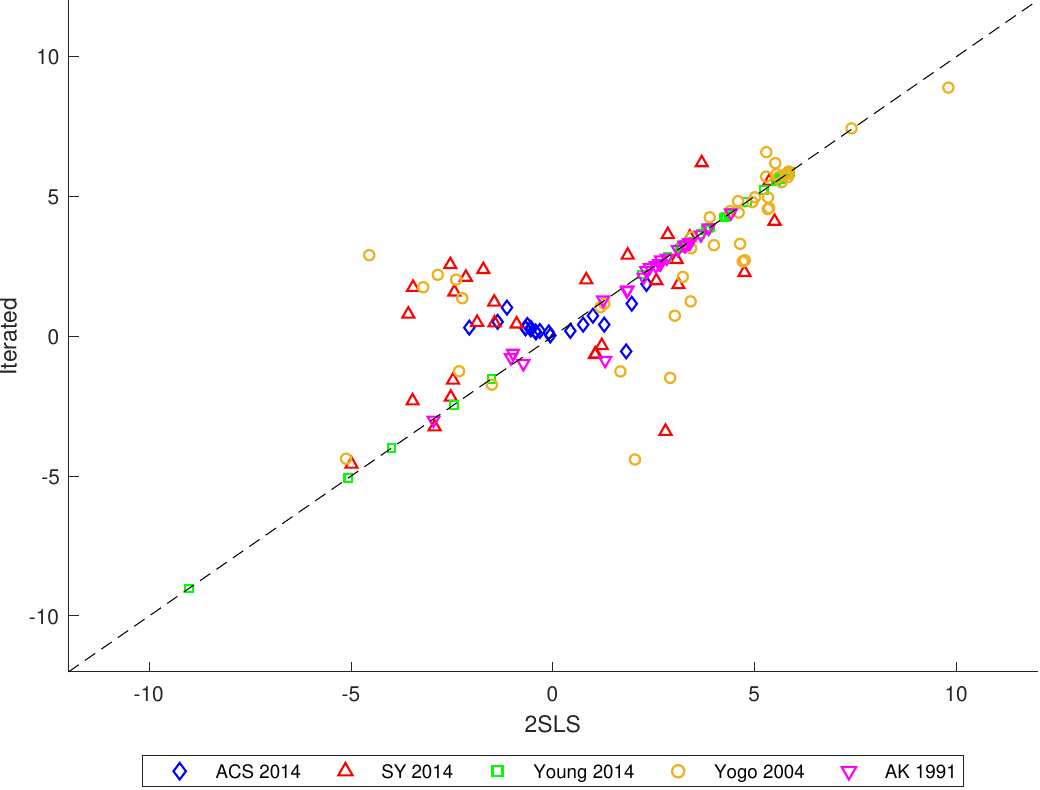}
	\end{subfigure}
	\caption{Cube-root-transformed percentage differences between CUE and alternative estimators}
	\label{pctdiff}
\end{figure}

Figure \ref{pctdiff} compares the percentage differences between CUE
and the other estimators across specifications. Each difference is
scaled by the average of the two estimates and expressed as a
percentage. We apply the real cube-root transformation to compress
large differences while preserving their signs and ordering.
The left panel plots the transformed differences for 2SLS against
those for LIML; the right panel compares 2SLS with iterated GMM.
Points near the origin show that the two estimators are close to the CUE, whereas points near the 45-degree line show that the estimators are similar in terms of their transformed percentage differences from the CUE. Because the percentage difference is calculated by dividing the difference between the estimates by their average, it can be large when that average is close to zero.

Blue diamonds denote \citet{AcconciaCorsettiSimonelli14}, red
upward triangles denote \citet{StephensYang14}, green squares denote
\citet{Young14}, orange circles denote \citet{Yogo04}, and magenta
downward triangles denote \citet{AngristKrueger91}.

In the left panel, LIML estimates are close to CUE in many
specifications from \citet{Young14} and \citet{AngristKrueger91},
while 2SLS estimates show larger percentage differences. Other
specifications lie near the diagonal but away from the origin,
indicating similar percentage differences for 2SLS and LIML relative
to CUE. The right panel shows that iterated GMM often produces
estimates close to 2SLS. As in the simulations, repeated weight
updates need not recover CUE.

\begin{table}
	\small
	\centering
	\renewcommand{\arraystretch}{1.1}
	\caption{Quartiles of absolute percentage differences between CUE and alternative estimators} \label{pdifftable}
	\begin{tabular}{ccccc}
		\hline\hline
		Application & Estimator & 25th percentile & Median & 75th percentile \\
		\hline
		& 2SLS & 0.0800 & 0.3107 & 2.3135 \\ 
		\cite{AcconciaCorsettiSimonelli14} & LIML & 0.0808 & 0.3482 & 2.1771 \\ 
		& Iterated GMM & 0.0068 & 0.0282 & 0.1479 \\ \hline
		& 2SLS & 5.1138 & 16.6242 & 41.5516 \\ 
		\cite{StephensYang14} & LIML & 4.7907 & 13.5528 & 23.1492 \\ 
		& Iterated GMM & 1.8198 & 9.8726 & 34.0177 \\ \hline
		& 2SLS & 38.5661 & 84.4015 & 178.7024 \\ 
		\cite{Young14} & LIML & 0.0003 & 0.0017 & 0.0163 \\ 
		& Iterated GMM & 38.5604 & 84.4022 & 178.6905 \\ \hline
		& 2SLS & 30.3433 & 98.6778 & 160.5497 \\ 
		\cite{Yogo04} & LIML & 12.3428 & 36.3257 & 136.0933 \\ 
		& Iterated GMM & 8.9277 & 84.7788 & 183.7710 \\ \hline
		& 2SLS & 6.2673 & 18.6199 & 35.1441 \\ 
		\cite{AngristKrueger91} & LIML & 0.5423 & 2.0802 & 3.1129 \\ 
		& Iterated GMM & 4.4153 & 17.5725 & 34.6890 \\ \hline
	\end{tabular}
\end{table}

Table \ref{pdifftable} reports the quartiles of the absolute
percentage differences within each application, before the cube-root
transformation. The median differences are below $0.4\%$ for all
three estimators in \citet{AcconciaCorsettiSimonelli14}. For
\citet{StephensYang14}, they range from $9.87\%$ to $16.62\%$,
and for \citet{Yogo04}, from $36.33\%$ to $98.68\%$.
Thus, substantial differences occur even in applications with few
instruments. In \citet{Young14}, the median difference is approximately
$84.4\%$ for both 2SLS and iterated GMM, compared with $0.0017\%$
for LIML. In \citet{AngristKrueger91}, the corresponding medians are
$18.62\%$, $17.57\%$, and $2.08\%$. These comparisons show that
estimator choice can really affect the reported coefficients. They do not by themselves identify which estimator is closer to the
true parameter or establish the source of the differences.

\subsection{Simulation studies: testing}\label{sec: MC-CLR}

We examine how the computation of the global minimum of the CUE
objective affects the implementation of the CLR test. Following
\citet{MoreiraRidderSharifvaghefi21a}, we simulate the standardized
statistic $R=\left(Z^{\prime}Z\right)^{-1/2}Z^{\prime}Y$ directly
from the Gaussian experiment
\begin{equation}\label{eq: dgp_r}
	\vect(R)\sim\mathcal{N}\left(
	\vect\left(\mu a^{\ast\prime}\right),\widetilde{\Sigma}\right),
\end{equation}
where $a^{\ast}=\left(\beta^{\ast},1\right)^{\prime}$,
$\mu\in\mathbb{R}^{k}$, and the variance matrix
\[
\widetilde{\Sigma}
=\left(I_2\otimes\left(Z^{\prime}Z/n\right)^{-1/2}\right)
\Sigma
\left(I_2\otimes\left(Z^{\prime}Z/n\right)^{-1/2}\right)
\]
is treated as known. Partition $\widetilde{\Sigma}$ into $k\times k$
blocks. We use the HAC design in \citet{MoreiraRidderSharifvaghefi21a}, with
\[
\widetilde{\Sigma}_{11}=c_{11}I_k,\qquad
\widetilde{\Sigma}_{12}=\widetilde{\Sigma}_{21}=c_{12}J_k,\qquad
\widetilde{\Sigma}_{22}=c_{22}I_k,
\]
where $J_k$ has ones on its anti-diagonal and zeros elsewhere.
We set $c_{11}=1$, $c_{12}=100$, and
$c_{22}=c_{12}^{2}+c_{12}^{-3}$. 

With $b(\beta)=\left(1,-\beta\right)^{\prime}$, the CUE objective is
\begin{equation}\label{eq:CU-GMM_obj_r}
	\begin{split}
		Q_n(\beta)
		&=\frac{1}{n}b(\beta)^{\prime}Y^{\prime}Z
		\left[\left(b(\beta)^{\prime}\otimes I_k\right)\Sigma
		\left(b(\beta)\otimes I_k\right)\right]^{-1}
		Z^{\prime}Yb(\beta)\\
		&=b(\beta)^{\prime}R^{\prime}
		\left[\left(b(\beta)^{\prime}\otimes I_k\right)\widetilde{\Sigma}
		\left(b(\beta)\otimes I_k\right)\right]^{-1}Rb(\beta).
	\end{split}
\end{equation}
For testing $H_0:\beta^{\ast}=\beta_0$ against
$H_1:\beta^{\ast}\neq\beta_0$, the LR statistic is
\[
LR_n(\beta_0)=Q_n(\beta_0)
-\min_{\beta\in\overline{\mathbb{R}}}Q_n(\beta).
\]
Let $a_0=\left(\beta_0,1\right)^{\prime}$ and $b_0=b(\beta_0)$, and define
\begin{align*}
	T&=\left[\left(a_0^{\prime}\otimes I_k\right)
	\widetilde{\Sigma}^{-1}\left(a_0\otimes I_k\right)\right]^{-1/2}
	\left(a_0^{\prime}\otimes I_k\right)\widetilde{\Sigma}^{-1}\vect(R),\\
	S&=\left[\left(b_0^{\prime}\otimes I_k\right)
	\widetilde{\Sigma}\left(b_0\otimes I_k\right)\right]^{-1/2}
	\left(b_0^{\prime}\otimes I_k\right)\vect(R).
\end{align*}
The mapping between $R$ and $\left(S,T\right)$ is one-to-one.
Under $H_0$, $T$ is sufficient for $\mu$, while
$S\sim\mathcal{N}\left(0,I_k\right)$ is independent of $T$.
To approximate the conditional null distribution described in
Section \ref{sec:setup}, we hold $T=t$ fixed and generate 1,000 draws
of $S$. For each draw, we reconstruct $R$ from $\left(S,t\right)$
and compute $LR_n(\beta_0)$. The empirical 95th percentile gives
the conditional critical value for a test at the 5\% level.

We compare the LR statistics, conditional critical values, and rejection
decisions obtained using the polynomial method with those obtained using
quasi-Newton methods. We use $\beta_0$ as a starting point and also
consider additional starting points. As discussed by
\citet{MoreiraRidderSharifvaghefi21b}, multiple starting points and a compact
parameterization can improve numerical performance, although they do not
guarantee global minimization. Specifically, setting $\beta=\tan\theta$
and $\overline{b}(\theta)=\left(\cos\theta,-\sin\theta\right)^{\prime}$ gives
\[
Q_n\left(\tan\theta\right)
=\overline{b}(\theta)^{\prime}R^{\prime}
\left[\left(\overline{b}(\theta)^{\prime}\otimes I_k\right)
\widetilde{\Sigma}
\left(\overline{b}(\theta)\otimes I_k\right)\right]^{-1}
R\overline{b}(\theta),
\qquad -\pi/2 \leq \theta \leq \pi/2.
\]
We denote the quasi-Newton implementations by
$\mathrm{QN}(n_d,n_r)$, where $n_d$ and $n_r$ are the numbers of
deterministic and random starting points, respectively.

Table \ref{tab:clr_test} reports results for $\beta_0=0$ and
$\mu=\sqrt{\lambda/k}\left(1,\ldots,1\right)^{\prime}$, so that
$\mu^{\prime}\mu=\lambda$. We set $\lambda/k = 2$ and consider $k = 2, 5 $ and $10$. We vary $\beta^{\ast}$ through the
noncentrality parameter of the AR statistic,
\[
\Delta_{\mathrm{AR}}^{2}
=\left(\beta^{\ast}-\beta_0\right)^2\mu^{\prime}
\left[\left(b_0^{\prime}\otimes I_k\right)\widetilde{\Sigma}
\left(b_0\otimes I_k\right)\right]^{-1}\mu.
\]
We consider $\Delta_{\mathrm{AR}}=0$ under the null and
$\Delta_{\mathrm{AR}}=1.5$ under the alternative.
Failure to find the global minimum overstates the minimized objective
and therefore understates the LR statistic. That is, if a numerical
procedure returns $\widetilde{m}\geq
\min_{\beta\in\overline{\mathbb{R}}}Q_n(\beta)$, then
\[
Q_n(\beta_0)-\widetilde{m}
\leq Q_n(\beta_0)
-\min_{\beta\in\overline{\mathbb{R}}}Q_n(\beta).
\]
When the same conditional simulation draws are used, the failure to find the global minimum also
implies that the empirical critical value cannot exceed the one based
on global minimization. While the polynomial method produces larger statistics
and critical values, their joint increase does not determine
the direction of the change in rejection probabilities.

Table \ref{tab:clr_test} shows differences in the LR statistics
in 39.0\% to 91.2\% of replications and differences in the
conditional critical values in every replication, where a
difference is recorded when its absolute magnitude exceeds
$0.001$. In the reported simulations, neither the LR statistic
nor the conditional critical value computed by the polynomial
method is smaller than its quasi-Newton counterpart. 
Rejection decisions differ in 3.6\% to 8.5\% of replications
under the null and 10.2\% to 20.1\% under the alternative.
These are frequencies of disagreement between implementations,
rather than rejection probabilities. Under the alternative,
the disagreement increases with the number of instruments
for each quasi-Newton implementation. Additional starting
points reduce the frequency of differences in the LR statistics,
but do not eliminate differences in rejection decisions.

\begin{table}[htbp]
	\centering
	\small
	\renewcommand{\arraystretch}{1.1}
	\setlength{\tabcolsep}{5pt}
	\caption{Differences from the polynomial implementation of the CLR test}
	\label{tab:clr_test}
	\begin{threeparttable}
		\begin{tabular}{llcccccc}
			\toprule\toprule
			& & \multicolumn{3}{c}{Null: $\Delta_{\mathrm{AR}}=0$}
			& \multicolumn{3}{c}{Alternative: $\Delta_{\mathrm{AR}}=1.5$}\\
			\cmidrule(lr){3-5}\cmidrule(l){6-8}
			Method & Quantity & $k=2$ & $k=5$ & $k=10$ & $k=2$ & $k=5$ & $k=10$\\
			\midrule
			\multirow{3}{*}{$\mathrm{QN}(1,0)$}
			& LR statistic & 75.6 & 78.9 & 85.3 & 82.0 & 88.8 & 91.2\\
			& Critical value & 100.0 & 100.0 & 100.0 & 100.0 & 100.0 & 100.0\\
			& Rejection decision & 6.6 & 8.5 & 7.8 & 10.4 & 14.2 & 20.1\\
			\addlinespace[0.6em]
			\multirow{3}{*}{$\mathrm{QN}(1,10)$}
			& LR statistic & 49.7 & 57.7 & 67.9 & 42.5 & 53.1 & 61.4\\
			& Critical value & 100.0 & 100.0 & 100.0 & 100.0 & 100.0 & 100.0\\
			& Rejection decision & 4.5 & 6.5 & 8.4 & 11.1 & 14.9 & 19.7\\
			\addlinespace[0.6em]
			\multirow{3}{*}{$\mathrm{QN}(1,50)$}
			& LR statistic & 45.7 & 52.5 & 61.9 & 39.0 & 50.6 & 59.5\\
			& Critical value & 100.0 & 100.0 & 100.0 & 100.0 & 100.0 & 100.0\\
			& Rejection decision & 3.6 & 4.7 & 7.2 & 10.2 & 13.0 & 16.7\\
			\bottomrule
		\end{tabular}
		\begin{tablenotes}[flushleft]
			\footnotesize
			\item \textit{Notes:} Entries are percentages of replications in which
			an implementation differs from the polynomial method. Differences
			in LR statistics and conditional critical values are recorded
			when the absolute difference exceeds $0.001$. Critical values
			are conditional 95th percentiles. $\mathrm{QN}(n_d,n_r)$ uses
			$n_d$ deterministic and $n_r$ random starting points.
		\end{tablenotes}
	\end{threeparttable}
\end{table}

\section{Conclusion}

\label{sec:conclusion}

This paper characterizes global minimization of the CUE objective in linear IV models.
With one endogenous regressor and a positive definite joint variance
estimate, the problem reduces to comparing at no more than $4k-2$ stationary
points and the value at infinity. Companion matrices express the root
calculation as an eigenvalue problem. The scalar extension allows singular
variance estimates and changes in rank. With several endogenous
regressors, polynomial systems identify finitely many stationary objective
values, including values at infinity; real-feasibility checks select the
global minimum.

Global minimization of the CUE objective matters for both estimation and testing. The minimizers determine the CUE estimates, while the minimum value defines the overidentification statistic and enters the observed and simulated LR statistics used for the CLR test. The reported numerical results show that failing to find the global minimum can considerably change the finite-sample behavior of CUE and the outcomes of the CLR test.

More generally, the algebraic approach developed here extends naturally
to objectives that are ratios of polynomials on each region of a finite
partition defined by polynomial equalities and inequalities. Applying the
method region by region requires appropriate treatment of boundaries,
denominator zeros, and behavior at infinity. Such criteria can arise in
GMM models and in other extremum estimation problems, including maximum
likelihood (MLE), nonlinear least squares (NLS), minimum distance (MD),
and two-step (TS) estimation. Approximating a nonlinear objective by
polynomials introduces approximation errors that must be controlled. These extensions are
separate from the linear-IV results proved here.

\section*{Acknowledgments}

Preliminary results of this paper were presented at the Rio--S\~{a}o
Paulo Econometrics Workshop, the IAAE Conference, the Conference
in Honor of Don Andrews, the Time Series and Econometrics Conference
(ESTE), the UCLA Mini Conference, the Rochester Conference in
Econometrics, and CFE Conference (A Tribute to Hashem Pesaran); and at seminars at Banco de
Portugal, Barcelona, CEMFI, Universidad Carlos III de Madrid,
Columbia University, the Federal Reserve Board, the University
of Georgia, NOVA University Lisbon, NYU, Paris, the University
of Pennsylvania, Universitat Pompeu Fabra, the University of
Porto, the University of Tennessee, Knoxville, Toulouse,
Duke University, Harvard--MIT, the University of Notre Dame,
the University of Pittsburgh, the University of Wisconsin,
Rutgers University, and the University of California, Riverside.
We thank Luan Borelli and Guilherme Exel for excellent research assistance.

This study was financed in part by the Coordena\c{c}\~{a}o de
Aperfei\c{c}oamento de Pessoal de N\'{i}vel Superior--Brasil
(CAPES), Finance Code 001. This research was also partially supported by NSF Grant 224247. Additional support was provided by resources from the University of Pittsburgh
Center for Research Computing and Data, RRID:SCR\_022735.
Specifically, this work used the H2P cluster, which is supported
by NSF award OAC-2117681.

\bibliographystyle{ecta}
\bibliography{References_combined}


\newpage

\setcounter{section}{0} \renewcommand{\thesection}{O-\arabic{section}}

\setcounter{table}{0} \renewcommand{\thetable}{O.\arabic{table}}

\setcounter{page}{1} \renewcommand{\thepage}{O.\arabic{page}}

\setcounter{equation}{0} \renewcommand{\theequation}{O.\arabic{equation}}

\setcounter{proposition}{0} \renewcommand{\theproposition}{O.%
	\arabic{proposition}}

\setcounter{corollary}{0} \renewcommand{\thecorollary}{O.\arabic{corollary}}

\setcounter{lemma}{0} \renewcommand{\thelemma}{O.\arabic{lemma}}

\quad \vspace{0.05in}

\begin{center}
	
	\textbf{\ {\large Simulation Supplement to} }\\[0pt]
	\textbf{{\large {``An Algebraic Approach to Continuously Updating GMM:\\[0pt]
				Estimation and Inference''}}} \\[0pt]
	
	\bigskip
	
	\begin{tabular}{@{}c@{\qquad}c@{\qquad}c@{}}
		Marcelo J. Moreira & Whitney K. Newey & Mahrad Sharifvaghefi \\
		FGV/EPGE           & MIT              & University of Pittsburgh
	\end{tabular}
	
	\bigskip
	
	\today
	
\end{center}

\section{Additional simulation results}
\label{sec:additional-simulation}

This section reports the full set of results for the univariate simulation
design in Section 5.1. Results are based on 10,000 Monte Carlo replications
with $n=800$ and $\beta^{\ast}=5$.

In the tables comparing numerical methods, QN denotes the
quasi-Newton method, with its starting value shown in parentheses
in the column heading; $\widehat{\beta}_{\mathrm{2SLS}}$ denotes
the 2SLS estimator. The polynomial method appears in the last column.

\subsection{Median bias and dispersion}

Tables \ref{median_12}--\ref{median_end2} compare the polynomial method
with quasi-Newton methods using different starting values.
Tables \ref{median_11}--\ref{median_end1} compare CUE computed using
the polynomial method with 2SLS, LIML, two-step GMM, and iterated GMM.
Entries report
median bias, with the 90\% interquantile range (the 95th percentile minus
the 5th percentile) in parentheses in the table body.

\begin{table}[p] 
	\centering
	\singlespacing
	\small
	\setlength{\tabcolsep}{4pt}
	\renewcommand{\arraystretch}{1}
	\captionsetup{font=small,justification=centering}
	\caption{Numerical computation of CUE: median bias and 90\% interquantile range for $w = 0$, $\rho = 0.3$.}  \label{median_12}
	\begin{adjustbox}{max width=\linewidth,max totalheight=0.86\textheight}
	\begin{tabular}{ccccccc}
\toprule
$\mu^2$ & $k$ & QN ($\widehat{\beta}_{\mathrm{2SLS}}$) & QN (5) & QN (4.5) & QN (0) & Polynomial \\ 
\midrule 
\multirow{ 2}{*}{1} & \multirow{ 2}{*}{1} & 0.1296 & 0.0077 & -0.1799 & -1.0730 & 0.1296 \\ 
 &  & (7.5424) & (236.7137) & (897.5008) & (1998.3973) & (7.5424) \\ 
\multirow{ 2}{*}{1} & \multirow{ 2}{*}{4} & 0.1844 & 0.0045 & -0.3084 & -2.0067 & 0.1753 \\ 
 &  & (9.2170) & (501.9348) & (1277.8696) & (3098.7852) & (9.5834) \\ 
\multirow{ 2}{*}{1} & \multirow{ 2}{*}{10} & 0.2260 & -0.0014 & -0.3414 & -3.0513 & 0.2265 \\ 
 &  & (10.0980) & (696.9385) & (1523.0757) & (3731.9796) & (10.3249) \\ 
\multirow{ 2}{*}{1} & \multirow{ 2}{*}{30} & 0.2717 & 0.0120 & -0.4137 & -3.4534 & 0.2635 \\
\bottomrule
\end{tabular}
	\end{adjustbox}
\end{table}

\begin{table}[p] 
	\centering
	\singlespacing
	\small
	\setlength{\tabcolsep}{4pt}
	\renewcommand{\arraystretch}{1}
	\captionsetup{font=small,justification=centering}
	\caption{Numerical computation of CUE: median bias and 90\% interquantile range for $w = \frac 1 2$, $\rho = 0.3$.}
	\begin{adjustbox}{max width=\linewidth,max totalheight=0.86\textheight}
	\begin{tabular}{ccccccc}
\toprule
$\mu^2$ & $k$ & QN ($\widehat{\beta}_{\mathrm{2SLS}}$) & QN (5) & QN (4.5) & QN (0) & Polynomial \\ 
\midrule 
\multirow{ 2}{*}{1} & \multirow{ 2}{*}{1} & 0.1154 & -0.0019 & -0.1804 & -1.2625 & 0.1154 \\ 
 &  & (9.0789) & (58.3688) & (817.7344) & (2296.3254) & (9.0789) \\ 
\multirow{ 2}{*}{1} & \multirow{ 2}{*}{4} & 0.1652 & -0.0526 & -0.4042 & -3.0598 & 0.1631 \\ 
 &  & (13.6082) & (596.7057) & (1362.0568) & (3637.0303) & (14.7277) \\ 
\multirow{ 2}{*}{1} & \multirow{ 2}{*}{10} & 0.2160 & -0.0573 & -0.4979 & -4.0212 & 0.2090 \\ 
 &  & (16.0780) & (814.9946) & (1589.6565) & (4111.0670) & (15.8038) \\ 
\multirow{ 2}{*}{1} & \multirow{ 2}{*}{30} & 0.2778 & -0.1256 & -0.7958 & -6.6051 & 0.2700 \\
\bottomrule
\end{tabular}
	\end{adjustbox}
\end{table}

\begin{table}[p] 
	\centering
	\singlespacing
	\small
	\setlength{\tabcolsep}{4pt}
	\renewcommand{\arraystretch}{1}
	\captionsetup{font=small,justification=centering}
	\caption{Numerical computation of CUE: median bias and 90\% interquantile range for $w = \frac{\sqrt{2}}{2}$, $\rho = 0.3$.}
	\begin{adjustbox}{max width=\linewidth,max totalheight=0.86\textheight}
	\begin{tabular}{ccccccc}
\toprule
$\mu^2$ & $k$ & QN ($\widehat{\beta}_{\mathrm{2SLS}}$) & QN (5) & QN (4.5) & QN (0) & Polynomial \\ 
\midrule 
\multirow{ 2}{*}{1} & \multirow{ 2}{*}{1} & 0.1550 & 0.0341 & -0.1598 & -1.4588 & 0.1550 \\ 
 &  & (10.5299) & (28.7354) & (844.8310) & (2562.3538) & (10.5299) \\ 
\multirow{ 2}{*}{1} & \multirow{ 2}{*}{4} & 0.1701 & -0.0310 & -0.3926 & -3.8706 & 0.1688 \\ 
 &  & (18.8739) & (511.1900) & (1420.0203) & (3791.4200) & (19.9826) \\ 
\multirow{ 2}{*}{1} & \multirow{ 2}{*}{10} & 0.1826 & -0.1365 & -0.6350 & -4.6119 & 0.1702 \\ 
 &  & (21.5264) & (753.1287) & (1633.2019) & (4358.1353) & (20.8610) \\ 
\multirow{ 2}{*}{1} & \multirow{ 2}{*}{30} & 0.2034 & -0.2529 & -1.0829 & -8.0647 & 0.2036 \\
\bottomrule
\end{tabular}
	\end{adjustbox}
\end{table}

\begin{table}[p] 
	\centering
	\singlespacing
	\small
	\setlength{\tabcolsep}{4pt}
	\renewcommand{\arraystretch}{1}
	\captionsetup{font=small,justification=centering}
	\caption{Numerical computation of CUE: median bias and 90\% interquantile range for $w = \frac{\sqrt{3}}{2}$, $\rho = 0.3$.}
	\begin{adjustbox}{max width=\linewidth,max totalheight=0.86\textheight}
	\begin{tabular}{ccccccc}
\toprule
$\mu^2$ & $k$ & QN ($\widehat{\beta}_{\mathrm{2SLS}}$) & QN (5) & QN (4.5) & QN (0) & Polynomial \\ 
\midrule 
\multirow{ 2}{*}{1} & \multirow{ 2}{*}{1} & 0.1137 & -0.0270 & -0.2091 & -1.5843 & 0.1137 \\ 
 &  & (12.1003) & (35.6154) & (802.8675) & (2669.5483) & (12.1003) \\ 
\multirow{ 2}{*}{1} & \multirow{ 2}{*}{4} & 0.1823 & -0.0623 & -0.4524 & -4.4039 & 0.1695 \\ 
 &  & (22.1512) & (727.2645) & (1451.4219) & (3976.9563) & (23.7401) \\ 
\multirow{ 2}{*}{1} & \multirow{ 2}{*}{10} & 0.2197 & -0.1110 & -0.6877 & -5.3893 & 0.2217 \\ 
 &  & (24.0699) & (995.5429) & (1942.3584) & (4480.4773) & (24.0388) \\ 
\multirow{ 2}{*}{1} & \multirow{ 2}{*}{30} & 0.2164 & -0.2924 & -1.1690 & -9.5205 & 0.2234 \\
\bottomrule
\end{tabular}
	\end{adjustbox}
\end{table}

\begin{table}[p] 
	\centering
	\singlespacing
	\small
	\setlength{\tabcolsep}{4pt}
	\renewcommand{\arraystretch}{1}
	\captionsetup{font=small,justification=centering}
	\caption{Numerical computation of CUE: median bias and 90\% interquantile range for $w = 1$, $\rho = 0.3$.}
	\begin{adjustbox}{max width=\linewidth,max totalheight=0.86\textheight}
	\begin{tabular}{ccccccc}
\toprule
$\mu^2$ & $k$ & QN ($\widehat{\beta}_{\mathrm{2SLS}}$) & QN (5) & QN (4.5) & QN (0) & Polynomial \\ 
\midrule 
\multirow{ 2}{*}{1} & \multirow{ 2}{*}{1} & 0.1274 & -0.0036 & -0.2128 & -1.6743 & 0.1274 \\ 
 &  & (13.0656) & (29.0677) & (825.7620) & (2738.5035) & (13.0656) \\ 
\multirow{ 2}{*}{1} & \multirow{ 2}{*}{4} & 0.1696 & -0.0494 & -0.4283 & -4.1551 & 0.1636 \\ 
 &  & (23.2012) & (569.4381) & (1582.3420) & (4121.3906) & (26.0727) \\ 
\multirow{ 2}{*}{1} & \multirow{ 2}{*}{10} & 0.2113 & -0.1059 & -0.6529 & -5.9476 & 0.2396 \\ 
 &  & (28.2342) & (927.1655) & (1858.7140) & (4740.2952) & (28.1077) \\ 
\multirow{ 2}{*}{1} & \multirow{ 2}{*}{30} & 0.2366 & -0.3369 & -1.3794 & -9.4377 & 0.1858 \\
\bottomrule
\end{tabular}
	\end{adjustbox}
\end{table}

\begin{table}[p] 
	\centering
	\singlespacing
	\small
	\setlength{\tabcolsep}{4pt}
	\renewcommand{\arraystretch}{1}
	\captionsetup{font=small,justification=centering}
	\caption{Numerical computation of CUE: median bias and 90\% interquantile range for $w = 0$, $\rho = 0.6$.}
	\begin{adjustbox}{max width=\linewidth,max totalheight=0.86\textheight}
	\begin{tabular}{ccccccc}
\toprule
$\mu^2$ & $k$ & QN ($\widehat{\beta}_{\mathrm{2SLS}}$) & QN (5) & QN (4.5) & QN (0) & Polynomial \\ 
\midrule 
\multirow{ 2}{*}{1} & \multirow{ 2}{*}{1} & 0.2267 & -0.0148 & -0.1255 & -0.5130 & 0.2267 \\ 
 &  & (8.2271) & (646.7126) & (953.2691) & (1640.2266) & (8.2271) \\ 
\multirow{ 2}{*}{1} & \multirow{ 2}{*}{4} & 0.3746 & 0.0058 & -0.2418 & -1.2174 & 0.3625 \\ 
 &  & (9.7953) & (958.9871) & (1455.7469) & (2887.9158) & (9.4026) \\ 
\multirow{ 2}{*}{1} & \multirow{ 2}{*}{10} & 0.4406 & 0.0080 & -0.2975 & -1.6367 & 0.4355 \\ 
 &  & (9.7228) & (1121.4194) & (1718.1751) & (3307.2011) & (9.2946) \\ 
\multirow{ 2}{*}{1} & \multirow{ 2}{*}{30} & 0.5160 & -0.0004 & -0.3702 & -2.5501 & 0.5177 \\
\bottomrule
\end{tabular}
	\end{adjustbox}
\end{table}

\begin{table}[p] 
	\centering
	\singlespacing
	\small
	\setlength{\tabcolsep}{4pt}
	\renewcommand{\arraystretch}{1}
	\captionsetup{font=small,justification=centering}
	\caption{Numerical computation of CUE: median bias and 90\% interquantile range for $w = \frac 1 2$, $\rho = 0.6$.}
	\begin{adjustbox}{max width=\linewidth,max totalheight=0.86\textheight}
	\begin{tabular}{ccccccc}
\toprule
$\mu^2$ & $k$ & QN ($\widehat{\beta}_{\mathrm{2SLS}}$) & QN (5) & QN (4.5) & QN (0) & Polynomial \\ 
\midrule 
\multirow{ 2}{*}{1} & \multirow{ 2}{*}{1} & 0.2551 & 0.0301 & -0.1165 & -0.7147 & 0.2551 \\ 
 &  & (8.7399) & (641.7577) & (1036.4666) & (1891.7732) & (8.7399) \\ 
\multirow{ 2}{*}{1} & \multirow{ 2}{*}{4} & 0.3441 & -0.0781 & -0.4195 & -2.2953 & 0.3451 \\ 
 &  & (14.2843) & (1077.9914) & (1569.0908) & (3429.7426) & (14.2591) \\ 
\multirow{ 2}{*}{1} & \multirow{ 2}{*}{10} & 0.4187 & -0.1034 & -0.5253 & -2.8671 & 0.4016 \\ 
 &  & (14.5322) & (1227.0818) & (1837.7235) & (3938.7390) & (14.1575) \\ 
\multirow{ 2}{*}{1} & \multirow{ 2}{*}{30} & 0.5034 & -0.3180 & -0.9707 & -5.2502 & 0.4804 \\
\bottomrule
\end{tabular}
	\end{adjustbox}
\end{table}

\begin{table}[p] 
	\centering
	\singlespacing
	\small
	\setlength{\tabcolsep}{4pt}
	\renewcommand{\arraystretch}{1}
	\captionsetup{font=small,justification=centering}
	\caption{Numerical computation of CUE: median bias and 90\% interquantile range for $w = \frac{\sqrt{2}}{2}$, $\rho = 0.6$.}
	\begin{adjustbox}{max width=\linewidth,max totalheight=0.86\textheight}
	\begin{tabular}{ccccccc}
\toprule
$\mu^2$ & $k$ & QN ($\widehat{\beta}_{\mathrm{2SLS}}$) & QN (5) & QN (4.5) & QN (0) & Polynomial \\ 
\midrule 
\multirow{ 2}{*}{1} & \multirow{ 2}{*}{1} & 0.2473 & 0.0042 & -0.1489 & -0.8504 & 0.2473 \\ 
 &  & (9.8226) & (632.2542) & (1078.7721) & (2012.5516) & (9.8226) \\ 
\multirow{ 2}{*}{1} & \multirow{ 2}{*}{4} & 0.3227 & -0.1288 & -0.4734 & -2.7840 & 0.3173 \\ 
 &  & (16.4931) & (1131.9235) & (1646.4076) & (3753.5143) & (17.8219) \\ 
\multirow{ 2}{*}{1} & \multirow{ 2}{*}{10} & 0.4222 & -0.1970 & -0.6837 & -3.9209 & 0.4208 \\ 
 &  & (17.6461) & (1345.8627) & (1933.7555) & (4146.0336) & (16.8657) \\ 
\multirow{ 2}{*}{1} & \multirow{ 2}{*}{30} & 0.5079 & -0.4592 & -1.2365 & -7.2206 & 0.4830 \\
\bottomrule
\end{tabular}
	\end{adjustbox}
\end{table}

\begin{table}[p] 
	\centering
	\singlespacing
	\small
	\setlength{\tabcolsep}{4pt}
	\renewcommand{\arraystretch}{1}
	\captionsetup{font=small,justification=centering}
	\caption{Numerical computation of CUE: median bias and 90\% interquantile range for $w = \frac{\sqrt{3}}{2}$, $\rho = 0.6$.}
	\begin{adjustbox}{max width=\linewidth,max totalheight=0.86\textheight}
	\begin{tabular}{ccccccc}
\toprule
$\mu^2$ & $k$ & QN ($\widehat{\beta}_{\mathrm{2SLS}}$) & QN (5) & QN (4.5) & QN (0) & Polynomial \\ 
\midrule 
\multirow{ 2}{*}{1} & \multirow{ 2}{*}{1} & 0.2365 & 0.0010 & -0.1508 & -0.9404 & 0.2365 \\ 
 &  & (10.6250) & (565.2446) & (1030.3272) & (2212.8566) & (10.6250) \\ 
\multirow{ 2}{*}{1} & \multirow{ 2}{*}{4} & 0.3104 & -0.1189 & -0.5033 & -3.1271 & 0.3191 \\ 
 &  & (18.4547) & (1181.4172) & (1729.4032) & (3827.6410) & (19.7548) \\ 
\multirow{ 2}{*}{1} & \multirow{ 2}{*}{10} & 0.4119 & -0.2444 & -0.7859 & -4.5885 & 0.3982 \\ 
 &  & (20.1378) & (1438.0982) & (2087.2570) & (4320.0445) & (19.7190) \\ 
\multirow{ 2}{*}{1} & \multirow{ 2}{*}{30} & 0.4351 & -0.6097 & -1.4504 & -6.9109 & 0.3595 \\
\bottomrule
\end{tabular}
	\end{adjustbox}
\end{table}

\begin{table}[p] 
	\centering
	\singlespacing
	\small
	\setlength{\tabcolsep}{4pt}
	\renewcommand{\arraystretch}{1}
	\captionsetup{font=small,justification=centering}
	\caption{Numerical computation of CUE: median bias and 90\% interquantile range for $w = 1$, $\rho = 0.6$.}
	\begin{adjustbox}{max width=\linewidth,max totalheight=0.86\textheight}
	\begin{tabular}{ccccccc}
\toprule
$\mu^2$ & $k$ & QN ($\widehat{\beta}_{\mathrm{2SLS}}$) & QN (5) & QN (4.5) & QN (0) & Polynomial \\ 
\midrule 
\multirow{ 2}{*}{1} & \multirow{ 2}{*}{1} & 0.2490 & 0.0066 & -0.1530 & -1.0133 & 0.2490 \\ 
 &  & (11.5197) & (544.7069) & (1075.3646) & (2359.9226) & (11.5197) \\ 
\multirow{ 2}{*}{1} & \multirow{ 2}{*}{4} & 0.3539 & -0.1207 & -0.4884 & -3.5899 & 0.3518 \\ 
 &  & (20.0285) & (1243.1746) & (1823.4042) & (3968.2766) & (22.3172) \\ 
\multirow{ 2}{*}{1} & \multirow{ 2}{*}{10} & 0.3503 & -0.2543 & -0.7496 & -4.5245 & 0.3441 \\ 
 &  & (23.5893) & (1382.6289) & (1997.3122) & (4416.9119) & (23.7697) \\ 
\multirow{ 2}{*}{1} & \multirow{ 2}{*}{30} & 0.4413 & -0.6971 & -1.5918 & -8.3128 & 0.3615 \\
\bottomrule
\end{tabular}
	\end{adjustbox}
\end{table}

\begin{table}[p] 
	\centering
	\singlespacing
	\small
	\setlength{\tabcolsep}{4pt}
	\renewcommand{\arraystretch}{1}
	\captionsetup{font=small,justification=centering}
	\caption{Numerical computation of CUE: median bias and 90\% interquantile range for $w = 0$, $\rho = 0.9$.}
	\begin{adjustbox}{max width=\linewidth,max totalheight=0.86\textheight}
	\begin{tabular}{ccccccc}
\toprule
$\mu^2$ & $k$ & QN ($\widehat{\beta}_{\mathrm{2SLS}}$) & QN (5) & QN (4.5) & QN (0) & Polynomial \\ 
\midrule 
\multirow{ 2}{*}{1} & \multirow{ 2}{*}{1} & 0.2949 & -0.0146 & -0.0200 & -0.0426 & 0.2949 \\ 
 &  & (8.7418) & (679.7259) & (849.0209) & (1207.3602) & (8.7418) \\ 
\multirow{ 2}{*}{1} & \multirow{ 2}{*}{4} & 0.4400 & 0.0039 & -0.0543 & -0.1642 & 0.4413 \\ 
 &  & (10.6161) & (939.8349) & (1263.1258) & (1982.3265) & (8.2675) \\ 
\multirow{ 2}{*}{1} & \multirow{ 2}{*}{10} & 0.5627 & -0.0231 & -0.1302 & -0.3352 & 0.5502 \\ 
 &  & (11.6355) & (1190.6475) & (1513.1314) & (2765.7692) & (7.5190) \\ 
\multirow{ 2}{*}{1} & \multirow{ 2}{*}{30} & 0.7087 & -0.0207 & -0.2368 & -0.8089 & 0.6989 \\
\bottomrule
\end{tabular}
	\end{adjustbox}
\end{table}

\begin{table}[p] 
	\centering
	\singlespacing
	\small
	\setlength{\tabcolsep}{4pt}
	\renewcommand{\arraystretch}{1}
	\captionsetup{font=small,justification=centering}
	\caption{Numerical computation of CUE: median bias and 90\% interquantile range for $w = \frac 1 2$, $\rho = 0.9$.}
	\begin{adjustbox}{max width=\linewidth,max totalheight=0.86\textheight}
	\begin{tabular}{ccccccc}
\toprule
$\mu^2$ & $k$ & QN ($\widehat{\beta}_{\mathrm{2SLS}}$) & QN (5) & QN (4.5) & QN (0) & Polynomial \\ 
\midrule 
\multirow{ 2}{*}{1} & \multirow{ 2}{*}{1} & 0.3103 & -0.0071 & -0.0322 & -0.0898 & 0.3103 \\ 
 &  & (8.5641) & (720.7898) & (889.2558) & (1270.3586) & (8.5641) \\ 
\multirow{ 2}{*}{1} & \multirow{ 2}{*}{4} & 0.4440 & -0.0877 & -0.2333 & -0.5778 & 0.4408 \\ 
 &  & (12.3915) & (1107.3884) & (1412.6194) & (2748.6742) & (10.1934) \\ 
\multirow{ 2}{*}{1} & \multirow{ 2}{*}{10} & 0.5350 & -0.1651 & -0.3797 & -0.8929 & 0.5276 \\ 
 &  & (16.9063) & (1241.2380) & (1586.9786) & (3010.7768) & (10.3731) \\ 
\multirow{ 2}{*}{1} & \multirow{ 2}{*}{30} & 0.6878 & -0.3805 & -0.7426 & -2.0366 & 0.6667 \\
\bottomrule
\end{tabular}
	\end{adjustbox}
\end{table}

\begin{table}[p] 
	\centering
	\singlespacing
	\small
	\setlength{\tabcolsep}{4pt}
	\renewcommand{\arraystretch}{1}
	\captionsetup{font=small,justification=centering}
	\caption{Numerical computation of CUE: median bias and 90\% interquantile range for $w = \frac{\sqrt{2}}{2}$, $\rho = 0.9$.}
	\begin{adjustbox}{max width=\linewidth,max totalheight=0.86\textheight}
	\begin{tabular}{ccccccc}
\toprule
$\mu^2$ & $k$ & QN ($\widehat{\beta}_{\mathrm{2SLS}}$) & QN (5) & QN (4.5) & QN (0) & Polynomial \\ 
\midrule 
\multirow{ 2}{*}{1} & \multirow{ 2}{*}{1} & 0.3209 & 0.0001 & -0.0334 & -0.1191 & 0.3209 \\ 
 &  & (8.9879) & (737.7618) & (915.4625) & (1322.8798) & (8.9879) \\ 
\multirow{ 2}{*}{1} & \multirow{ 2}{*}{4} & 0.4320 & -0.1487 & -0.3422 & -0.9298 & 0.4416 \\ 
 &  & (13.6249) & (1152.8817) & (1514.3304) & (2975.0290) & (12.0229) \\ 
\multirow{ 2}{*}{1} & \multirow{ 2}{*}{10} & 0.5497 & -0.2281 & -0.4814 & -1.2884 & 0.5502 \\ 
 &  & (18.3470) & (1364.7240) & (1717.6819) & (3236.5741) & (11.6551) \\ 
\multirow{ 2}{*}{1} & \multirow{ 2}{*}{30} & 0.6770 & -0.6593 & -1.1556 & -2.8158 & 0.6644 \\
\bottomrule
\end{tabular}
	\end{adjustbox}
\end{table}

\begin{table}[p] 
	\centering
	\singlespacing
	\small
	\setlength{\tabcolsep}{4pt}
	\renewcommand{\arraystretch}{1}
	\captionsetup{font=small,justification=centering}
	\caption{Numerical computation of CUE: median bias and 90\% interquantile range for $w = \frac{\sqrt{3}}{2}$, $\rho = 0.9$.}
	\begin{adjustbox}{max width=\linewidth,max totalheight=0.86\textheight}
	\begin{tabular}{ccccccc}
\toprule
$\mu^2$ & $k$ & QN ($\widehat{\beta}_{\mathrm{2SLS}}$) & QN (5) & QN (4.5) & QN (0) & Polynomial \\ 
\midrule 
\multirow{ 2}{*}{1} & \multirow{ 2}{*}{1} & 0.3384 & 0.0153 & -0.0408 & -0.1819 & 0.3384 \\ 
 &  & (9.1576) & (775.5571) & (959.0774) & (1442.9726) & (9.1576) \\ 
\multirow{ 2}{*}{1} & \multirow{ 2}{*}{4} & 0.4396 & -0.1249 & -0.3413 & -1.1834 & 0.4541 \\ 
 &  & (15.0923) & (1218.3117) & (1582.1966) & (3137.8553) & (13.3289) \\ 
\multirow{ 2}{*}{1} & \multirow{ 2}{*}{10} & 0.5435 & -0.2935 & -0.6185 & -1.7285 & 0.5334 \\ 
 &  & (16.0574) & (1403.8756) & (1822.5272) & (3410.5816) & (12.9085) \\ 
\multirow{ 2}{*}{1} & \multirow{ 2}{*}{30} & 0.6844 & -0.7807 & -1.3578 & -3.7922 & 0.6284 \\
\bottomrule
\end{tabular}
	\end{adjustbox}
\end{table}

\begin{table}[p] 
	\centering
	\singlespacing
	\small
	\setlength{\tabcolsep}{4pt}
	\renewcommand{\arraystretch}{1}
	\captionsetup{font=small,justification=centering}
	\caption{Numerical computation of CUE: median bias and 90\% interquantile range for $w = 1$, $\rho = 0.9$.}  \label{median_end2}
	\begin{adjustbox}{max width=\linewidth,max totalheight=0.86\textheight}
	\begin{tabular}{ccccccc}
\toprule
$\mu^2$ & $k$ & QN ($\widehat{\beta}_{\mathrm{2SLS}}$) & QN (5) & QN (4.5) & QN (0) & Polynomial \\ 
\midrule 
\multirow{ 2}{*}{1} & \multirow{ 2}{*}{1} & 0.3317 & 0.0003 & -0.0621 & -0.2232 & 0.3317 \\ 
 &  & (9.1984) & (792.3518) & (969.0726) & (1472.4553) & (9.1984) \\ 
\multirow{ 2}{*}{1} & \multirow{ 2}{*}{4} & 0.4327 & -0.1893 & -0.4351 & -1.5303 & 0.4235 \\ 
 &  & (15.9394) & (1229.0672) & (1617.5857) & (3249.0077) & (15.2158) \\ 
\multirow{ 2}{*}{1} & \multirow{ 2}{*}{10} & 0.5505 & -0.3476 & -0.6990 & -2.3341 & 0.5561 \\ 
 &  & (18.2521) & (1478.9529) & (1838.3032) & (3599.3891) & (14.3812) \\ 
\multirow{ 2}{*}{1} & \multirow{ 2}{*}{30} & 0.6629 & -1.0346 & -1.6933 & -4.5958 & 0.6285 \\
\bottomrule
\end{tabular}
	\end{adjustbox}
\end{table}

\begin{table}[p]
	\centering
	\singlespacing
	\small
	\setlength{\tabcolsep}{4pt}
	\renewcommand{\arraystretch}{1}
	\captionsetup{font=small,justification=centering}
	\caption{Comparison of estimators: median bias and 90\% interquantile range for $w = 0$, $\rho = 0.3$.} \label{median_11} 
	\begin{adjustbox}{max width=\linewidth,max totalheight=0.86\textheight}
	\begin{tabular}{ccccccc}
\toprule
$\mu^2$ & $k$ & 2SLS & LIML & Two-step & Iterated & CUE \\ 
\midrule 
\multirow{ 2}{*}{1} & \multirow{ 2}{*}{1} & 0.1296 & 0.1296 & 0.1296 & 0.1296 & 0.1296 \\ 
 &  & (7.5424) & (7.5424) & (7.5424) & (7.5424) & (7.5424) \\ 
\multirow{ 2}{*}{1} & \multirow{ 2}{*}{4} & 0.2320 & 0.1733 & 0.2382 & 0.2377 & 0.1753 \\ 
 &  & (1.8869) & (8.9119) & (2.0222) & (2.0241) & (9.5834) \\ 
\multirow{ 2}{*}{1} & \multirow{ 2}{*}{10} & 0.2751 & 0.2084 & 0.2759 & 0.2766 & 0.2265 \\ 
 &  & (1.0528) & (10.0334) & (1.1266) & (1.1238) & (10.3249) \\ 
\multirow{ 2}{*}{1} & \multirow{ 2}{*}{30} & 0.2937 & 0.2500 & 0.2921 & 0.2922 & 0.2635 \\
\bottomrule
\end{tabular}
	\end{adjustbox}
\end{table}

\begin{table}[p] 
	\centering
	\singlespacing
	\small
	\setlength{\tabcolsep}{4pt}
	\renewcommand{\arraystretch}{1}
	\captionsetup{font=small,justification=centering}
	\caption{Comparison of estimators: median bias and 90\% interquantile range for $w = \frac 1 2$, $\rho = 0.3$.}
	\begin{adjustbox}{max width=\linewidth,max totalheight=0.86\textheight}
	\begin{tabular}{ccccccc}
\toprule
$\mu^2$ & $k$ & 2SLS & LIML & Two-step & Iterated & CUE \\ 
\midrule 
\multirow{ 2}{*}{1} & \multirow{ 2}{*}{1} & 0.1154 & 0.1154 & 0.1154 & 0.1154 & 0.1154 \\ 
 &  & (9.0789) & (9.0789) & (9.0789) & (9.0789) & (9.0789) \\ 
\multirow{ 2}{*}{1} & \multirow{ 2}{*}{4} & 0.2245 & 0.0705 & 0.2266 & 0.2268 & 0.1631 \\ 
 &  & (2.6545) & (19.3054) & (2.8984) & (2.9356) & (14.7277) \\ 
\multirow{ 2}{*}{1} & \multirow{ 2}{*}{10} & 0.2721 & 0.1043 & 0.2728 & 0.2727 & 0.2090 \\ 
 &  & (1.3587) & (25.4254) & (1.4810) & (1.4862) & (15.8038) \\ 
\multirow{ 2}{*}{1} & \multirow{ 2}{*}{30} & 0.2915 & 0.1204 & 0.2974 & 0.2976 & 0.2700 \\
\bottomrule
\end{tabular}
	\end{adjustbox}
\end{table}

\begin{table}[p] 
	\centering
	\singlespacing
	\small
	\setlength{\tabcolsep}{4pt}
	\renewcommand{\arraystretch}{1}
	\captionsetup{font=small,justification=centering}
	\caption{Comparison of estimators: median bias and 90\% interquantile range for $w = \frac{\sqrt{2}}{2}$, $\rho = 0.3$.}
	\begin{adjustbox}{max width=\linewidth,max totalheight=0.86\textheight}
	\begin{tabular}{ccccccc}
\toprule
$\mu^2$ & $k$ & 2SLS & LIML & Two-step & Iterated & CUE \\ 
\midrule 
\multirow{ 2}{*}{1} & \multirow{ 2}{*}{1} & 0.1550 & 0.1550 & 0.1550 & 0.1550 & 0.1550 \\ 
 &  & (10.5299) & (10.5299) & (10.5299) & (10.5299) & (10.5299) \\ 
\multirow{ 2}{*}{1} & \multirow{ 2}{*}{4} & 0.2320 & 0.0224 & 0.2289 & 0.2287 & 0.1688 \\ 
 &  & (3.2748) & (27.9688) & (3.5293) & (3.5184) & (19.9826) \\ 
\multirow{ 2}{*}{1} & \multirow{ 2}{*}{10} & 0.2664 & -0.1305 & 0.2680 & 0.2697 & 0.1702 \\ 
 &  & (1.5910) & (38.1278) & (1.7215) & (1.7279) & (20.8610) \\ 
\multirow{ 2}{*}{1} & \multirow{ 2}{*}{30} & 0.2825 & -0.8937 & 0.2862 & 0.2855 & 0.2036 \\
\bottomrule
\end{tabular}
	\end{adjustbox}
\end{table}

\begin{table}[p] 
	\centering
	\singlespacing
	\small
	\setlength{\tabcolsep}{4pt}
	\renewcommand{\arraystretch}{1}
	\captionsetup{font=small,justification=centering}
	\caption{Comparison of estimators: median bias and 90\% interquantile range for $w = \frac{\sqrt{3}}{2}$, $\rho = 0.3$.}
	\begin{adjustbox}{max width=\linewidth,max totalheight=0.86\textheight}
	\begin{tabular}{ccccccc}
\toprule
$\mu^2$ & $k$ & 2SLS & LIML & Two-step & Iterated & CUE \\ 
\midrule 
\multirow{ 2}{*}{1} & \multirow{ 2}{*}{1} & 0.1137 & 0.1137 & 0.1137 & 0.1137 & 0.1137 \\ 
 &  & (12.1003) & (12.1003) & (12.1003) & (12.1003) & (12.1003) \\ 
\multirow{ 2}{*}{1} & \multirow{ 2}{*}{4} & 0.2431 & -0.0329 & 0.2277 & 0.2342 & 0.1695 \\ 
 &  & (3.7121) & (37.0789) & (4.0941) & (4.1189) & (23.7401) \\ 
\multirow{ 2}{*}{1} & \multirow{ 2}{*}{10} & 0.2718 & -0.2919 & 0.2795 & 0.2765 & 0.2217 \\ 
 &  & (1.8159) & (48.9697) & (1.9560) & (1.9787) & (24.0388) \\ 
\multirow{ 2}{*}{1} & \multirow{ 2}{*}{30} & 0.2929 & -1.0822 & 0.2892 & 0.2903 & 0.2234 \\
\bottomrule
\end{tabular}
	\end{adjustbox}
\end{table}

\begin{table}[p] 
	\centering
	\singlespacing
	\small
	\setlength{\tabcolsep}{4pt}
	\renewcommand{\arraystretch}{1}
	\captionsetup{font=small,justification=centering}
	\caption{Comparison of estimators: median bias and 90\% interquantile range for $w = 1$, $\rho = 0.3$.}
	\begin{adjustbox}{max width=\linewidth,max totalheight=0.86\textheight}
	\begin{tabular}{ccccccc}
\toprule
$\mu^2$ & $k$ & 2SLS & LIML & Two-step & Iterated & CUE \\ 
\midrule 
\multirow{ 2}{*}{1} & \multirow{ 2}{*}{1} & 0.1274 & 0.1274 & 0.1274 & 0.1274 & 0.1274 \\ 
 &  & (13.0656) & (13.0656) & (13.0656) & (13.0656) & (13.0656) \\ 
\multirow{ 2}{*}{1} & \multirow{ 2}{*}{4} & 0.2271 & -0.1191 & 0.2218 & 0.2241 & 0.1636 \\ 
 &  & (4.1635) & (42.8037) & (4.5265) & (4.5059) & (26.0727) \\ 
\multirow{ 2}{*}{1} & \multirow{ 2}{*}{10} & 0.2819 & -0.2586 & 0.2764 & 0.2709 & 0.2396 \\ 
 &  & (1.9908) & (59.3000) & (2.1248) & (2.1219) & (28.1077) \\ 
\multirow{ 2}{*}{1} & \multirow{ 2}{*}{30} & 0.2864 & -1.8516 & 0.2867 & 0.2849 & 0.1858 \\
\bottomrule
\end{tabular}
	\end{adjustbox}
\end{table}

\begin{table}[p] 
	\centering
	\singlespacing
	\small
	\setlength{\tabcolsep}{4pt}
	\renewcommand{\arraystretch}{1}
	\captionsetup{font=small,justification=centering}
	\caption{Comparison of estimators: median bias and 90\% interquantile range for $w = 0$, $\rho = 0.6$.}
	\begin{adjustbox}{max width=\linewidth,max totalheight=0.86\textheight}
	\begin{tabular}{ccccccc}
\toprule
$\mu^2$ & $k$ & 2SLS & LIML & Two-step & Iterated & CUE \\ 
\midrule 
\multirow{ 2}{*}{1} & \multirow{ 2}{*}{1} & 0.2267 & 0.2267 & 0.2267 & 0.2267 & 0.2267 \\ 
 &  & (8.2271) & (8.2271) & (8.2271) & (8.2271) & (8.2271) \\ 
\multirow{ 2}{*}{1} & \multirow{ 2}{*}{4} & 0.4705 & 0.3254 & 0.4865 & 0.4866 & 0.3625 \\ 
 &  & (1.6278) & (8.7250) & (1.7519) & (1.7498) & (9.4026) \\ 
\multirow{ 2}{*}{1} & \multirow{ 2}{*}{10} & 0.5456 & 0.3988 & 0.5488 & 0.5487 & 0.4355 \\ 
 &  & (0.9053) & (9.0568) & (0.9804) & (0.9790) & (9.2946) \\ 
\multirow{ 2}{*}{1} & \multirow{ 2}{*}{30} & 0.5845 & 0.5046 & 0.5824 & 0.5826 & 0.5177 \\
\bottomrule
\end{tabular}
	\end{adjustbox}
\end{table}

\begin{table}[p] 
	\centering
	\singlespacing
	\small
	\setlength{\tabcolsep}{4pt}
	\renewcommand{\arraystretch}{1}
	\captionsetup{font=small,justification=centering}
	\caption{Comparison of estimators: median bias and 90\% interquantile range for $w = \frac 1 2$, $\rho = 0.6$.}
	\begin{adjustbox}{max width=\linewidth,max totalheight=0.86\textheight}
	\begin{tabular}{ccccccc}
\toprule
$\mu^2$ & $k$ & 2SLS & LIML & Two-step & Iterated & CUE \\ 
\midrule 
\multirow{ 2}{*}{1} & \multirow{ 2}{*}{1} & 0.2551 & 0.2551 & 0.2551 & 0.2551 & 0.2551 \\ 
 &  & (8.7399) & (8.7399) & (8.7399) & (8.7399) & (8.7399) \\ 
\multirow{ 2}{*}{1} & \multirow{ 2}{*}{4} & 0.4678 & 0.1888 & 0.4606 & 0.4621 & 0.3451 \\ 
 &  & (2.2682) & (18.4818) & (2.4622) & (2.4728) & (14.2591) \\ 
\multirow{ 2}{*}{1} & \multirow{ 2}{*}{10} & 0.5394 & 0.1792 & 0.5423 & 0.5416 & 0.4016 \\ 
 &  & (1.1550) & (20.6269) & (1.2649) & (1.2699) & (14.1575) \\ 
\multirow{ 2}{*}{1} & \multirow{ 2}{*}{30} & 0.5851 & 0.1767 & 0.5827 & 0.5838 & 0.4804 \\
\bottomrule
\end{tabular}
	\end{adjustbox}
\end{table}

\begin{table}[p] 
	\centering
	\singlespacing
	\small
	\setlength{\tabcolsep}{4pt}
	\renewcommand{\arraystretch}{1}
	\captionsetup{font=small,justification=centering}
	\caption{Comparison of estimators: median bias and 90\% interquantile range for $w = \frac{\sqrt{2}}{2}$, $\rho = 0.6$.}
	\begin{adjustbox}{max width=\linewidth,max totalheight=0.86\textheight}
	\begin{tabular}{ccccccc}
\toprule
$\mu^2$ & $k$ & 2SLS & LIML & Two-step & Iterated & CUE \\ 
\midrule 
\multirow{ 2}{*}{1} & \multirow{ 2}{*}{1} & 0.2473 & 0.2473 & 0.2473 & 0.2473 & 0.2473 \\ 
 &  & (9.8226) & (9.8226) & (9.8226) & (9.8226) & (9.8226) \\ 
\multirow{ 2}{*}{1} & \multirow{ 2}{*}{4} & 0.4644 & 0.0664 & 0.4479 & 0.4488 & 0.3173 \\ 
 &  & (2.7985) & (26.0358) & (3.0197) & (3.0210) & (17.8219) \\ 
\multirow{ 2}{*}{1} & \multirow{ 2}{*}{10} & 0.5574 & 0.0483 & 0.5445 & 0.5456 & 0.4208 \\ 
 &  & (1.3939) & (31.9866) & (1.5126) & (1.5077) & (16.8657) \\ 
\multirow{ 2}{*}{1} & \multirow{ 2}{*}{30} & 0.5816 & -0.5245 & 0.5817 & 0.5806 & 0.4830 \\
\bottomrule
\end{tabular}
	\end{adjustbox}
\end{table}

\begin{table}[p] 
	\centering
	\singlespacing
	\small
	\setlength{\tabcolsep}{4pt}
	\renewcommand{\arraystretch}{1}
	\captionsetup{font=small,justification=centering}
	\caption{Comparison of estimators: median bias and 90\% interquantile range for $w = \frac{\sqrt{3}}{2}$, $\rho = 0.6$.}
	\begin{adjustbox}{max width=\linewidth,max totalheight=0.86\textheight}
	\begin{tabular}{ccccccc}
\toprule
$\mu^2$ & $k$ & 2SLS & LIML & Two-step & Iterated & CUE \\ 
\midrule 
\multirow{ 2}{*}{1} & \multirow{ 2}{*}{1} & 0.2365 & 0.2365 & 0.2365 & 0.2365 & 0.2365 \\ 
 &  & (10.6250) & (10.6250) & (10.6250) & (10.6250) & (10.6250) \\ 
\multirow{ 2}{*}{1} & \multirow{ 2}{*}{4} & 0.4520 & -0.0909 & 0.4232 & 0.4309 & 0.3191 \\ 
 &  & (3.2017) & (30.3308) & (3.4900) & (3.5100) & (19.7548) \\ 
\multirow{ 2}{*}{1} & \multirow{ 2}{*}{10} & 0.5489 & -0.3290 & 0.5414 & 0.5406 & 0.3982 \\ 
 &  & (1.5760) & (42.1405) & (1.6769) & (1.6761) & (19.7190) \\ 
\multirow{ 2}{*}{1} & \multirow{ 2}{*}{30} & 0.5768 & -1.1690 & 0.5738 & 0.5738 & 0.3595 \\
\bottomrule
\end{tabular}
	\end{adjustbox}
\end{table}

\begin{table}[p] 
	\centering
	\singlespacing
	\small
	\setlength{\tabcolsep}{4pt}
	\renewcommand{\arraystretch}{1}
	\captionsetup{font=small,justification=centering}
	\caption{Comparison of estimators: median bias and 90\% interquantile range for $w = 1$, $\rho = 0.6$.}
	\begin{adjustbox}{max width=\linewidth,max totalheight=0.86\textheight}
	\begin{tabular}{ccccccc}
\toprule
$\mu^2$ & $k$ & 2SLS & LIML & Two-step & Iterated & CUE \\ 
\midrule 
\multirow{ 2}{*}{1} & \multirow{ 2}{*}{1} & 0.2490 & 0.2490 & 0.2490 & 0.2490 & 0.2490 \\ 
 &  & (11.5197) & (11.5197) & (11.5197) & (11.5197) & (11.5197) \\ 
\multirow{ 2}{*}{1} & \multirow{ 2}{*}{4} & 0.4712 & -0.0900 & 0.4456 & 0.4505 & 0.3518 \\ 
 &  & (3.5081) & (37.4738) & (3.7901) & (3.8372) & (22.3172) \\ 
\multirow{ 2}{*}{1} & \multirow{ 2}{*}{10} & 0.5314 & -0.7301 & 0.5215 & 0.5229 & 0.3441 \\ 
 &  & (1.7187) & (52.0547) & (1.8375) & (1.8371) & (23.7697) \\ 
\multirow{ 2}{*}{1} & \multirow{ 2}{*}{30} & 0.5768 & -1.6532 & 0.5757 & 0.5759 & 0.3615 \\
\bottomrule
\end{tabular}
	\end{adjustbox}
\end{table}

\begin{table}[p] 
	\centering
	\singlespacing
	\small
	\setlength{\tabcolsep}{4pt}
	\renewcommand{\arraystretch}{1}
	\captionsetup{font=small,justification=centering}
	\caption{Comparison of estimators: median bias and 90\% interquantile range for $w = 0$, $\rho = 0.9$.}
	\begin{adjustbox}{max width=\linewidth,max totalheight=0.86\textheight}
	\begin{tabular}{ccccccc}
\toprule
$\mu^2$ & $k$ & 2SLS & LIML & Two-step & Iterated & CUE \\ 
\midrule 
\multirow{ 2}{*}{1} & \multirow{ 2}{*}{1} & 0.2949 & 0.2949 & 0.2949 & 0.2949 & 0.2949 \\ 
 &  & (8.7418) & (8.7418) & (8.7418) & (8.7418) & (8.7418) \\ 
\multirow{ 2}{*}{1} & \multirow{ 2}{*}{4} & 0.6831 & 0.3877 & 0.7061 & 0.7050 & 0.4413 \\ 
 &  & (1.1957) & (8.2465) & (1.2134) & (1.2080) & (8.2675) \\ 
\multirow{ 2}{*}{1} & \multirow{ 2}{*}{10} & 0.8138 & 0.5157 & 0.8202 & 0.8192 & 0.5502 \\ 
 &  & (0.5660) & (8.0513) & (0.6018) & (0.6039) & (7.5190) \\ 
\multirow{ 2}{*}{1} & \multirow{ 2}{*}{30} & 0.8709 & 0.6582 & 0.8701 & 0.8697 & 0.6989 \\
\bottomrule
\end{tabular}
	\end{adjustbox}
\end{table}

\begin{table}[p] 
	\centering
	\singlespacing
	\small
	\setlength{\tabcolsep}{4pt}
	\renewcommand{\arraystretch}{1}
	\captionsetup{font=small,justification=centering}
	\caption{Comparison of estimators: median bias and 90\% interquantile range for $w = \frac 1 2$, $\rho = 0.9$.}
	\begin{adjustbox}{max width=\linewidth,max totalheight=0.86\textheight}
	\begin{tabular}{ccccccc}
\toprule
$\mu^2$ & $k$ & 2SLS & LIML & Two-step & Iterated & CUE \\ 
\midrule 
\multirow{ 2}{*}{1} & \multirow{ 2}{*}{1} & 0.3103 & 0.3103 & 0.3103 & 0.3103 & 0.3103 \\ 
 &  & (8.5641) & (8.5641) & (8.5641) & (8.5641) & (8.5641) \\ 
\multirow{ 2}{*}{1} & \multirow{ 2}{*}{4} & 0.6861 & 0.2668 & 0.6833 & 0.6846 & 0.4408 \\ 
 &  & (1.4840) & (12.1677) & (1.5837) & (1.5969) & (10.1934) \\ 
\multirow{ 2}{*}{1} & \multirow{ 2}{*}{10} & 0.8090 & 0.2684 & 0.8095 & 0.8091 & 0.5276 \\ 
 &  & (0.7005) & (13.9185) & (0.7582) & (0.7632) & (10.3731) \\ 
\multirow{ 2}{*}{1} & \multirow{ 2}{*}{30} & 0.8709 & 0.2902 & 0.8699 & 0.8703 & 0.6667 \\
\bottomrule
\end{tabular}
	\end{adjustbox}
\end{table}

\begin{table}[p] 
	\centering
	\singlespacing
	\small
	\setlength{\tabcolsep}{4pt}
	\renewcommand{\arraystretch}{1}
	\captionsetup{font=small,justification=centering}
	\caption{Comparison of estimators: median bias and 90\% interquantile range for $w = \frac{\sqrt{2}}{2}$, $\rho = 0.9$.}
	\begin{adjustbox}{max width=\linewidth,max totalheight=0.86\textheight}
	\begin{tabular}{ccccccc}
\toprule
$\mu^2$ & $k$ & 2SLS & LIML & Two-step & Iterated & CUE \\ 
\midrule 
\multirow{ 2}{*}{1} & \multirow{ 2}{*}{1} & 0.3209 & 0.3209 & 0.3209 & 0.3209 & 0.3209 \\ 
 &  & (8.9879) & (8.9879) & (8.9879) & (8.9879) & (8.9879) \\ 
\multirow{ 2}{*}{1} & \multirow{ 2}{*}{4} & 0.6815 & 0.1456 & 0.6605 & 0.6669 & 0.4416 \\ 
 &  & (1.7058) & (16.1851) & (1.8543) & (1.8622) & (12.0229) \\ 
\multirow{ 2}{*}{1} & \multirow{ 2}{*}{10} & 0.8142 & 0.1155 & 0.8047 & 0.8068 & 0.5502 \\ 
 &  & (0.7837) & (20.0194) & (0.8606) & (0.8592) & (11.6551) \\ 
\multirow{ 2}{*}{1} & \multirow{ 2}{*}{30} & 0.8708 & -0.1419 & 0.8706 & 0.8706 & 0.6644 \\
\bottomrule
\end{tabular}
	\end{adjustbox}
\end{table}

\begin{table}[p] 
	\centering
	\singlespacing
	\small
	\setlength{\tabcolsep}{4pt}
	\renewcommand{\arraystretch}{1}
	\captionsetup{font=small,justification=centering}
	\caption{Comparison of estimators: median bias and 90\% interquantile range for $w = \frac{\sqrt{3}}{2}$, $\rho = 0.9$.}
	\begin{adjustbox}{max width=\linewidth,max totalheight=0.86\textheight}
	\begin{tabular}{ccccccc}
\toprule
$\mu^2$ & $k$ & 2SLS & LIML & Two-step & Iterated & CUE \\ 
\midrule 
\multirow{ 2}{*}{1} & \multirow{ 2}{*}{1} & 0.3384 & 0.3384 & 0.3384 & 0.3384 & 0.3384 \\ 
 &  & (9.1576) & (9.1576) & (9.1576) & (9.1576) & (9.1576) \\ 
\multirow{ 2}{*}{1} & \multirow{ 2}{*}{4} & 0.6886 & 0.0817 & 0.6444 & 0.6579 & 0.4541 \\ 
 &  & (1.9095) & (19.3402) & (2.0621) & (2.0658) & (13.3289) \\ 
\multirow{ 2}{*}{1} & \multirow{ 2}{*}{10} & 0.8115 & -0.0531 & 0.8025 & 0.8033 & 0.5334 \\ 
 &  & (0.8907) & (24.6671) & (0.9521) & (0.9525) & (12.9085) \\ 
\multirow{ 2}{*}{1} & \multirow{ 2}{*}{30} & 0.8718 & -0.4316 & 0.8693 & 0.8690 & 0.6284 \\
\bottomrule
\end{tabular}
	\end{adjustbox}
\end{table}

\begin{table}[p] 
	\centering
	\singlespacing
	\small
	\setlength{\tabcolsep}{4pt}
	\renewcommand{\arraystretch}{1}
	\captionsetup{font=small,justification=centering}
	\caption{Comparison of estimators: median bias and 90\% interquantile range for $w = 1$, $\rho = 0.9$.}  \label{median_end1}
	\begin{adjustbox}{max width=\linewidth,max totalheight=0.86\textheight}
	\begin{tabular}{ccccccc}
\toprule
$\mu^2$ & $k$ & 2SLS & LIML & Two-step & Iterated & CUE \\ 
\midrule 
\multirow{ 2}{*}{1} & \multirow{ 2}{*}{1} & 0.3317 & 0.3317 & 0.3317 & 0.3317 & 0.3317 \\ 
 &  & (9.1984) & (9.1984) & (9.1984) & (9.1984) & (9.1984) \\ 
\multirow{ 2}{*}{1} & \multirow{ 2}{*}{4} & 0.6799 & -0.0009 & 0.6337 & 0.6487 & 0.4235 \\ 
 &  & (2.1015) & (22.4346) & (2.2982) & (2.2727) & (15.2158) \\ 
\multirow{ 2}{*}{1} & \multirow{ 2}{*}{10} & 0.8174 & -0.2164 & 0.7990 & 0.7988 & 0.5561 \\ 
 &  & (0.9835) & (27.9529) & (1.0539) & (1.0418) & (14.3812) \\ 
\multirow{ 2}{*}{1} & \multirow{ 2}{*}{30} & 0.8718 & -0.7381 & 0.8665 & 0.8670 & 0.6285 \\
\bottomrule
\end{tabular}
	\end{adjustbox}
\end{table}

\clearpage

\subsection{Quantile tables}

Tables \ref{quantile_1}--\ref{quantile_end} report the 5th, 25th, 50th,
75th, and 95th percentiles of the estimates themselves, rather than their
estimation errors; the true parameter is $\beta^{\ast}=5$. The first
15 tables compare numerical methods for computing CUE, and the remaining
15 compare estimators.

\clearpage
\begin{landscape}
\begin{table}[p]
\centering
\singlespacing
\fontsize{9}{11}\selectfont
\setlength{\tabcolsep}{2pt}
\renewcommand{\arraystretch}{1}
\captionsetup{font=small,justification=centering}
\caption{Numerical computation of CUE: quantiles for $w = 0$, $\rho = 0.3$.}
\label{quantile_1}
\smallskip
\begin{minipage}[t]{0.325\linewidth}
\centering
$\mu^2=1$\par\smallskip
\begin{adjustbox}{width=\linewidth}
\begin{minipage}{270pt}
\centering
\begin{tabular}{@{}cccccc@{}}
\toprule
Percentile & QN ($\widehat{\beta}_{\mathrm{2SLS}}$) & QN (5) & QN (4.5) & QN (0) & Poly. \\
\midrule
\multicolumn{6}{c}{$k=1$} \\
5\% & 1.2898 & -229.9119 & -891.4493 & -1993.0417 & 1.2898 \\
25\% & 4.4664 & 4.2022 & 3.4605 & -755.6692 & 4.4664 \\
50\% & 5.1296 & 5.0077 & 4.8201 & 3.9270 & 5.1296 \\
75\% & 5.7538 & 5.5633 & 5.3724 & 4.9269 & 5.7538 \\
95\% & 8.8322 & 6.8019 & 6.0514 & 5.3557 & 8.8322 \\
\addlinespace[3pt]
\multicolumn{6}{c}{$k=4$} \\
5\% & 0.5929 & -494.7328 & -1271.7311 & -3093.3515 & 0.2711 \\
25\% & 4.4742 & 4.0753 & 2.5904 & -1611.6252 & 4.4127 \\
50\% & 5.1844 & 5.0045 & 4.6916 & 2.9933 & 5.1753 \\
75\% & 5.8763 & 5.6098 & 5.3290 & 4.8172 & 5.9088 \\
95\% & 9.8100 & 7.2019 & 6.1385 & 5.4338 & 9.8546 \\
\addlinespace[3pt]
\multicolumn{6}{c}{$k=10$} \\
5\% & 0.0718 & -689.7811 & -1516.9243 & -3726.5452 & -0.0730 \\
25\% & 4.4757 & 3.9665 & 1.8808 & -2369.0584 & 4.4041 \\
50\% & 5.2260 & 4.9986 & 4.6586 & 1.9487 & 5.2265 \\
75\% & 5.9690 & 5.6671 & 5.3426 & 4.7209 & 6.0364 \\
95\% & 10.1698 & 7.1574 & 6.1514 & 5.4344 & 10.2519 \\
\addlinespace[3pt]
\multicolumn{6}{c}{$k=30$} \\
5\% & 0.2734 & -719.4270 & -1582.3832 & -4574.9845 & 0.1730 \\
25\% & 4.5232 & 3.9790 & 1.7594 & -3107.9942 & 4.4083 \\
50\% & 5.2717 & 5.0120 & 4.5863 & 1.5466 & 5.2635 \\
75\% & 6.0164 & 5.6765 & 5.3238 & 4.6544 & 6.1179 \\
95\% & 10.9227 & 7.3886 & 6.1865 & 5.5024 & 10.9957 \\
\addlinespace[3pt]
\multicolumn{6}{c}{$k=60$} \\
5\% & -0.9904 & -673.6234 & -1675.3518 & -5697.5635 & -0.3967 \\
25\% & 4.5260 & 4.0712 & 2.2704 & -3750.9079 & 4.3983 \\
50\% & 5.2631 & 5.0100 & 4.5520 & 1.0659 & 5.2703 \\
75\% & 6.0111 & 5.6638 & 5.2904 & 4.5998 & 6.2088 \\
95\% & 10.3910 & 7.5347 & 6.2521 & 5.5627 & 10.9796 \\
\bottomrule
\end{tabular}
\end{minipage}
\end{adjustbox}
\end{minipage}
\hfill
\begin{minipage}[t]{0.325\linewidth}
\centering
$\mu^2=4$\par\smallskip
\begin{adjustbox}{width=\linewidth}
\begin{minipage}{270pt}
\centering
\begin{tabular}{@{}cccccc@{}}
\toprule
Percentile & QN ($\widehat{\beta}_{\mathrm{2SLS}}$) & QN (5) & QN (4.5) & QN (0) & Poly. \\
\midrule
\multicolumn{6}{c}{$k=1$} \\
5\% & 3.4059 & 3.0805 & 2.1254 & -2227.6528 & 3.4059 \\
25\% & 4.6357 & 4.6114 & 4.5719 & 3.8582 & 4.6357 \\
50\% & 5.0158 & 5.0028 & 4.9858 & 4.7852 & 5.0158 \\
75\% & 5.3357 & 5.3163 & 5.2884 & 5.1049 & 5.3357 \\
95\% & 5.9976 & 5.8803 & 5.7624 & 5.3746 & 5.9976 \\
\addlinespace[3pt]
\multicolumn{6}{c}{$k=4$} \\
5\% & 2.6651 & 0.0541 & -794.4387 & -3134.3735 & 2.4891 \\
25\% & 4.5895 & 4.4953 & 4.2543 & -878.3500 & 4.5633 \\
50\% & 5.0626 & 5.0012 & 4.9041 & 4.4762 & 5.0591 \\
75\% & 5.4696 & 5.3984 & 5.2967 & 5.0193 & 5.4797 \\
95\% & 6.8643 & 6.2979 & 5.8968 & 5.4178 & 7.1674 \\
\addlinespace[3pt]
\multicolumn{6}{c}{$k=10$} \\
5\% & 2.0682 & -6.7225 & -1083.5868 & -3664.5037 & 1.7698 \\
25\% & 4.5430 & 4.3859 & 3.9670 & -1658.6743 & 4.5108 \\
50\% & 5.0950 & 4.9976 & 4.8604 & 4.1944 & 5.0868 \\
75\% & 5.5873 & 5.4681 & 5.3167 & 4.9690 & 5.6031 \\
95\% & 7.6139 & 6.5748 & 5.9739 & 5.4660 & 7.9689 \\
\addlinespace[3pt]
\multicolumn{6}{c}{$k=30$} \\
5\% & 1.5229 & -263.8416 & -1374.4262 & -4653.2119 & 1.1320 \\
25\% & 4.5427 & 4.2472 & 3.4424 & -2817.5807 & 4.4462 \\
50\% & 5.1648 & 4.9938 & 4.7581 & 3.4093 & 5.1486 \\
75\% & 5.7349 & 5.5307 & 5.3116 & 4.8553 & 5.7895 \\
95\% & 8.6808 & 6.8212 & 6.0776 & 5.5131 & 9.2302 \\
\addlinespace[3pt]
\multicolumn{6}{c}{$k=60$} \\
5\% & 1.2817 & -140.7023 & -1392.2648 & -5705.3195 & 0.6459 \\
25\% & 4.5616 & 4.2623 & 3.3524 & -3540.3045 & 4.4239 \\
50\% & 5.1947 & 5.0077 & 4.7146 & 2.8570 & 5.1847 \\
75\% & 5.7722 & 5.5651 & 5.3107 & 4.7528 & 5.8917 \\
95\% & 8.7882 & 7.0846 & 6.0967 & 5.5624 & 9.8068 \\
\bottomrule
\end{tabular}
\end{minipage}
\end{adjustbox}
\end{minipage}
\hfill
\begin{minipage}[t]{0.325\linewidth}
\centering
$\mu^2=16$\par\smallskip
\begin{adjustbox}{width=\linewidth}
\begin{minipage}{270pt}
\centering
\begin{tabular}{@{}cccccc@{}}
\toprule
Percentile & QN ($\widehat{\beta}_{\mathrm{2SLS}}$) & QN (5) & QN (4.5) & QN (0) & Poly. \\
\midrule
\multicolumn{6}{c}{$k=1$} \\
5\% & 4.4825 & 4.4825 & 4.4818 & 4.2823 & 4.4825 \\
25\% & 4.8191 & 4.8191 & 4.8189 & 4.7915 & 4.8191 \\
50\% & 4.9973 & 4.9973 & 4.9971 & 4.9780 & 4.9973 \\
75\% & 5.1589 & 5.1589 & 5.1587 & 5.1378 & 5.1589 \\
95\% & 5.3980 & 5.3980 & 5.3976 & 5.3362 & 5.3980 \\
\addlinespace[3pt]
\multicolumn{6}{c}{$k=4$} \\
5\% & 4.3169 & 4.2983 & 4.2152 & -2805.4940 & 4.2952 \\
25\% & 4.7877 & 4.7839 & 4.7714 & 4.6708 & 4.7851 \\
50\% & 5.0016 & 4.9997 & 4.9910 & 4.9367 & 5.0003 \\
75\% & 5.1946 & 5.1917 & 5.1842 & 5.1286 & 5.1939 \\
95\% & 5.4967 & 5.4860 & 5.4643 & 5.3588 & 5.4967 \\
\addlinespace[3pt]
\multicolumn{6}{c}{$k=10$} \\
5\% & 4.1161 & 4.0524 & 3.8485 & -3398.9767 & 4.0788 \\
25\% & 4.7460 & 4.7329 & 4.7047 & 4.4938 & 4.7394 \\
50\% & 5.0053 & 4.9971 & 4.9823 & 4.8888 & 5.0017 \\
75\% & 5.2293 & 5.2169 & 5.2000 & 5.1148 & 5.2272 \\
95\% & 5.6448 & 5.6065 & 5.5596 & 5.3854 & 5.6619 \\
\addlinespace[3pt]
\multicolumn{6}{c}{$k=30$} \\
5\% & 3.6972 & 3.3341 & 1.8468 & -4559.7485 & 3.5076 \\
25\% & 4.7052 & 4.6631 & 4.5706 & 3.3329 & 4.6809 \\
50\% & 5.0362 & 5.0057 & 4.9594 & 4.7535 & 5.0231 \\
75\% & 5.3278 & 5.2900 & 5.2488 & 5.0907 & 5.3258 \\
95\% & 6.0255 & 5.8326 & 5.6937 & 5.4615 & 6.1479 \\
\addlinespace[3pt]
\multicolumn{6}{c}{$k=60$} \\
5\% & 3.3904 & 2.6704 & -4.8369 & -5618.4281 & 2.9068 \\
25\% & 4.6835 & 4.5968 & 4.3925 & -3.9864 & 4.6208 \\
50\% & 5.0684 & 4.9989 & 4.9112 & 4.5399 & 5.0398 \\
75\% & 5.4120 & 5.3355 & 5.2588 & 5.0358 & 5.4225 \\
95\% & 6.3601 & 6.0509 & 5.8036 & 5.5461 & 6.8125 \\
\bottomrule
\end{tabular}
\end{minipage}
\end{adjustbox}
\end{minipage}
\par\medskip
\begin{minipage}{\linewidth}
\footnotesize
\centering
\textit{Notes:} QN denotes the quasi-Newton method, with its starting value in parentheses; $\widehat{\beta}_{\mathrm{2SLS}}$ denotes the 2SLS estimator. Poly. denotes the polynomial method.
\end{minipage}
\end{table}
\end{landscape}
\clearpage

\clearpage
\begin{landscape}
\begin{table}[p]
\centering
\singlespacing
\fontsize{9}{11}\selectfont
\setlength{\tabcolsep}{2pt}
\renewcommand{\arraystretch}{1}
\captionsetup{font=small,justification=centering}
\caption{Numerical computation of CUE: quantiles for $w = \frac 1 2$, $\rho = 0.3$.}

\smallskip
\begin{minipage}[t]{0.325\linewidth}
\centering
$\mu^2=1$\par\smallskip
\begin{adjustbox}{width=\linewidth}
\begin{minipage}{270pt}
\centering
\begin{tabular}{@{}cccccc@{}}
\toprule
Percentile & QN ($\widehat{\beta}_{\mathrm{2SLS}}$) & QN (5) & QN (4.5) & QN (0) & Poly. \\
\midrule
\multicolumn{6}{c}{$k=1$} \\
5\% & 0.2508 & -51.0557 & -811.3461 & -2290.9189 & 0.2508 \\
25\% & 4.3231 & 4.0839 & 3.4742 & -811.5667 & 4.3231 \\
50\% & 5.1154 & 4.9981 & 4.8196 & 3.7375 & 5.1154 \\
75\% & 5.8572 & 5.6890 & 5.4795 & 4.8872 & 5.8572 \\
95\% & 9.3297 & 7.3131 & 6.3883 & 5.4066 & 9.3297 \\
\addlinespace[3pt]
\multicolumn{6}{c}{$k=4$} \\
5\% & -1.7271 & -588.1996 & -1355.1345 & -3631.4684 & -2.0230 \\
25\% & 4.1181 & 3.5911 & 2.1436 & -1835.1932 & 4.0150 \\
50\% & 5.1652 & 4.9474 & 4.5958 & 1.9402 & 5.1631 \\
75\% & 6.1714 & 5.8030 & 5.4547 & 4.6302 & 6.2868 \\
95\% & 11.8812 & 8.5062 & 6.9223 & 5.5619 & 12.7046 \\
\addlinespace[3pt]
\multicolumn{6}{c}{$k=10$} \\
5\% & -2.8372 & -806.1758 & -1582.9037 & -4105.5381 & -2.8641 \\
25\% & 4.1548 & 3.5524 & 1.3516 & -2486.7223 & 4.0075 \\
50\% & 5.2160 & 4.9427 & 4.5021 & 0.9788 & 5.2090 \\
75\% & 6.2663 & 5.8483 & 5.4091 & 4.5322 & 6.4044 \\
95\% & 13.2408 & 8.8188 & 6.7528 & 5.5289 & 12.9396 \\
\addlinespace[3pt]
\multicolumn{6}{c}{$k=30$} \\
5\% & -5.5996 & -998.7032 & -1890.2154 & -5106.9966 & -3.6532 \\
25\% & 4.1486 & 3.1958 & -0.7112 & -3301.8706 & 3.8744 \\
50\% & 5.2778 & 4.8744 & 4.2042 & -1.6051 & 5.2700 \\
75\% & 6.4038 & 5.8260 & 5.2797 & 4.1362 & 6.7280 \\
95\% & 16.1190 & 8.8005 & 6.6506 & 5.4813 & 14.5665 \\
\addlinespace[3pt]
\multicolumn{6}{c}{$k=60$} \\
5\% & -11.9336 & -1022.7859 & -1967.7298 & -5951.4704 & -4.6258 \\
25\% & 4.0354 & 2.8584 & -3.2899 & -3847.6284 & 3.5829 \\
50\% & 5.3049 & 4.7668 & 3.8381 & -3.4371 & 5.3117 \\
75\% & 6.5606 & 5.8257 & 5.1066 & 3.6943 & 7.0926 \\
95\% & 23.2156 & 9.3945 & 6.6274 & 5.5486 & 16.0594 \\
\bottomrule
\end{tabular}
\end{minipage}
\end{adjustbox}
\end{minipage}
\hfill
\begin{minipage}[t]{0.325\linewidth}
\centering
$\mu^2=4$\par\smallskip
\begin{adjustbox}{width=\linewidth}
\begin{minipage}{270pt}
\centering
\begin{tabular}{@{}cccccc@{}}
\toprule
Percentile & QN ($\widehat{\beta}_{\mathrm{2SLS}}$) & QN (5) & QN (4.5) & QN (0) & Poly. \\
\midrule
\multicolumn{6}{c}{$k=1$} \\
5\% & 3.1914 & 2.8393 & 1.7890 & -2594.3555 & 3.1914 \\
25\% & 4.5616 & 4.5296 & 4.4836 & 3.5226 & 4.5616 \\
50\% & 5.0091 & 4.9902 & 4.9640 & 4.7034 & 5.0091 \\
75\% & 5.4019 & 5.3805 & 5.3488 & 5.1121 & 5.4019 \\
95\% & 6.2758 & 6.1065 & 5.9790 & 5.4399 & 6.2758 \\
\addlinespace[3pt]
\multicolumn{6}{c}{$k=4$} \\
5\% & 0.9066 & -5.9513 & -740.3147 & -3625.6987 & 0.4673 \\
25\% & 4.4196 & 4.3070 & 3.9995 & -1149.4003 & 4.3787 \\
50\% & 5.0382 & 4.9725 & 4.8647 & 4.1580 & 5.0332 \\
75\% & 5.5656 & 5.4699 & 5.3482 & 4.9694 & 5.5999 \\
95\% & 8.4162 & 7.0710 & 6.3171 & 5.5057 & 9.2795 \\
\addlinespace[3pt]
\multicolumn{6}{c}{$k=10$} \\
5\% & 0.0827 & -30.3895 & -1169.0492 & -4117.2272 & -0.4647 \\
25\% & 4.3095 & 4.0663 & 3.4192 & -2031.5088 & 4.2324 \\
50\% & 5.0535 & 4.9328 & 4.7215 & 3.5277 & 5.0427 \\
75\% & 5.7278 & 5.5474 & 5.3289 & 4.8443 & 5.7966 \\
95\% & 9.1753 & 7.5048 & 6.4493 & 5.5127 & 10.1518 \\
\addlinespace[3pt]
\multicolumn{6}{c}{$k=30$} \\
5\% & -1.9573 & -653.3090 & -1535.9607 & -5140.1379 & -2.0672 \\
25\% & 4.2248 & 3.6741 & 2.1364 & -3106.1200 & 4.0008 \\
50\% & 5.1395 & 4.8798 & 4.4798 & 1.5961 & 5.1037 \\
75\% & 6.0343 & 5.6770 & 5.2904 & 4.5221 & 6.1939 \\
95\% & 12.2462 & 8.1206 & 6.4882 & 5.5572 & 12.6597 \\
\addlinespace[3pt]
\multicolumn{6}{c}{$k=60$} \\
5\% & -6.7801 & -774.3442 & -1665.1743 & -5898.7514 & -4.3262 \\
25\% & 4.0551 & 3.2435 & 0.5346 & -3575.4144 & 3.6078 \\
50\% & 5.1463 & 4.7817 & 4.1171 & 0.1274 & 5.0572 \\
75\% & 6.2149 & 5.6705 & 5.1120 & 4.0767 & 6.5809 \\
95\% & 15.3883 & 8.5233 & 6.4891 & 5.4667 & 14.6042 \\
\bottomrule
\end{tabular}
\end{minipage}
\end{adjustbox}
\end{minipage}
\hfill
\begin{minipage}[t]{0.325\linewidth}
\centering
$\mu^2=16$\par\smallskip
\begin{adjustbox}{width=\linewidth}
\begin{minipage}{270pt}
\centering
\begin{tabular}{@{}cccccc@{}}
\toprule
Percentile & QN ($\widehat{\beta}_{\mathrm{2SLS}}$) & QN (5) & QN (4.5) & QN (0) & Poly. \\
\midrule
\multicolumn{6}{c}{$k=1$} \\
5\% & 4.4028 & 4.4028 & 4.4028 & 4.0971 & 4.4028 \\
25\% & 4.7883 & 4.7883 & 4.7883 & 4.7497 & 4.7883 \\
50\% & 5.0042 & 5.0042 & 5.0042 & 4.9786 & 5.0042 \\
75\% & 5.1986 & 5.1986 & 5.1986 & 5.1696 & 5.1986 \\
95\% & 5.4829 & 5.4829 & 5.4829 & 5.4038 & 5.4829 \\
\addlinespace[3pt]
\multicolumn{6}{c}{$k=4$} \\
5\% & 4.0233 & 3.9892 & 3.8321 & -3452.8750 & 3.9756 \\
25\% & 4.7366 & 4.7330 & 4.7171 & 4.4796 & 4.7338 \\
50\% & 4.9866 & 4.9835 & 4.9734 & 4.8830 & 4.9855 \\
75\% & 5.2203 & 5.2167 & 5.2053 & 5.1182 & 5.2212 \\
95\% & 5.6371 & 5.6035 & 5.5698 & 5.3978 & 5.6776 \\
\addlinespace[3pt]
\multicolumn{6}{c}{$k=10$} \\
5\% & 3.7407 & 3.6429 & 3.2339 & -4019.2197 & 3.6024 \\
25\% & 4.6850 & 4.6701 & 4.6317 & 4.1431 & 4.6727 \\
50\% & 4.9930 & 4.9839 & 4.9582 & 4.8154 & 4.9886 \\
75\% & 5.2615 & 5.2489 & 5.2220 & 5.0977 & 5.2619 \\
95\% & 5.9181 & 5.8236 & 5.7074 & 5.4330 & 6.0193 \\
\addlinespace[3pt]
\multicolumn{6}{c}{$k=30$} \\
5\% & 2.8791 & 2.1754 & -4.7642 & -5203.8494 & 2.1834 \\
25\% & 4.5260 & 4.4459 & 4.2329 & -2.1621 & 4.4575 \\
50\% & 5.0197 & 4.9644 & 4.8638 & 4.4252 & 4.9921 \\
75\% & 5.4040 & 5.3469 & 5.2583 & 5.0190 & 5.4115 \\
95\% & 6.7093 & 6.3000 & 5.9223 & 5.5531 & 7.2196 \\
\addlinespace[3pt]
\multicolumn{6}{c}{$k=60$} \\
5\% & 1.6815 & -3.4546 & -849.4032 & -6035.7133 & 0.0016 \\
25\% & 4.3689 & 4.1535 & 3.6428 & -2155.8238 & 4.1888 \\
50\% & 5.0164 & 4.8660 & 4.6285 & 3.5470 & 4.9607 \\
75\% & 5.5980 & 5.4187 & 5.1960 & 4.7422 & 5.6598 \\
95\% & 8.2659 & 6.9014 & 6.0589 & 5.5787 & 9.2917 \\
\bottomrule
\end{tabular}
\end{minipage}
\end{adjustbox}
\end{minipage}
\par\medskip
\begin{minipage}{\linewidth}
\footnotesize
\centering
\textit{Notes:} QN denotes the quasi-Newton method, with its starting value in parentheses; $\widehat{\beta}_{\mathrm{2SLS}}$ denotes the 2SLS estimator. Poly. denotes the polynomial method.
\end{minipage}
\end{table}
\end{landscape}
\clearpage

\clearpage
\begin{landscape}
\begin{table}[p]
\centering
\singlespacing
\fontsize{9}{11}\selectfont
\setlength{\tabcolsep}{2pt}
\renewcommand{\arraystretch}{1}
\captionsetup{font=small,justification=centering}
\caption{Numerical computation of CUE: quantiles for $w = \frac{\sqrt{2}}{2}$, $\rho = 0.3$.}

\smallskip
\begin{minipage}[t]{0.325\linewidth}
\centering
$\mu^2=1$\par\smallskip
\begin{adjustbox}{width=\linewidth}
\begin{minipage}{270pt}
\centering
\begin{tabular}{@{}cccccc@{}}
\toprule
Percentile & QN ($\widehat{\beta}_{\mathrm{2SLS}}$) & QN (5) & QN (4.5) & QN (0) & Poly. \\
\midrule
\multicolumn{6}{c}{$k=1$} \\
5\% & -0.2681 & -20.8976 & -838.0908 & -2556.8816 & -0.2681 \\
25\% & 4.2567 & 3.9881 & 3.3848 & -907.9822 & 4.2567 \\
50\% & 5.1550 & 5.0341 & 4.8402 & 3.5412 & 5.1550 \\
75\% & 6.0307 & 5.8122 & 5.5892 & 4.8877 & 6.0307 \\
95\% & 10.2617 & 7.8378 & 6.7403 & 5.4722 & 10.2617 \\
\addlinespace[3pt]
\multicolumn{6}{c}{$k=4$} \\
5\% & -4.8059 & -500.5649 & -1412.1490 & -3785.7130 & -5.4498 \\
25\% & 3.8641 & 3.3504 & 1.8916 & -1967.5312 & 3.6937 \\
50\% & 5.1701 & 4.9690 & 4.6074 & 1.1294 & 5.1688 \\
75\% & 6.3711 & 5.9495 & 5.5231 & 4.6027 & 6.5296 \\
95\% & 14.0680 & 10.6251 & 7.8713 & 5.7070 & 14.5327 \\
\addlinespace[3pt]
\multicolumn{6}{c}{$k=10$} \\
5\% & -5.0337 & -743.4390 & -1625.9753 & -4352.4932 & -5.6191 \\
25\% & 3.8806 & 3.1689 & 1.1570 & -2512.8884 & 3.6372 \\
50\% & 5.1826 & 4.8635 & 4.3650 & 0.3881 & 5.1702 \\
75\% & 6.4418 & 5.9287 & 5.4547 & 4.3403 & 6.6465 \\
95\% & 16.4927 & 9.6897 & 7.2266 & 5.6421 & 15.2419 \\
\addlinespace[3pt]
\multicolumn{6}{c}{$k=30$} \\
5\% & -16.1981 & -1052.2461 & -1974.7178 & -5312.7600 & -8.2798 \\
25\% & 3.7205 & 2.6587 & -2.3107 & -3300.1424 & 3.2914 \\
50\% & 5.2034 & 4.7471 & 3.9171 & -3.0647 & 5.2036 \\
75\% & 6.6827 & 5.9303 & 5.2188 & 3.7733 & 7.1585 \\
95\% & 21.4039 & 10.0744 & 7.1407 & 5.5032 & 18.2889 \\
\addlinespace[3pt]
\multicolumn{6}{c}{$k=60$} \\
5\% & -42.3204 & -1186.6732 & -2051.9008 & -6070.0185 & -10.2283 \\
25\% & 3.5334 & 1.6937 & -8.9685 & -3873.3940 & 2.7301 \\
50\% & 5.2516 & 4.5191 & 3.3153 & -6.8073 & 5.1823 \\
75\% & 6.9914 & 5.8721 & 4.9346 & 3.0311 & 7.7037 \\
95\% & 59.9347 & 11.0419 & 7.0321 & 5.3167 & 20.1033 \\
\bottomrule
\end{tabular}
\end{minipage}
\end{adjustbox}
\end{minipage}
\hfill
\begin{minipage}[t]{0.325\linewidth}
\centering
$\mu^2=4$\par\smallskip
\begin{adjustbox}{width=\linewidth}
\begin{minipage}{270pt}
\centering
\begin{tabular}{@{}cccccc@{}}
\toprule
Percentile & QN ($\widehat{\beta}_{\mathrm{2SLS}}$) & QN (5) & QN (4.5) & QN (0) & Poly. \\
\midrule
\multicolumn{6}{c}{$k=1$} \\
5\% & 2.9524 & 2.6311 & 1.3729 & -2846.7123 & 2.9524 \\
25\% & 4.5149 & 4.4838 & 4.4323 & 3.1397 & 4.5149 \\
50\% & 5.0194 & 5.0029 & 4.9710 & 4.6570 & 5.0194 \\
75\% & 5.4789 & 5.4531 & 5.4121 & 5.1158 & 5.4789 \\
95\% & 6.5453 & 6.4012 & 6.1987 & 5.5042 & 6.5453 \\
\addlinespace[3pt]
\multicolumn{6}{c}{$k=4$} \\
5\% & 0.0065 & -9.3532 & -721.0657 & -3843.7624 & -0.5530 \\
25\% & 4.3103 & 4.1671 & 3.8037 & -1205.0303 & 4.2598 \\
50\% & 5.0199 & 4.9606 & 4.8407 & 3.8754 & 5.0197 \\
75\% & 5.6422 & 5.5356 & 5.3985 & 4.9417 & 5.6906 \\
95\% & 9.3506 & 7.9518 & 6.8111 & 5.5974 & 10.5765 \\
\addlinespace[3pt]
\multicolumn{6}{c}{$k=10$} \\
5\% & -1.7284 & -123.8503 & -1199.0245 & -4389.7569 & -1.7928 \\
25\% & 4.2187 & 3.9634 & 3.2456 & -2082.6626 & 4.1226 \\
50\% & 5.0701 & 4.9480 & 4.7284 & 3.1610 & 5.0686 \\
75\% & 5.8594 & 5.6300 & 5.3986 & 4.8210 & 5.9760 \\
95\% & 10.7590 & 8.2718 & 6.7541 & 5.6133 & 12.2771 \\
\addlinespace[3pt]
\multicolumn{6}{c}{$k=30$} \\
5\% & -4.6389 & -665.0791 & -1664.0946 & -5388.3753 & -4.2490 \\
25\% & 3.9375 & 3.3329 & 1.3591 & -3076.2559 & 3.5633 \\
50\% & 5.1210 & 4.8250 & 4.2771 & 0.8286 & 5.0677 \\
75\% & 6.2055 & 5.7581 & 5.2664 & 4.2628 & 6.5389 \\
95\% & 14.6512 & 9.0831 & 6.9213 & 5.5635 & 14.7236 \\
\addlinespace[3pt]
\multicolumn{6}{c}{$k=60$} \\
5\% & -20.9617 & -997.6856 & -1893.4834 & -6127.8645 & -8.0214 \\
25\% & 3.7012 & 2.4473 & -2.2816 & -3646.6568 & 3.0188 \\
50\% & 5.1706 & 4.6474 & 3.7024 & -2.3877 & 5.0779 \\
75\% & 6.5525 & 5.7903 & 5.0191 & 3.5087 & 7.2712 \\
95\% & 31.0183 & 10.6977 & 6.8138 & 5.4288 & 18.9836 \\
\bottomrule
\end{tabular}
\end{minipage}
\end{adjustbox}
\end{minipage}
\hfill
\begin{minipage}[t]{0.325\linewidth}
\centering
$\mu^2=16$\par\smallskip
\begin{adjustbox}{width=\linewidth}
\begin{minipage}{270pt}
\centering
\begin{tabular}{@{}cccccc@{}}
\toprule
Percentile & QN ($\widehat{\beta}_{\mathrm{2SLS}}$) & QN (5) & QN (4.5) & QN (0) & Poly. \\
\midrule
\multicolumn{6}{c}{$k=1$} \\
5\% & 4.3086 & 4.3086 & 4.3086 & 3.9573 & 4.3086 \\
25\% & 4.7669 & 4.7669 & 4.7669 & 4.7175 & 4.7669 \\
50\% & 5.0071 & 5.0071 & 5.0071 & 4.9765 & 5.0071 \\
75\% & 5.2291 & 5.2291 & 5.2291 & 5.1946 & 5.2291 \\
95\% & 5.5614 & 5.5614 & 5.5614 & 5.4716 & 5.5614 \\
\addlinespace[3pt]
\multicolumn{6}{c}{$k=4$} \\
5\% & 3.9307 & 3.8756 & 3.6918 & -3712.1504 & 3.8002 \\
25\% & 4.7111 & 4.7071 & 4.6858 & 4.3977 & 4.7010 \\
50\% & 4.9946 & 4.9917 & 4.9813 & 4.8742 & 4.9915 \\
75\% & 5.2426 & 5.2384 & 5.2258 & 5.1298 & 5.2425 \\
95\% & 5.7302 & 5.7007 & 5.6385 & 5.4410 & 5.7749 \\
\addlinespace[3pt]
\multicolumn{6}{c}{$k=10$} \\
5\% & 3.5217 & 3.3866 & 2.7808 & -4285.9864 & 3.3191 \\
25\% & 4.6230 & 4.6021 & 4.5608 & 3.8374 & 4.6073 \\
50\% & 4.9794 & 4.9678 & 4.9380 & 4.7616 & 4.9754 \\
75\% & 5.2854 & 5.2714 & 5.2407 & 5.1037 & 5.2903 \\
95\% & 6.1206 & 5.9639 & 5.8008 & 5.5075 & 6.3926 \\
\addlinespace[3pt]
\multicolumn{6}{c}{$k=30$} \\
5\% & 1.9179 & 0.7932 & -12.8626 & -5484.6859 & 1.0447 \\
25\% & 4.3971 & 4.2834 & 4.0300 & -6.5630 & 4.2839 \\
50\% & 4.9783 & 4.9138 & 4.7969 & 4.1538 & 4.9533 \\
75\% & 5.4677 & 5.3845 & 5.2707 & 4.9514 & 5.4923 \\
95\% & 7.4006 & 6.7643 & 6.1763 & 5.6199 & 8.5635 \\
\addlinespace[3pt]
\multicolumn{6}{c}{$k=60$} \\
5\% & -0.4044 & -11.5629 & -1052.0954 & -6237.9162 & -2.2992 \\
25\% & 4.1143 & 3.8184 & 3.0645 & -2711.7511 & 3.7711 \\
50\% & 4.9983 & 4.8294 & 4.4645 & 2.5145 & 4.9170 \\
75\% & 5.7123 & 5.4945 & 5.1809 & 4.4938 & 5.9005 \\
95\% & 9.4962 & 7.6294 & 6.3898 & 5.6608 & 12.0871 \\
\bottomrule
\end{tabular}
\end{minipage}
\end{adjustbox}
\end{minipage}
\par\medskip
\begin{minipage}{\linewidth}
\footnotesize
\centering
\textit{Notes:} QN denotes the quasi-Newton method, with its starting value in parentheses; $\widehat{\beta}_{\mathrm{2SLS}}$ denotes the 2SLS estimator. Poly. denotes the polynomial method.
\end{minipage}
\end{table}
\end{landscape}
\clearpage

\clearpage
\begin{landscape}
\begin{table}[p]
\centering
\singlespacing
\fontsize{9}{11}\selectfont
\setlength{\tabcolsep}{2pt}
\renewcommand{\arraystretch}{1}
\captionsetup{font=small,justification=centering}
\caption{Numerical computation of CUE: quantiles for $w = \frac{\sqrt{3}}{2}$, $\rho = 0.3$.}

\smallskip
\begin{minipage}[t]{0.325\linewidth}
\centering
$\mu^2=1$\par\smallskip
\begin{adjustbox}{width=\linewidth}
\begin{minipage}{270pt}
\centering
\begin{tabular}{@{}cccccc@{}}
\toprule
Percentile & QN ($\widehat{\beta}_{\mathrm{2SLS}}$) & QN (5) & QN (4.5) & QN (0) & Poly. \\
\midrule
\multicolumn{6}{c}{$k=1$} \\
5\% & -1.0667 & -27.3173 & -795.8325 & -2664.0130 & -1.0667 \\
25\% & 4.0941 & 3.7966 & 3.1794 & -873.3476 & 4.0941 \\
50\% & 5.1137 & 4.9730 & 4.7909 & 3.4157 & 5.1137 \\
75\% & 6.0907 & 5.8634 & 5.6471 & 4.8487 & 6.0907 \\
95\% & 11.0337 & 8.2980 & 7.0350 & 5.5353 & 11.0337 \\
\addlinespace[3pt]
\multicolumn{6}{c}{$k=4$} \\
5\% & -6.2459 & -714.9894 & -1443.0229 & -3971.1348 & -6.7998 \\
25\% & 3.7562 & 3.1512 & 1.5074 & -1954.8775 & 3.5363 \\
50\% & 5.1823 & 4.9377 & 4.5476 & 0.5961 & 5.1695 \\
75\% & 6.6310 & 6.0675 & 5.6129 & 4.4973 & 6.8567 \\
95\% & 15.9053 & 12.2751 & 8.3990 & 5.8215 & 16.9404 \\
\addlinespace[3pt]
\multicolumn{6}{c}{$k=10$} \\
5\% & -6.2598 & -984.0132 & -1934.5223 & -4474.7553 & -6.1630 \\
25\% & 3.7308 & 2.9483 & 0.3152 & -2561.2410 & 3.4654 \\
50\% & 5.2197 & 4.8890 & 4.3123 & -0.3893 & 5.2217 \\
75\% & 6.7594 & 6.1540 & 5.5031 & 4.2695 & 7.0855 \\
95\% & 17.8100 & 11.5297 & 7.8361 & 5.7220 & 17.8758 \\
\addlinespace[3pt]
\multicolumn{6}{c}{$k=30$} \\
5\% & -18.1434 & -1131.7247 & -2006.6702 & -5486.4425 & -9.3630 \\
25\% & 3.4618 & 2.1206 & -2.8437 & -3286.2600 & 2.9205 \\
50\% & 5.2164 & 4.7076 & 3.8310 & -4.5205 & 5.2234 \\
75\% & 6.8653 & 6.0539 & 5.2736 & 3.3966 & 7.5100 \\
95\% & 26.5636 & 12.2771 & 7.6348 & 5.5358 & 20.5552 \\
\addlinespace[3pt]
\multicolumn{6}{c}{$k=60$} \\
5\% & -256.5906 & -1297.8767 & -2089.1581 & -6188.8803 & -14.9936 \\
25\% & 3.1202 & 0.7945 & -14.5486 & -3854.9883 & 2.0829 \\
50\% & 5.2504 & 4.4341 & 2.8817 & -9.3727 & 5.2261 \\
75\% & 7.3054 & 5.9434 & 4.8763 & 2.4089 & 8.4796 \\
95\% & 142.9584 & 13.3399 & 7.4061 & 5.3068 & 25.8023 \\
\bottomrule
\end{tabular}
\end{minipage}
\end{adjustbox}
\end{minipage}
\hfill
\begin{minipage}[t]{0.325\linewidth}
\centering
$\mu^2=4$\par\smallskip
\begin{adjustbox}{width=\linewidth}
\begin{minipage}{270pt}
\centering
\begin{tabular}{@{}cccccc@{}}
\toprule
Percentile & QN ($\widehat{\beta}_{\mathrm{2SLS}}$) & QN (5) & QN (4.5) & QN (0) & Poly. \\
\midrule
\multicolumn{6}{c}{$k=1$} \\
5\% & 2.8367 & 2.5297 & 1.8835 & -2987.8370 & 2.8367 \\
25\% & 4.4644 & 4.4370 & 4.3994 & 3.1073 & 4.4644 \\
50\% & 5.0190 & 5.0015 & 4.9802 & 4.6292 & 5.0190 \\
75\% & 5.5252 & 5.5054 & 5.4761 & 5.1267 & 5.5252 \\
95\% & 6.6459 & 6.5105 & 6.3527 & 5.5679 & 6.6459 \\
\addlinespace[3pt]
\multicolumn{6}{c}{$k=4$} \\
5\% & -0.7410 & -6.9656 & -643.2014 & -4141.7715 & -1.8107 \\
25\% & 4.2961 & 4.1621 & 3.7944 & -1088.6673 & 4.2200 \\
50\% & 5.0372 & 4.9724 & 4.8532 & 3.8107 & 5.0337 \\
75\% & 5.6903 & 5.5684 & 5.4340 & 4.9454 & 5.7473 \\
95\% & 10.1152 & 8.4563 & 7.1934 & 5.7109 & 11.3894 \\
\addlinespace[3pt]
\multicolumn{6}{c}{$k=10$} \\
5\% & -3.0339 & -52.8828 & -1280.8910 & -4598.8296 & -3.7110 \\
25\% & 4.1105 & 3.7992 & 3.0305 & -2190.3059 & 3.9714 \\
50\% & 5.0775 & 4.9477 & 4.7265 & 2.5355 & 5.0831 \\
75\% & 5.9725 & 5.7075 & 5.4454 & 4.7635 & 6.1471 \\
95\% & 10.9493 & 9.0127 & 7.2763 & 5.6985 & 13.3830 \\
\addlinespace[3pt]
\multicolumn{6}{c}{$k=30$} \\
5\% & -8.3610 & -748.5205 & -1720.1454 & -5610.8107 & -6.5813 \\
25\% & 3.7584 & 3.0973 & 0.9791 & -3146.9703 & 3.3671 \\
50\% & 5.1176 & 4.8019 & 4.2236 & -0.2783 & 5.0878 \\
75\% & 6.4053 & 5.8596 & 5.3267 & 4.0146 & 6.8912 \\
95\% & 16.4428 & 10.3591 & 7.3515 & 5.5705 & 17.1771 \\
\addlinespace[3pt]
\multicolumn{6}{c}{$k=60$} \\
5\% & -26.6961 & -1139.8928 & -1921.3370 & -6307.4410 & -12.5147 \\
25\% & 3.3595 & 1.7715 & -4.1245 & -3782.4442 & 2.3503 \\
50\% & 5.1806 & 4.5590 & 3.4012 & -4.6064 & 5.0382 \\
75\% & 6.8319 & 5.8832 & 5.0147 & 2.9958 & 7.7783 \\
95\% & 46.0663 & 12.7679 & 7.3960 & 5.3878 & 22.2013 \\
\bottomrule
\end{tabular}
\end{minipage}
\end{adjustbox}
\end{minipage}
\hfill
\begin{minipage}[t]{0.325\linewidth}
\centering
$\mu^2=16$\par\smallskip
\begin{adjustbox}{width=\linewidth}
\begin{minipage}{270pt}
\centering
\begin{tabular}{@{}cccccc@{}}
\toprule
Percentile & QN ($\widehat{\beta}_{\mathrm{2SLS}}$) & QN (5) & QN (4.5) & QN (0) & Poly. \\
\midrule
\multicolumn{6}{c}{$k=1$} \\
5\% & 4.2489 & 4.2489 & 4.2489 & 3.6949 & 4.2489 \\
25\% & 4.7260 & 4.7260 & 4.7260 & 4.6719 & 4.7260 \\
50\% & 5.0010 & 5.0010 & 5.0010 & 4.9615 & 5.0010 \\
75\% & 5.2576 & 5.2576 & 5.2576 & 5.2119 & 5.2576 \\
95\% & 5.6475 & 5.6475 & 5.6475 & 5.5192 & 5.6475 \\
\addlinespace[3pt]
\multicolumn{6}{c}{$k=4$} \\
5\% & 3.7574 & 3.7440 & 3.4564 & -3899.3751 & 3.6038 \\
25\% & 4.6877 & 4.6855 & 4.6626 & 4.3503 & 4.6801 \\
50\% & 4.9932 & 4.9912 & 4.9803 & 4.8632 & 4.9919 \\
75\% & 5.2532 & 5.2495 & 5.2343 & 5.1306 & 5.2563 \\
95\% & 5.7789 & 5.7489 & 5.6873 & 5.4763 & 5.8532 \\
\addlinespace[3pt]
\multicolumn{6}{c}{$k=10$} \\
5\% & 3.1827 & 3.0305 & 2.2723 & -4470.8702 & 2.7290 \\
25\% & 4.6297 & 4.6161 & 4.5683 & 3.4043 & 4.6122 \\
50\% & 4.9926 & 4.9848 & 4.9592 & 4.7556 & 4.9886 \\
75\% & 5.3095 & 5.2986 & 5.2687 & 5.1070 & 5.3146 \\
95\% & 6.2531 & 6.1188 & 5.8876 & 5.5490 & 6.5849 \\
\addlinespace[3pt]
\multicolumn{6}{c}{$k=30$} \\
5\% & 1.3156 & -0.0323 & -21.2624 & -5687.1535 & 0.3391 \\
25\% & 4.2970 & 4.1845 & 3.8486 & -14.8480 & 4.1646 \\
50\% & 4.9654 & 4.9017 & 4.7666 & 3.9156 & 4.9354 \\
75\% & 5.5135 & 5.4236 & 5.2754 & 4.8971 & 5.5628 \\
95\% & 7.8421 & 7.0796 & 6.4453 & 5.6515 & 9.5115 \\
\addlinespace[3pt]
\multicolumn{6}{c}{$k=60$} \\
5\% & -3.2530 & -103.0675 & -1152.4049 & -6396.8204 & -4.4382 \\
25\% & 3.9168 & 3.5481 & 2.6287 & -2756.4147 & 3.4736 \\
50\% & 4.9629 & 4.7534 & 4.3617 & 1.7949 & 4.8645 \\
75\% & 5.7684 & 5.5003 & 5.1564 & 4.3085 & 6.0917 \\
95\% & 10.7700 & 8.1033 & 6.5189 & 5.7488 & 13.6556 \\
\bottomrule
\end{tabular}
\end{minipage}
\end{adjustbox}
\end{minipage}
\par\medskip
\begin{minipage}{\linewidth}
\footnotesize
\centering
\textit{Notes:} QN denotes the quasi-Newton method, with its starting value in parentheses; $\widehat{\beta}_{\mathrm{2SLS}}$ denotes the 2SLS estimator. Poly. denotes the polynomial method.
\end{minipage}
\end{table}
\end{landscape}
\clearpage

\clearpage
\begin{landscape}
\begin{table}[p]
\centering
\singlespacing
\fontsize{9}{11}\selectfont
\setlength{\tabcolsep}{2pt}
\renewcommand{\arraystretch}{1}
\captionsetup{font=small,justification=centering}
\caption{Numerical computation of CUE: quantiles for $w = 1$, $\rho = 0.3$.}

\smallskip
\input{tabs_v10a/tabquant2_w5rho3_panels}
\par\medskip
\begin{minipage}{\linewidth}
\footnotesize
\centering
\textit{Notes:} QN denotes the quasi-Newton method, with its starting value in parentheses; $\widehat{\beta}_{\mathrm{2SLS}}$ denotes the 2SLS estimator. Poly. denotes the polynomial method.
\end{minipage}
\end{table}
\end{landscape}
\clearpage

\clearpage
\begin{landscape}
\begin{table}[p]
\centering
\singlespacing
\fontsize{9}{11}\selectfont
\setlength{\tabcolsep}{2pt}
\renewcommand{\arraystretch}{1}
\captionsetup{font=small,justification=centering}
\caption{Numerical computation of CUE: quantiles for $w = 0$, $\rho = 0.6$.}

\smallskip
\input{tabs_v10a/tabquant2_w1rho6_panels}
\par\medskip
\begin{minipage}{\linewidth}
\footnotesize
\centering
\textit{Notes:} QN denotes the quasi-Newton method, with its starting value in parentheses; $\widehat{\beta}_{\mathrm{2SLS}}$ denotes the 2SLS estimator. Poly. denotes the polynomial method.
\end{minipage}
\end{table}
\end{landscape}
\clearpage

\clearpage
\begin{landscape}
\begin{table}[p]
\centering
\singlespacing
\fontsize{9}{11}\selectfont
\setlength{\tabcolsep}{2pt}
\renewcommand{\arraystretch}{1}
\captionsetup{font=small,justification=centering}
\caption{Numerical computation of CUE: quantiles for $w = \frac 1 2$, $\rho = 0.6$.}

\smallskip
\input{tabs_v10a/tabquant2_w2rho6_panels}
\par\medskip
\begin{minipage}{\linewidth}
\footnotesize
\centering
\textit{Notes:} QN denotes the quasi-Newton method, with its starting value in parentheses; $\widehat{\beta}_{\mathrm{2SLS}}$ denotes the 2SLS estimator. Poly. denotes the polynomial method.
\end{minipage}
\end{table}
\end{landscape}
\clearpage

\clearpage
\begin{landscape}
\begin{table}[p]
\centering
\singlespacing
\fontsize{9}{11}\selectfont
\setlength{\tabcolsep}{2pt}
\renewcommand{\arraystretch}{1}
\captionsetup{font=small,justification=centering}
\caption{Numerical computation of CUE: quantiles for $w = \frac{\sqrt{2}}{2}$, $\rho = 0.6$.}

\smallskip
\input{tabs_v10a/tabquant2_w3rho6_panels}
\par\medskip
\begin{minipage}{\linewidth}
\footnotesize
\centering
\textit{Notes:} QN denotes the quasi-Newton method, with its starting value in parentheses; $\widehat{\beta}_{\mathrm{2SLS}}$ denotes the 2SLS estimator. Poly. denotes the polynomial method.
\end{minipage}
\end{table}
\end{landscape}
\clearpage

\clearpage
\begin{landscape}
\begin{table}[p]
\centering
\singlespacing
\fontsize{9}{11}\selectfont
\setlength{\tabcolsep}{2pt}
\renewcommand{\arraystretch}{1}
\captionsetup{font=small,justification=centering}
\caption{Numerical computation of CUE: quantiles for $w = \frac{\sqrt{3}}{2}$, $\rho = 0.6$.}

\smallskip
\input{tabs_v10a/tabquant2_w4rho6_panels}
\par\medskip
\begin{minipage}{\linewidth}
\footnotesize
\centering
\textit{Notes:} QN denotes the quasi-Newton method, with its starting value in parentheses; $\widehat{\beta}_{\mathrm{2SLS}}$ denotes the 2SLS estimator. Poly. denotes the polynomial method.
\end{minipage}
\end{table}
\end{landscape}
\clearpage

\clearpage
\begin{landscape}
\begin{table}[p]
\centering
\singlespacing
\fontsize{9}{11}\selectfont
\setlength{\tabcolsep}{2pt}
\renewcommand{\arraystretch}{1}
\captionsetup{font=small,justification=centering}
\caption{Numerical computation of CUE: quantiles for $w = 1$, $\rho = 0.6$.}

\smallskip
\input{tabs_v10a/tabquant2_w5rho6_panels}
\par\medskip
\begin{minipage}{\linewidth}
\footnotesize
\centering
\textit{Notes:} QN denotes the quasi-Newton method, with its starting value in parentheses; $\widehat{\beta}_{\mathrm{2SLS}}$ denotes the 2SLS estimator. Poly. denotes the polynomial method.
\end{minipage}
\end{table}
\end{landscape}
\clearpage

\clearpage
\begin{landscape}
\begin{table}[p]
\centering
\singlespacing
\fontsize{9}{11}\selectfont
\setlength{\tabcolsep}{2pt}
\renewcommand{\arraystretch}{1}
\captionsetup{font=small,justification=centering}
\caption{Numerical computation of CUE: quantiles for $w = 0$, $\rho = 0.9$.}

\smallskip
\input{tabs_v10a/tabquant2_w1rho9_panels}
\par\medskip
\begin{minipage}{\linewidth}
\footnotesize
\centering
\textit{Notes:} QN denotes the quasi-Newton method, with its starting value in parentheses; $\widehat{\beta}_{\mathrm{2SLS}}$ denotes the 2SLS estimator. Poly. denotes the polynomial method.
\end{minipage}
\end{table}
\end{landscape}
\clearpage

\clearpage
\begin{landscape}
\begin{table}[p]
\centering
\singlespacing
\fontsize{9}{11}\selectfont
\setlength{\tabcolsep}{2pt}
\renewcommand{\arraystretch}{1}
\captionsetup{font=small,justification=centering}
\caption{Numerical computation of CUE: quantiles for $w = \frac 1 2$, $\rho = 0.9$.}

\smallskip
\input{tabs_v10a/tabquant2_w2rho9_panels}
\par\medskip
\begin{minipage}{\linewidth}
\footnotesize
\centering
\textit{Notes:} QN denotes the quasi-Newton method, with its starting value in parentheses; $\widehat{\beta}_{\mathrm{2SLS}}$ denotes the 2SLS estimator. Poly. denotes the polynomial method.
\end{minipage}
\end{table}
\end{landscape}
\clearpage

\clearpage
\begin{landscape}
\begin{table}[p]
\centering
\singlespacing
\fontsize{9}{11}\selectfont
\setlength{\tabcolsep}{2pt}
\renewcommand{\arraystretch}{1}
\captionsetup{font=small,justification=centering}
\caption{Numerical computation of CUE: quantiles for $w = \frac{\sqrt{2}}{2}$, $\rho = 0.9$.}

\smallskip
\input{tabs_v10a/tabquant2_w3rho9_panels}
\par\medskip
\begin{minipage}{\linewidth}
\footnotesize
\centering
\textit{Notes:} QN denotes the quasi-Newton method, with its starting value in parentheses; $\widehat{\beta}_{\mathrm{2SLS}}$ denotes the 2SLS estimator. Poly. denotes the polynomial method.
\end{minipage}
\end{table}
\end{landscape}
\clearpage

\clearpage
\begin{landscape}
\begin{table}[p]
\centering
\singlespacing
\fontsize{9}{11}\selectfont
\setlength{\tabcolsep}{2pt}
\renewcommand{\arraystretch}{1}
\captionsetup{font=small,justification=centering}
\caption{Numerical computation of CUE: quantiles for $w = \frac{\sqrt{3}}{2}$, $\rho = 0.9$.}

\smallskip
\input{tabs_v10a/tabquant2_w4rho9_panels}
\par\medskip
\begin{minipage}{\linewidth}
\footnotesize
\centering
\textit{Notes:} QN denotes the quasi-Newton method, with its starting value in parentheses; $\widehat{\beta}_{\mathrm{2SLS}}$ denotes the 2SLS estimator. Poly. denotes the polynomial method.
\end{minipage}
\end{table}
\end{landscape}
\clearpage

\clearpage
\begin{landscape}
\begin{table}[p]
\centering
\singlespacing
\fontsize{9}{11}\selectfont
\setlength{\tabcolsep}{2pt}
\renewcommand{\arraystretch}{1}
\captionsetup{font=small,justification=centering}
\caption{Numerical computation of CUE: quantiles for $w = 1$, $\rho = 0.9$.}

\smallskip
\input{tabs_v10a/tabquant2_w5rho9_panels}
\par\medskip
\begin{minipage}{\linewidth}
\footnotesize
\centering
\textit{Notes:} QN denotes the quasi-Newton method, with its starting value in parentheses; $\widehat{\beta}_{\mathrm{2SLS}}$ denotes the 2SLS estimator. Poly. denotes the polynomial method.
\end{minipage}
\end{table}
\end{landscape}
\clearpage

\clearpage
\begin{landscape}
\begin{table}[p]
\centering
\singlespacing
\fontsize{9}{11}\selectfont
\setlength{\tabcolsep}{2pt}
\renewcommand{\arraystretch}{1}
\captionsetup{font=small,justification=centering}
\caption{Comparison of estimators: quantiles for $w = 0$, $\rho = 0.3$.}
\smallskip
\input{tabs_v10a/tabquant1_w1rho3_panels}
\par\medskip
\begin{minipage}{\linewidth}
\footnotesize
\centering
\textit{Notes:} Two-step and Iterated denote two-step and iterated GMM. CUE is computed using the polynomial method.
\end{minipage}
\end{table}
\end{landscape}
\clearpage

\clearpage
\begin{landscape}
\begin{table}[p]
\centering
\singlespacing
\fontsize{9}{11}\selectfont
\setlength{\tabcolsep}{2pt}
\renewcommand{\arraystretch}{1}
\captionsetup{font=small,justification=centering}
\caption{Comparison of estimators: quantiles for $w = \frac 1 2$, $\rho = 0.3$.}

\smallskip
\input{tabs_v10a/tabquant1_w2rho3_panels}
\par\medskip
\begin{minipage}{\linewidth}
\footnotesize
\centering
\textit{Notes:} Two-step and Iterated denote two-step and iterated GMM. CUE is computed using the polynomial method.
\end{minipage}
\end{table}
\end{landscape}
\clearpage

\clearpage
\begin{landscape}
\begin{table}[p]
\centering
\singlespacing
\fontsize{9}{11}\selectfont
\setlength{\tabcolsep}{2pt}
\renewcommand{\arraystretch}{1}
\captionsetup{font=small,justification=centering}
\caption{Comparison of estimators: quantiles for $w = \frac{\sqrt{2}}{2}$, $\rho = 0.3$.}

\smallskip
\input{tabs_v10a/tabquant1_w3rho3_panels}
\par\medskip
\begin{minipage}{\linewidth}
\footnotesize
\centering
\textit{Notes:} Two-step and Iterated denote two-step and iterated GMM. CUE is computed using the polynomial method.
\end{minipage}
\end{table}
\end{landscape}
\clearpage

\clearpage
\begin{landscape}
\begin{table}[p]
\centering
\singlespacing
\fontsize{9}{11}\selectfont
\setlength{\tabcolsep}{2pt}
\renewcommand{\arraystretch}{1}
\captionsetup{font=small,justification=centering}
\caption{Comparison of estimators: quantiles for $w = \frac{\sqrt{3}}{2}$, $\rho = 0.3$.}

\smallskip
\input{tabs_v10a/tabquant1_w4rho3_panels}
\par\medskip
\begin{minipage}{\linewidth}
\footnotesize
\centering
\textit{Notes:} Two-step and Iterated denote two-step and iterated GMM. CUE is computed using the polynomial method.
\end{minipage}
\end{table}
\end{landscape}
\clearpage

\clearpage
\begin{landscape}
\begin{table}[p]
\centering
\singlespacing
\fontsize{9}{11}\selectfont
\setlength{\tabcolsep}{2pt}
\renewcommand{\arraystretch}{1}
\captionsetup{font=small,justification=centering}
\caption{Comparison of estimators: quantiles for $w = 1$, $\rho = 0.3$.}

\smallskip
\input{tabs_v10a/tabquant1_w5rho3_panels}
\par\medskip
\begin{minipage}{\linewidth}
\footnotesize
\centering
\textit{Notes:} Two-step and Iterated denote two-step and iterated GMM. CUE is computed using the polynomial method.
\end{minipage}
\end{table}
\end{landscape}
\clearpage

\clearpage
\begin{landscape}
\begin{table}[p]
\centering
\singlespacing
\fontsize{9}{11}\selectfont
\setlength{\tabcolsep}{2pt}
\renewcommand{\arraystretch}{1}
\captionsetup{font=small,justification=centering}
\caption{Comparison of estimators: quantiles for $w = 0$, $\rho = 0.6$.}

\smallskip
\input{tabs_v10a/tabquant1_w1rho6_panels}
\par\medskip
\begin{minipage}{\linewidth}
\footnotesize
\centering
\textit{Notes:} Two-step and Iterated denote two-step and iterated GMM. CUE is computed using the polynomial method.
\end{minipage}
\end{table}
\end{landscape}
\clearpage

\clearpage
\begin{landscape}
\begin{table}[p]
\centering
\singlespacing
\fontsize{9}{11}\selectfont
\setlength{\tabcolsep}{2pt}
\renewcommand{\arraystretch}{1}
\captionsetup{font=small,justification=centering}
\caption{Comparison of estimators: quantiles for $w = \frac 1 2$, $\rho = 0.6$.}

\smallskip
\input{tabs_v10a/tabquant1_w2rho6_panels}
\par\medskip
\begin{minipage}{\linewidth}
\footnotesize
\centering
\textit{Notes:} Two-step and Iterated denote two-step and iterated GMM. CUE is computed using the polynomial method.
\end{minipage}
\end{table}
\end{landscape}
\clearpage

\clearpage
\begin{landscape}
\begin{table}[p]
\centering
\singlespacing
\fontsize{9}{11}\selectfont
\setlength{\tabcolsep}{2pt}
\renewcommand{\arraystretch}{1}
\captionsetup{font=small,justification=centering}
\caption{Comparison of estimators: quantiles for $w = \frac{\sqrt{2}}{2}$, $\rho = 0.6$.}

\smallskip
\input{tabs_v10a/tabquant1_w3rho6_panels}
\par\medskip
\begin{minipage}{\linewidth}
\footnotesize
\centering
\textit{Notes:} Two-step and Iterated denote two-step and iterated GMM. CUE is computed using the polynomial method.
\end{minipage}
\end{table}
\end{landscape}
\clearpage

\clearpage
\begin{landscape}
\begin{table}[p]
\centering
\singlespacing
\fontsize{9}{11}\selectfont
\setlength{\tabcolsep}{2pt}
\renewcommand{\arraystretch}{1}
\captionsetup{font=small,justification=centering}
\caption{Comparison of estimators: quantiles for $w = \frac{\sqrt{3}}{2}$, $\rho = 0.6$.}

\smallskip
\input{tabs_v10a/tabquant1_w4rho6_panels}
\par\medskip
\begin{minipage}{\linewidth}
\footnotesize
\centering
\textit{Notes:} Two-step and Iterated denote two-step and iterated GMM. CUE is computed using the polynomial method.
\end{minipage}
\end{table}
\end{landscape}
\clearpage

\clearpage
\begin{landscape}
\begin{table}[p]
\centering
\singlespacing
\fontsize{9}{11}\selectfont
\setlength{\tabcolsep}{2pt}
\renewcommand{\arraystretch}{1}
\captionsetup{font=small,justification=centering}
\caption{Comparison of estimators: quantiles for $w = 1$, $\rho = 0.6$.}

\smallskip
\input{tabs_v10a/tabquant1_w5rho6_panels}
\par\medskip
\begin{minipage}{\linewidth}
\footnotesize
\centering
\textit{Notes:} Two-step and Iterated denote two-step and iterated GMM. CUE is computed using the polynomial method.
\end{minipage}
\end{table}
\end{landscape}
\clearpage

\clearpage
\begin{landscape}
\begin{table}[p]
\centering
\singlespacing
\fontsize{9}{11}\selectfont
\setlength{\tabcolsep}{2pt}
\renewcommand{\arraystretch}{1}
\captionsetup{font=small,justification=centering}
\caption{Comparison of estimators: quantiles for $w = 0$, $\rho = 0.9$.}

\smallskip
\input{tabs_v10a/tabquant1_w1rho9_panels}
\par\medskip
\begin{minipage}{\linewidth}
\footnotesize
\centering
\textit{Notes:} Two-step and Iterated denote two-step and iterated GMM. CUE is computed using the polynomial method.
\end{minipage}
\end{table}
\end{landscape}
\clearpage

\clearpage
\begin{landscape}
\begin{table}[p]
\centering
\singlespacing
\fontsize{9}{11}\selectfont
\setlength{\tabcolsep}{2pt}
\renewcommand{\arraystretch}{1}
\captionsetup{font=small,justification=centering}
\caption{Comparison of estimators: quantiles for $w = \frac 1 2$, $\rho = 0.9$.}

\smallskip
\input{tabs_v10a/tabquant1_w2rho9_panels}
\par\medskip
\begin{minipage}{\linewidth}
\footnotesize
\centering
\textit{Notes:} Two-step and Iterated denote two-step and iterated GMM. CUE is computed using the polynomial method.
\end{minipage}
\end{table}
\end{landscape}
\clearpage

\clearpage
\begin{landscape}
\begin{table}[p]
\centering
\singlespacing
\fontsize{9}{11}\selectfont
\setlength{\tabcolsep}{2pt}
\renewcommand{\arraystretch}{1}
\captionsetup{font=small,justification=centering}
\caption{Comparison of estimators: quantiles for $w = \frac{\sqrt{2}}{2}$, $\rho = 0.9$.}

\smallskip
\input{tabs_v10a/tabquant1_w3rho9_panels}
\par\medskip
\begin{minipage}{\linewidth}
\footnotesize
\centering
\textit{Notes:} Two-step and Iterated denote two-step and iterated GMM. CUE is computed using the polynomial method.
\end{minipage}
\end{table}
\end{landscape}
\clearpage

\clearpage
\begin{landscape}
\begin{table}[p]
\centering
\singlespacing
\fontsize{9}{11}\selectfont
\setlength{\tabcolsep}{2pt}
\renewcommand{\arraystretch}{1}
\captionsetup{font=small,justification=centering}
\caption{Comparison of estimators: quantiles for $w = \frac{\sqrt{3}}{2}$, $\rho = 0.9$.}

\smallskip
\input{tabs_v10a/tabquant1_w4rho9_panels}
\par\medskip
\begin{minipage}{\linewidth}
\footnotesize
\centering
\textit{Notes:} Two-step and Iterated denote two-step and iterated GMM. CUE is computed using the polynomial method.
\end{minipage}
\end{table}
\end{landscape}
\clearpage

\clearpage
\begin{landscape}
\begin{table}[p]
\centering
\singlespacing
\fontsize{9}{11}\selectfont
\setlength{\tabcolsep}{2pt}
\renewcommand{\arraystretch}{1}
\captionsetup{font=small,justification=centering}
\caption{Comparison of estimators: quantiles for $w = 1$, $\rho = 0.9$.}
\label{quantile_end}
\smallskip
\input{tabs_v10a/tabquant1_w5rho9_panels}
\par\medskip
\begin{minipage}{\linewidth}
\footnotesize
\centering
\textit{Notes:} Two-step and Iterated denote two-step and iterated GMM. CUE is computed using the polynomial method.
\end{minipage}
\end{table}
\end{landscape}
\clearpage

\end{document}